\documentclass[preprint,12pt,authoryear]{elsarticle}
\usepackage{amssymb,amsfonts,amsthm}
\usepackage{mathrsfs}
\usepackage{geometry}
\usepackage{graphics,subfigure}
\usepackage{booktabs}
\usepackage{hyperref}
\usepackage{pifont}
\usepackage{float}
\usepackage{natbib}
\usepackage[fleqn]{amsmath}
\usepackage{color}
\usepackage{lineno}
\usepackage{bm}
\numberwithin{equation}{section}
\journal{}

\begin{document}
	\title{Impact waves in soft bilayer tissues}

	\author[add1]{Zhenwei Liu}
    \author[add1,add2]{Yang Liu\corref{cor1}}
    \ead{tracy\_liu@tju.edu.cn}
    \author[add3]{Michel Destrade}
    \author[add1]{Yue-Sheng Wang}

	\address[add1]{Department of Mechanics, School of Mechanical Engineering, Tianjin University, Tianjin 300350, China}
    \address[add2]{National Key Laboratory of Vehicle Power System, Tianjin 300350, China}
    \address[add3]{School of Mathematical and Statistical Sciences, University of Galway, University Road, Galway, H91 TK33, Ireland}
	\cortext[cor1]{Corresponding author.}
	

\begin{abstract}
Elastography has emerged as a promising approach for the non-invasive mechanical characterization of soft tissues, with increasing attention devoted to layered tissues through transient-wave measurements. Understanding how multi-layering, pre-stress, and wave dispersion govern transient wavefields is therefore essential for interpreting measured wave responses. In this study, we investigate impact wave propagation in a pre-stressed compressible hyperelastic bilayer resting on a frictionless rigid substrate within the framework of nonlinear elasticity. Semi-analytical solutions for the transient displacement fields induced by localized surface impulses are derived using the Fourier–Laplace transform, and their long-time asymptotic behavior is obtained through the method of stationary phase. Finite element simulations are further performed to validate the theoretical predictions. The results establish a direct connection between characteristic features of the transient displacement fields and the group velocities of the guided-wave modes, revealing how dispersive propagation and modal contributions shape the observed waveforms. Three representative bilayer models corresponding to human articular cartilage, gingiva, and skin are then considered to examine the effects of variations in layer stiffness and thickness. The results show that structural and material variations can substantially modify the spatial distribution, arrival positions, and characteristic velocities of impact waves. In particular, higher-order modes can make significant contributions to the displacement response in certain parameter regimes, such that the dominant features of the wavefield cannot always be characterized by the fundamental mode alone. These findings demonstrate that transient-wave measurements contain mechanical information beyond that captured by a single characteristic surface wave velocity and provide a theoretical basis for interpreting wave signals in multilayered soft materials and for future characterization of tissue properties.

\end{abstract}
	
 \maketitle  
\section{Introduction}\label{Introduction}

Biological tissues are ubiquitously organized as multilayered structures, enabling a wide range of specialized physiological functions \citep{dhandapani2026translating,hao2026conjugate,peng2026synchronizing}. The functionality of these tissues depends critically on the structural and mechanical integrity of individual layers. Pathological processes often induce layer-specific alterations in geometry and mechanical properties, which may compromise tissue function even at early stages \citep{tschaikowsky2022articular,elzayat2023mri}. However, conventional imaging modalities frequently lack sufficient sensitivity to detect such subtle mechanical changes before pronounced structural damage occurs. Consequently, there is an increasing demand for non-invasive techniques capable of quantitatively characterizing tissue mechanical properties. Elastography has emerged as an effective diagnostic approach by mapping tissue stiffness through elastic wave propagation and has been widely applied to the assessment of various diseases, including liver cirrhosis, periodontitis, and osteoarthritis \citep{Cao2019elastodiagnpsis,yokucs2021evaluation,xue2023use,widmann2026current}. Most conventional elastography approaches model biological tissues as infinite media or elastic half-spaces, assuming wave propagation within a single homogeneous material \citep{jiang2015characterization,deng2022quantifying}. Although these idealized models provide useful approximations, they neglect the intrinsic multilayered architecture of biological tissues. To overcome this limitation, recent studies have incorporated multilayer structures into elastography models, enabling more realistic descriptions of wave propagation in heterogeneous biological media \citep{xu2017near,feng2022vivo,feng2023ultra,ma2024guided,Jiang2025simultaneous,ghaderi2026multilayered,liu2026nondestructive}.

The bilayer resting on a rigid substrate provides a canonical mechanical representation of multilayered biological tissues. This configuration is related to the symmetric component of wave propagation in pre-stressed sandwich structures \citep{rogerson1996small}, allowing the corresponding dispersion problem to be investigated through the established framework of symmetric sandwich plates. Extensive theoretical studies have characterized elastic wave dispersion in layered structures under different constitutive assumptions and interfacial conditions. For incompressible hyperelastic materials, closed-form dispersion relations have been derived for extensional and flexural modes, together with asymptotic solutions in both short- and long-wavelength regimes \citep{rogerson1996small,rogerson1997flexual,rogerson2000effect}. These analyses were subsequently extended to compressible hyperelastic materials, enabling the investigation of both symmetric and antisymmetric wave modes \citep{lashhab2015dispersion,lashhab2016antisymmetric}. The influence of imperfect bonding has also been incorporated to describe how interface degradation modifies wave speeds and cutoff frequencies over different wavenumber ranges \citep{leungvichcharoen2003dispersion,leungvichcharoen2004anti,wijeyewickrema2009wave}. Furthermore, advanced theoretical frameworks have been developed to account for additional physical effects. The Three-Dimensional Linearized Theory of Elastic Waves in Initially Stressed Bodies (TLTEWISB) has been employed to examine the influence of pre-strain \citep{akbarov2011wave}, while the Stroh formalism has revealed modal curve intersections in anisotropic symmetric sandwich plates \citep{kuznetsov2024lamb}. Since dispersion relations provide the fundamental basis for evaluating dynamic responses in layered structures, accurate characterization of these relations is essential for understanding transient mechanical behavior in biological tissues.

Computing transient responses requires solving time-domain wave equations in layered media. Under impulsive loading, such problems generally involve integral transforms and subsequent inversion procedures \citep{lamb1904propagation,gakenheimer1969transient,miklowitz2012theory,van2016propagation}. Three major analytical approaches have been developed for transient wave analysis. The normal mode method constructs transient fields through orthogonal eigenfunction expansions, reducing governing partial differential equations to modal coefficients described by time-domain Duhamel integrals \citep{santosa1989transient,zeng2024theoretical}. The generalized ray method tracks individual wavefronts generated from the source, enabling accurate determination of arrival times and wave amplitudes \citep{pao1977generalized,shan2018analytical}. The transform method converts the governing equations into ordinary differential equations through forward transforms, solves the transformed problem, and obtains physical responses through inverse transformations \citep{erbay2000effects,dai2000uniform,pol2013modeling}. Building upon these classical approaches, subsequent studies have developed advanced analytical formulations, numerical strategies, and engineering applications for transient wave problems.

Recent analytical and semi-analytical developments have extended transient wave theory to increasingly complex material systems. \citet{zhao2025semianalytical} coupled a fractional-order Poynting–Thomson viscoelastic model with Biot-type multi-fluid equations for stratified unsaturated porous media, capturing viscosity-induced attenuation and dispersion of multiple compressional waves. \citet{fortunati2026multiscale} proposed a spectro-hierarchical Fourier perturbative framework for transient wave solutions in periodic media, overcoming limitations associated with classical asymptotic homogenization. \citet{taherraftar2026surface} extended the Cagniard–De Hoop method through a Galilean transformation to derive time-domain solutions for transient surface waves in transversely isotropic half-spaces subjected to moving forces.

In parallel, numerical methodologies for transient wave simulation have advanced rapidly. \citet{dimitriou20252d} developed a multi-resolution finite wavelet domain method combined with mass-decoupled explicit integration for transient waves in composite plates, allowing selective representation of different wave modes. \citet{jeong2026generalized} incorporated generalized element-wise mass scaling into the method of finite spheres with metaheuristic optimization to achieve stable and efficient explicit simulations. \citet{miao2026simulation} combined frequency-domain meshless backward substitution with fast Fourier transform and symmetric extension to improve computational accuracy while reducing boundary truncation effects. These analytical and numerical developments have enabled transient wave analysis in diverse engineering and natural systems, including Arctic ice–water structures \citep{zeng2024theoretical}, soil-embedded piles \citep{lu2025propagation}, ice shelves under ocean loading \citep{bennetts2026modelling}, and layered pavement systems \citep{li2026semi}.

Recent advances in optical coherence elastography (OCE) and surface wave elastography have further highlighted the potential of transient-wave approaches for biological tissue characterization. \citet{chen2025noncontact} demonstrated that laser-profiler-based surface wave elastography can identify stiffness interfaces and quantify localized elastic moduli in bilayer heterogeneous phantoms. \citet{liu2026nondestructive} developed a handheld OCE system using dual-pulse excitation and time-of-flight analysis to estimate the equivalent Young's modulus of stratified soft tissues. \citet{chawla2026assessment} employed transient air-coupled ultrasound excitation and broadband surface waves to evaluate deep-tissue stiffness and validate elastographic measurements against histological fibrosis. Collectively, these studies demonstrate the increasing importance of transient-wave measurements in quantitative tissue characterization.

Despite substantial progress in dispersion analysis and transient wave computation for layered structures, the relationship between transient wave responses, propagating modes, and underlying material or structural parameters in pre-stressed multilayer biological tissues remains insufficiently understood. In a recent study, \citet{liu2026wave} focused primarily on wave dispersion in hyperelastic bilayers and the corresponding asymptotic wave velocities. In contrast, the present study shifts the focus from dispersion relations to the propagation of impact waves in soft bilayer tissues, with particular emphasis on elucidating the relationship between the dispersion characteristics and the resulting wave signals. The developed semi-analytical solution paves a convenient way for unraveling the effects of the modulus ratio, layer thickness, and pre-stretch on transient wave propagation. These results may provide a theoretical basis for interpreting wave signals in relevant experiments and facilitate future parameter identification.

This paper is organized as follows. In Section~\ref{Problem formulation}, the governing equations for impact waves in pre-stressed bilayers are formulated within the framework of incremental elasticity. In Section~\ref{Impact waves}, Fourier–Laplace transforms are applied to the governing equations and boundary conditions, and the corresponding inverse transforms are derived analytically to obtain the transient responses. In Section~\ref{Asymptotic solutions}, the method of stationary phase is employed to derive asymptotic solutions near the wavefront and to elucidate the contributions of different wave modes. In Section~\ref{Finite element simulations}, finite element simulations are performed to validate the theoretical predictions, and the applicability and limitations of the numerical approach are discussed. In Section~\ref{Applications to soft tissues}, bilayer models representing articular cartilage, gingiva, and skin are investigated to demonstrate the effects of pathological and structural variations on group velocities and transient displacement responses. Finally, concluding remarks and the implications of the present findings are summarized in Section~\ref{Conclusion}.

\section{Problem formulation}\label{Problem formulation}

We investigate the propagation of impact waves in an infinitely long, pre-stressed, compressible hyperelastic bilayer resting on a smooth rigid foundation in a plane-strain setting. The primary objective of this section is to provide a brief summary of the governing equations for impact wave propagation based on the general incremental theory of nonlinear elasticity. A comprehensive account of the general theory of nonlinear elasticity can be found in \citet{ogden1997non,ogden2007incremental} and \citet{goriely2017mathematics}. For brevity, we only present the derivations that are valid for both layers.

\begin{figure}[H]
		\centering
        \includegraphics[width=1\linewidth]{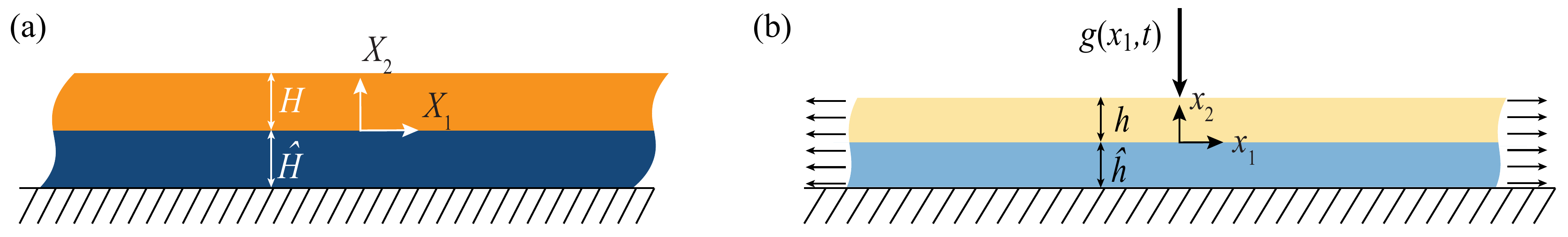}
        \caption{The geometry of an impacted bilayer in (a) the reference configuration $\mathcal{B}_0$ and (b) the current configuration $\mathcal{B}_1$.  An axial stretch is exerted on the bilayer and an impact load is then applied to generate impact waves.}\label{fig:bilayer}
	\end{figure}

The initial geometry, defined as the reference configuration $\mathcal{B}_0$, is depicted in Figure \ref{fig:bilayer}(a). The thicknesses of the upper and lower layers in $\mathcal{B}_0$ are denoted as $H$ and $\hat{H}$. Both layers share the same initial density $\rho_0$. A uniform stretch is applied to the initial configuration and generates the current configuration $\mathcal{B}_1$, as shown in Figure \ref{fig:bilayer}(b). The thicknesses of the upper and lower layer in $\mathcal{B}_1$ are $h$ and $\hat{h}$. The common density of both layers becomes $\rho $. Cartesian coordinate systems are adopted in both configurations. The position vectors of the same material particle in $\mathcal{B}_0$ and $\mathcal{B}_1$ are denoted by  $(X_1,X_2)$ and $(x_1,x_2)$, respectively. An impact load $g(x_1,t)$ is exerted at the upper surface to produce impact waves in the bilayer structure, as shown in Figure \ref{fig:bilayer}(b).

The deformation gradient associated with the deformation $\mathcal{B}_0 \rightarrow \mathcal{B}_1$ is denoted by $\bar{\mathbf{F}}$. Let $W$ denote the strain-energy function. The corresponding nominal stress tensor is given by
\begin{align}\label{eq:nominal stress}
\bar{\mathbf{S}}=\frac{\partial W(\mathbf{F})}{\partial\mathbf{F}}\bigg|_{\mathbf{F}=\bar{\mathbf{F}}},
\end{align}
where $\mathbf{F}$ is the general expression of the deformation gradient. When the bilayer is subjected to a uniform stretch, the deformation gradient $\bar{\mathbf{F}}$ is constant and diagonal, with principal stretches $\lambda_1$ and $\lambda_2$, where the indices $1$ and $2$ correspond to the $x_1$- and $x_2$-directions, respectively. Consequently, the nominal stress tensor $\bar{\mathbf{S}}$ is also constant and diagonal. In the absence of body forces, the equilibrium equation is satisfied automatically, and it follows from the traction-free condition on the surface normal to the $x_2$-direction that
\begin{align}\label{eq:S22}
\bar{S}_{22}=0,
\end{align}
which can be used to identify the relationship between $\lambda_1$ and $\lambda_2$.

To explore elastic waves caused by impact loading, we superimpose a time-dependent incremental displacement $\mathbf{u}$ in $\mathcal{B}_1$ to induce the perturbed configuration $\mathcal{B}_2$. The corresponding deformation gradient associated with the deformation $\mathcal{B}_0\rightarrow\mathcal{B}_2$ is indicated by $\mathbf{F}$. The associated nominal stress tensor $\mathbf{S}$ is defined analogously to \eqref{eq:nominal stress}. To formulate linearized incremental equations, we introduce the incremental stress tensor $\boldsymbol{\chi}$ through
\begin{align}\label{eq:inc-str}
\boldsymbol{\chi}^{\mathrm{T}}=\bar{J}^{-1}\bar{\mathbf{F}}(\mathbf{S}-\bar{\mathbf{S}}),
\end{align}
where $\bar{J}=\det{\bar{\mathbf{F}}}$. The incremental equation of motion can then be written in the form \citep{fu1999nonlinear}
\begin{align}\label{eq:eq-motion}
\mathrm{div}~\boldsymbol{\chi}^{\mathrm{T}}=\rho \ddot{\mathbf{u}},
\end{align}
where `$\operatorname{div}$' denotes the divergence operator with respect to the coordinates in $\mathcal{B}_1$, $\rho $ is the mass density in the pre-deformed configuration, and an overdot denotes differentiation with respect to time $t$.

The incremental surface traction $\mathbf{T}$ reads
\begin{align}\label{eq:traction}
\mathbf{T}=\boldsymbol{\chi}\mathbf{n},
\end{align}
where $\mathbf{n}$ denotes the outward unit normal to the surface. Assuming that the magnitude of $\mathbf{u}$ is small, we expand $\boldsymbol{\chi}$ in a Taylor series with respect to $\mathbf{u}$. Keeping only the linear terms yields
\begin{align}\label{eq:linear-str}
\chi_{ij}=\mathcal{A}_{jilk}u_{k,l},
\end{align}
where $\mathcal{A}$ is the instantaneous modulus tensor that can be calculated through
\begin{align}\label{eq:Ajilk}
\mathcal{A}_{jilk}=\bar{J}^{-1}\bar{F}_{jA}\bar{F}_{lB}\frac{\partial^2W}{\partial F_{iA}\partial F_{kB}}\bigg|_{\mathbf{F}=\bar{\mathbf{F}}}.
\end{align}

It should be noted that, under a uniform stretch, the instantaneous modulus tensor $\boldsymbol{\mathcal{A}}$ is constant in Cartesian coordinates. Substituting \eqref{eq:linear-str} into \eqref{eq:eq-motion} gives rise to the linearized incremental equations governing elastic wave propagation in the bilayer:
\begin{equation}\label{eq:gov-eq}
\mathcal{A}_{jilk}u_{k,lj}=\rho \ddot{u}_i,\quad\hat{\mathcal{A}}_{jilk}\hat{u}_{k,lj}=\rho \ddot{\hat{u}}_i,
\end{equation}
where, here and hereafter, quantities marked with a hat refer to the lower layer. 

We assume that the upper surface is subjected to an impact loading in the normal direction, the interface is perfectly bonded, and the lower surface is constrained to have zero vertical displacement. Under these assumptions, the corresponding linearized boundary and continuity conditions are
\begin{equation}\label{eq:BC}
\begin{aligned}
&\mathcal{A}_{21lk}u_{k,l}=0,\quad \mathcal{A}_{22lk}u_{k,l}=g(x_1,t),\quad \mbox{at }x_2=h,\\&\mathcal{A}_{21lk}u_{k,l}=\hat{\mathcal{A}}_{21lk}\hat{u}_{k,l},\quad\mathcal{A}_{22lk}u_{k,l}=\hat{\mathcal{A}}_{22lk}\hat{u}_{k,l},\quad \mbox{at }x_2=0,\\
&u_{1}=\hat{u}_1,\quad u_{2}=\hat{u}_2,\quad \mbox{at }x_2=0,\\&
\hat{u}_2=0,\quad\hat{\mathcal{A}}_{21lk}\hat{u}_{k,l}=0,\quad \mbox{at }x_2=-\hat{h}.
\end{aligned}
\end{equation}
For the counterpart bilayer with a traction-free upper surface, the prescribed traction $g(x_1,t)$ in $\eqref{eq:BC}_1$ shall be replaced by zero.

Currently, the general governing equations and the associated boundary conditions for impact waves propagating in compressible bilayers are established. In the next section, we shall follow the methodology used in \citet{erbay2000effects} to analytically derive the solution of impact waves.

Although the derivations presented in this section and the solution procedure described in the next section are applicable to general compressible hyperelastic materials, we specialize to the compressible neo-Hookean model for both layers. The corresponding strain-energy function is given by \citep{pence2015compressible}
\begin{equation}
W=\frac{\mu}{2}(\,\operatorname{tr}\mathbf{B}-2-2\log (\det \mathbf{F})\,)+\frac{\mu\nu}{1-2\nu}(\det \mathbf{F}-1)^2,
\label{eq:neo-Hookean}
\end{equation}
where $\mathbf{B}=\mathbf{F}\mathbf{F}^\mathrm{T}$ stands for the left Cauchy–Green deformation tensor, $\mu$ is the ground-state shear modulus and $\nu$ denotes Poisson's ratio. For simplicity, we assume that the two layers have the same Poisson's ratio, so that only a single parameter $\nu$ appears in the subsequent analysis. Substituting the strain-energy function \eqref{eq:neo-Hookean} into equation \eqref{eq:S22}, we obtain the following relation between $\lambda_1$ and $\lambda_2$:
\begin{align}\label{eq:lambda2}
\lambda_2=\frac{\lambda_1\nu+\sqrt{1-4\nu+2\lambda_1^2\nu+4\nu^2-3\lambda_1^2\nu^2}}{1-2\nu+2\lambda_1^2\nu}.
\end{align}

We emphasize that, in order to characterize the impact wave solutions, the dispersion curves of the corresponding model without impact loading are required. Therefore, we summarize the procedure for obtaining the dispersion curves in \ref{Dispersion relation}.

\section{Analytical solution of impact waves}\label{Impact waves}

In general, the impact loading $g(x_1,t)$ is both localized in space and transient in time. To solve the initial-boundary-value problem defined by \eqref{eq:gov-eq} and \eqref{eq:BC}, we employ the Fourier-Laplace transform. For convenience, all quantities with dimension of length are normalized by $h$, while all quantities with dimension of stress are normalized by $\mu$. For simplicity, all scaled quantities retain their original expressions. This nondimensionalization introduces the following two ratios:
\begin{equation}
    \zeta=\frac{\hat{h}}{h}=\frac{\hat{H}}{H},\quad r=\frac{\hat{\mu}}{\mu}.
\end{equation}
Recall that $g(x_1,t)$ represents an impact line load acting in a plane-strain setting. Consequently, $g(x_1,t)$ has the stress dimension (force per unit length). Furthermore, we define the dimensionless time $t_1$, dimensionless wavenumber $k_1$ and frequency $\omega_1$ as follows
\begin{equation}
    t_1=\frac{t}{h}\sqrt{\frac{\mu}{\rho }},\quad k_1=kh,\quad
\omega_1=\omega h\sqrt{\frac{\rho }{\mu}},
\end{equation}
where $k$ and $\omega$ denote the wavenumber and angular frequency, respectively.

Similarly, we only present the general derivations
that are applicable to both layers. Applying the Fourier–Laplace transform to the displacement components $u_i$ $(i=1,2)$ and the impulsive load $g$ results in
\begin{equation}
\begin{aligned}\label{eq:transform}
&\tilde{u}_1(s,\xi,x_2)=\int_0^{\infty}\left(\int_{-\infty}^{+\infty}u_1(x_1,x_2,t_1)\mathrm{e}^{-\mathrm{i}\xi x_1}\mathrm{d}x_1\right)\mathrm{e}^{-s t_1}\mathrm{d}t_1,\\
&\tilde{u}_2(s,\xi,x_2)=\int_0^{\infty}\left(\int_{-\infty}^{+\infty}u_2(x_1,x_2,t_1)\mathrm{e}^{-\mathrm{i}\xi x_1}\mathrm{d}x_1\right)\mathrm{e}^{-s t_1}\mathrm{d}t_1,\\
&\tilde{g}(s,\xi)=\int_0^{\infty}\left(\int_{-\infty}^{+\infty}g(x_1,t_1)\mathrm{e}^{-\mathrm{i}\xi x_1}\mathrm{d}x_1\right)\mathrm{e}^{-s t_1}\mathrm{d}t_1,
\end{aligned}
\end{equation}
where $\xi$ and $s$ denote the Fourier and Laplace transform variables, respectively.

Inserting \eqref{eq:transform} into $\eqref{eq:gov-eq}_1$ yields
\begin{equation}
\begin{aligned}\label{eq:eq-motion-t}
&\left(-s^2-\xi^2\mathcal{A}_{1111}\right)\tilde{u}_1+\mathrm{i}\left(\mathcal{A}_{1122}+\mathcal{A}_{2112}\right)\xi\tilde{u}_2'+\mathcal{A}_{2121}\tilde{u}_1''=0,\\
&\left(-s^2-\xi^2\mathcal{A}_{1212}\right)\tilde{u}_2+\mathrm{i}\left(\mathcal{A}_{2211}+\mathcal{A}_{1221}\right)\xi\tilde{u}_1'+\mathcal{A}_{2222}\tilde{u}_2''=0,
\end{aligned}
\end{equation}
where a prime indicates differentiation with respect to $x_2$ and we only retain the nonzero components of $\mathcal{A}_{jilk}$. We express $\tilde{u}_2$ in terms of $\tilde{u}_1$ according to equation $\eqref{eq:eq-motion-t}_1$ and then eliminate $\tilde{u}_2$ in $\eqref{eq:eq-motion-t}_2$ to obtain
\begin{align}\label{eq:ODE-fourth}
d_4\tilde{u}_1''''+d_2\xi^2\tilde{u}_1''+d_0\xi^4\tilde{u}_1=0,
\end{align}
where $d_4$, $d_2$, and $d_0$ are given by
\begin{equation}
\begin{aligned}\label{eq:d0-d4}
d_4&=\mathcal{A}_{2121}\mathcal{A}_{2222},\\
d_2&=\left(\mathcal{A}_{1122}+\mathcal{A}_{2112}\right)\left(\mathcal{A}_{2211}+\mathcal{A}_{1221}\right)-\mathcal{A}_{2121}\left(\frac{s^2}{\xi^2}+\mathcal{A}_{1212}\right)-\mathcal{A}_{2222}\left(\frac{s^2}{\xi^2}+\mathcal{A}_{1111}\right),\\
d_0&=\left(\frac{s^2}{\xi^2}+\mathcal{A}_{1212}\right)\left(\frac{s^2}{\xi^2}+\mathcal{A}_{1111}\right).
\end{aligned}
\end{equation}

We seek a general solution of the form $\tilde{u}_1=\mathrm{e}^{p \xi x_2}$. Substituting this expression into \eqref{eq:ODE-fourth} shows that the eigenvalue $p$ has four roots, which can be written as $\pm p_1$ and $\pm p_2$, where $p_1$ and $p_2$ are given by
\begin{align}\label{eq:eigenvalue}
p_{1,2}=\sqrt{\frac{-d_2\pm\sqrt{d_2^2-4d_0d_4}}{2d_4}}.
\end{align}
As a result, the general solutions of $\tilde{u}_1$ and $\tilde{\hat{u}}_1$ can be written as
\begin{equation}
\begin{aligned}\label{eq:sol-general}
\tilde{u}_1&=C_1\cosh{(p_1\xi x_2)}+C_2\sinh{(p_1\xi x_2)}+C_3\cosh{(p_2\xi x_2)}+C_4\sinh{(p_2\xi x_2)},\\
\tilde{\hat{u}}_1&=C_5\cosh{(\hat{p}_1\xi x_2)}+C_6\sinh{(\hat{p}_1\xi x_2)}+C_7\cosh{(\hat{p}_2\xi x_2)}+C_8\sinh{(\hat{p}_2\xi x_2)},
\end{aligned}
\end{equation}
where $C_i$ $(i=1,2,\cdots,8)$ are arbitrary constants.

With the use of \eqref{eq:transform}, together with its counterpart for the bottom layer, the boundary and continuity conditions in \eqref{eq:BC} can be rewritten as
\begin{equation}
\begin{aligned}\label{eq:trans-BC}
&\mathrm{i}\xi\mathcal{A}_{2112}\tilde{u}_{2}+\mathcal{A}_{2121}\tilde{u}_{1}^{'}=0,\quad \mathrm{i}\xi\mathcal{A}_{2211}\tilde{u}_{1}+\mathcal{A}_{2222}\tilde{u}_{2}^{'}=\tilde{g}(s,\xi),\quad \mbox{at }x_2=1,\\&\mathrm{i}\xi\mathcal{A}_{2112}\tilde{u}_{2}+\mathcal{A}_{2121}\tilde{u}_{1}^{'}=\mathrm{i}\xi\hat{\mathcal{A}}_{2112}\tilde{\hat{u}}_{2}+\hat{\mathcal{A}}_{2121}\tilde{\hat{u}}_{1}^{'},\quad \mbox{at }x_2=0,\\&\mathrm{i}\xi\mathcal{A}_{2211}\tilde{u}_{1}+\mathcal{A}_{2222}\tilde{u}_{2}^{'}=\mathrm{i}\xi\hat{\mathcal{A}}_{2211}\tilde{\hat{u}}_{1}+\hat{\mathcal{A}}_{2222}\tilde{\hat{u}}_{2}^{'},\quad \mbox{at }x_2=0,\\
&\tilde{u}_{1}=\tilde{\hat{u}}_1,\quad \tilde{u}_{2}=\tilde{\hat{u}}_2,\quad \mbox{at }x_2=0,\\&
\tilde{\hat{u}}_2=0,\quad\mathrm{i}\xi\hat{\mathcal{A}}_{2112}\tilde{\hat{u}}_{2}+\hat{\mathcal{A}}_{2121}\tilde{\hat{u}}_{1}^{'}=0,\quad \mbox{at }x_2=-\zeta.
\end{aligned}
\end{equation}
Upon inserting the general solution \eqref{eq:sol-general} into \eqref{eq:trans-BC} (note that $\tilde{u}_2$ and $\tilde{\hat{u}}_2$ are expressed in terms of $\tilde{u}_1$ and $\tilde{\hat{u}}_1$), we arrive at a linear system
\begin{align}\label{eq:MC}
\mathbf{M}\mathbf{C}=\mathbf{D},
\end{align}
where $\mathbf{C}=(C_1,\ldots,C_8)^{\mathrm{T}}$ and $\mathbf{D}=(0,\tilde{g},0,0,0,0,0,0)^{\mathrm{T}}$. The explicit form of the coefficient matrix $\mathbf{M}$ is omitted for brevity.

To proceed further, we define 
\begin{equation}
    \Gamma(s,\xi)=\det\mathbf{M}.
\end{equation}
Assuming that $\Gamma(s,\xi)$ is not identically zero, the matrix $\mathbf{M}$ is invertible except at the zeros of $\Gamma$, and hence the solution of \eqref{eq:MC} is given by
\begin{equation}
    \mathbf{C}=\mathbf{M}^{-1}\mathbf{D}=\frac{\mathbf{M}^*}{\Gamma(s,\xi)}\mathbf{D},
    \label{eq:sol-C}
\end{equation}
where $\mathbf{M}^*$ denotes the adjugate matrix of $\mathbf{M}$.

In view of \eqref{eq:sol-C}, we are able to rewrite $\tilde{u}_1$ and $\tilde{u}_2$ as follows
\begin{equation}
\begin{aligned}\label{eq:u1tu2t}
\tilde{u}_1=\frac{\tilde{g}(s,\xi)\Gamma_1(s,\xi,x_2)}{\Gamma(s,\xi)},\quad
\tilde{u}_2=\frac{\tilde{g}(s,\xi)\Gamma_2(s,\xi,x_2)}{\Gamma(s,\xi)},
\end{aligned}
\end{equation}
where $\Gamma_1$ and $\Gamma_2$ are functions related to $\mathbf{M}$. The original displacement components $u_1$ and $u_2$ can then be derived by the inverse Fourier-Laplace transform:
\begin{equation}
\begin{aligned}\label{eq:u1u2}
u_1(x_1,x_2,t_1)&=\frac{1}{2\pi}\int_{-\infty}^{\infty}\left(\frac{1}{2\pi \mathrm{i}}\int_{\beta-\mathrm{i}\infty}^{\beta+\mathrm{i}\infty}\frac{\tilde{g}(s,\xi)\Gamma_1(s,\xi,x_2)}{\Gamma(s,\xi)}\mathrm{e}^{s t_1}\mathrm{d}s\right)\mathrm{e}^{\mathrm{i}\xi x_1}\mathrm{d}\xi,\\
u_2(x_1,x_2,t_1)&=\frac{1}{2\pi}\int_{-\infty}^{\infty}\left(\frac{1}{2\pi \mathrm{i}}\int_{\beta-\mathrm{i}\infty}^{\beta+\mathrm{i}\infty}\frac{\tilde{g}(s,\xi)\Gamma_2(s,\xi,x_2)}{\Gamma(s,\xi)}\mathrm{e}^{s t_1}\mathrm{d}s\right)\mathrm{e}^{\mathrm{i}\xi x_1}\mathrm{d}\xi,
\end{aligned}
\end{equation}
where the real number $\beta$ must be chosen such that the vertical line $s=\beta$ lies entirely to the right of all singular points of the integrand. We note that $\Gamma(\mathrm{i}\omega_1,k_1)=0$ recovers the dispersion relation determined by \eqref{eq:dispersion}.  

Next, we examine the zeros of $\Gamma(s,\xi)$ in detail. From the preceding analysis, we know that, for any fixed $\xi$, the equation $\Gamma(s,\xi)=0$ possesses infinitely many pairs of roots, denoted by $\pm s^{(m)}$ $(m=1,2,\ldots)$. In addition, $\tilde{g}(s,\xi)$ may contribute $N$ further singularities, denoted by $s_{(n)}$. Consequently, the integrals with respect to $s$ in \eqref{eq:u1u2} can be evaluated by means of the residue theorem. Provided that all poles associated with $\pm s^{(m)}$ and $s_{(n)}$ are simple, we obtain
\begin{equation}
\begin{aligned}\label{eq:sol-u1u2}
u_1(x_1,x_2,t_1)&=u_1^{(0)}+\frac{1}{2\pi}\int_{-\infty}^{+\infty}\mathrm{e}^{\mathrm{i}\xi x_1}\sum_{m=1}^{\infty}\left[\left(\frac{\tilde{g}\Gamma_1\mathrm{e}^{st_1}}{\partial\Gamma/\partial s}\right)\bigg|_{s=s^{(m)}}+\left(\frac{\tilde{g}\Gamma_1\mathrm{e}^{st_1}}{\partial\Gamma/\partial s}\right)\bigg|_{s=-s^{(m)}}\right]\mathrm{d}\xi,\\
u_2(x_1,x_2,t_1)&=u_2^{(0)}+\frac{1}{2\pi}\int_{-\infty}^{+\infty}\mathrm{e}^{\mathrm{i}\xi x_1}\sum_{m=1}^{\infty}\left[\left(\frac{\tilde{g}\Gamma_2\mathrm{e}^{st_1}}{\partial\Gamma/\partial s}\right)\bigg|_{s=s^{(m)}}+\left(\frac{\tilde{g}\Gamma_2\mathrm{e}^{st_1}}{\partial\Gamma/\partial s}\right)\bigg|_{s=-s^{(m)}}\right]\mathrm{d}\xi,
\end{aligned}
\end{equation}
where $u_1^{(0)}$ and $u_2^{(0)}$ represent the contributions from the singularities $s_{(n)}$ and are given by
\begin{equation}
    \begin{aligned}
        u_1^{(0)}&=\frac{1}{2\pi}\int_{-\infty}^{+\infty}\mathrm{e}^{\mathrm{i}\xi x_1}\sum_{n=1}^{N}\left(\frac{G_1\Gamma_1\mathrm{e}^{st_1}}{\Gamma\partial G_2/\partial s}\right)\bigg|_{s=s_{(n)}}\mathrm{d}\xi,\\
        u_2^{(0)}&=\frac{1}{2\pi}\int_{-\infty}^{+\infty}\mathrm{e}^{\mathrm{i}\xi x_1}\sum_{n=1}^{N}\left(\frac{G_1\Gamma_2\mathrm{e}^{st_1}}{\Gamma\partial G_2/\partial s}\right)\bigg|_{s=s_{(n)}}\mathrm{d}\xi.
    \end{aligned}
\end{equation}
Here, $\tilde{g}(s,\xi)$ has been expressed as $\tilde{g}(s,\xi)=G_1(s,\xi)/G_2(s,\xi)$, where $G_1(s,\xi)\neq0$, and $s_{(n)}$ are assumed to be simple zeros of $G_2(s,\xi)$. The corresponding solutions for $\tilde{\hat{u}}_1$ and $\tilde{\hat{u}}_2$ can be derived in an analogous manner, and the details are omitted for brevity.

So far, the displacement fields generated by an impulsive load in a compressible bilayer have been derived. Throughout the preceding analysis, the impact loading function $g(x_1,t_1)$ has been kept generic. In our illustrative examples, we consider a concentrated impulsive load applied at the center of the upper surface, given by
\begin{align}\label{eq:impact}
g(x_1,t_1)=-\delta(x_1)\delta(t_1),
\end{align}
where $\delta(\cdot)$ is the Dirac delta function, and the negative sign indicates that the load is applied in the downward direction. The corresponding Fourier-Laplace transform is
\begin{align}
\tilde{g}=-1.
\end{align}
Since $\tilde{g}$ is constant, the terms associated with $u_1^{\mathrm{(0)}}$ and $u_2^{\mathrm{(0)}}$ vanish. In contrast, if the loading function is taken as $g(x_1,t_1)=-\delta(x_1)\mathcal{H}(t_1)$, where $\mathcal{H}(t_1)$ denotes the Heaviside step function, its Fourier-Laplace transform is $-1/s$ and an additional singular point at $s_{(1)}=0$ is produced. This type of loading was adopted by \citet{erbay2000effects} to investigate impact wave propagation in an incompressible layer. Note that when \eqref{eq:impact} is adopted, we obtain
\begin{align}\label{eq:load function}
\int_{0}^{\infty}\int_{-\infty}^{\infty}(-\delta(x_1)\delta(t_1))\mathrm{d}x_1\mathrm{d}t_1=-1.
\end{align}

\section{Asymptotic solutions}\label{Asymptotic solutions}

We note that, in order to determine the solutions $u_1$ and $u_2$ given in \eqref{eq:sol-u1u2}, as well as their counterparts for the bottom layer, the exact dispersion curves are required. However, explicit expressions for the dispersion curves generally do not exist. Therefore, numerical schemes are necessary to compute the displacement field associated with impact waves. Nevertheless, the method of stationary phase offers an effective approach to derive asymptotic solutions for highly oscillatory integrals, especially for wave problems in the preceding section \citep{Wong1989asymptotic,erbay2000effects,achenbach2012wave}. In this section, we adopt this method to deal with the integrations formulated in Section \ref{Impact waves}.

We first truncate the infinite summations in equation \eqref{eq:sol-u1u2} by retaining only the first $M$ terms. The resulting approximations for the displacement fields $u_1$ and $u_2$ are given by
\begin{equation}
\begin{aligned}\label{eq:sol-u1u21}
u_1&\approx\frac{1}{2\pi}\left(\int_{-\infty}^{\infty}\sum_{m=1}^{M}\left[ \Phi^{(m)}_1\mathrm{e}^{\mathrm{i}\alpha_1\left(\xi\right)t_1}+ \Psi^{(m)}_1\mathrm{e}^{\mathrm{i}\alpha_2\left(\xi\right)t_1}\right]\mathrm{d}\xi\right),\\
u_2&\approx\frac{1}{2\pi}\left(\int_{-\infty}^{\infty}\sum_{m=1}^{M}\left[\Phi^{(m)}_2\mathrm{e}^{\mathrm{i}\alpha_1\left(\xi\right)t_1}+\Psi^{(m)}_2\mathrm{e}^{\mathrm{i}\alpha_2\left(\xi\right)t_1}\right]\mathrm{d}\xi\right),
\end{aligned}
\end{equation}
where 
\begin{align}
    &\alpha_1(\xi)=\frac{x_1\xi}{t_1}+\frac{s^{(m)}}{\mathrm{i}},\quad \alpha_2(\xi)=\frac{x_1\xi}{t_1}-\frac{s^{(m)}}{\mathrm{i}},\label{eq:a1a2}\\&
    \Phi^{(m)}_1=-\frac{\Gamma_1}{\partial\Gamma/\partial s}\bigg|_{s=-s^{(m)}},\quad \Psi^{(m)}_1=-\frac{\Gamma_1}{\partial\Gamma/\partial s}\bigg|_{s=s^{(m)}},\label{eq:Phi}\\&
     \Phi^{(m)}_2=-\frac{\Gamma_2}{\partial\Gamma/\partial s}\bigg|_{s=-s^{(m)}},\quad \Psi^{(m)}_2=-\frac{\Gamma_2}{\partial\Gamma/\partial s}\bigg|_{s=s^{(m)}}.\label{eq:Psi}
\end{align}
For large values of $t_1$, the exponential phase terms oscillate at increasingly high frequencies, causing the integrands to alternate rapidly between positive and negative values. Over most of the integration domain, these rapid oscillations produce nearly complete destructive interference, so that the net contribution to the integral is negligible. Consequently, the dominant contributions arise only from the neighbourhood of the stationary phase points, which satisfy $\alpha_1'(\xi)=0$ or $\alpha_2'(\xi)=0$. To simplify the notation in this section, we define $\alpha_1'=\mathrm{d}\alpha_1/\mathrm{d}\xi$, $\alpha_2'=\mathrm{d}\alpha_2/\mathrm{d}\xi$, $\alpha_1''=\mathrm{d}^2\alpha_1/\mathrm{d}\xi^2$, and so on.

We present the derivation by considering the stationary phase point associated with $\alpha_2$. The analysis for $\alpha_1$ is entirely analogous. It follows from $\eqref{eq:a1a2}_2$ that the stationary phase point is given by
\begin{equation}\label{eq:group velocity}
c_g^{(m)}=\frac{x_1}{t_1},
\end{equation}
where $c_g^{(m)}=\mathrm{d}s^{(m)}/(\mathrm{i}\,\mathrm{d}\xi)$ denotes the group velocity of the $m$th mode. Consequently, each stationary point must lie on a dispersion curve. We denote by $\xi=\varsigma$ the stationary wavenumber associated with the $m$th mode, where $(s^{(m)},\varsigma)$ satisfies the dispersion relation $\Gamma(s^{(m)},\varsigma)=0$. 

For fixed dimensionless time $t_1$, the function $\Psi_1^{(m)}$ is evaluated at $\xi=\varsigma$. Furthermore, in a neighbourhood of the stationary point, $\alpha_2(\xi)$ is approximated by its second-order Taylor expansion $\alpha_2(\varsigma)+\alpha_2''(\varsigma)(\xi-\varsigma)^2/2$. Introducing the change of variables $\xi=\varsigma+\eta\mathrm{e}^{\pm\mathrm{i}\pi/4}$ $(-\infty<\eta<\infty)$ and retaining only the leading-order contribution, we obtain
\begin{equation}
\int_{-\infty}^{+\infty}\Psi_1^{(m)}\mathrm{e}^{\mathrm{i}\alpha_2(\xi)t_1}\mathrm{d}\xi\approx\left(\Psi_1^{(m)}(s^{(m)},\varsigma,x_2)\sqrt{\frac{2\pi}{ t_1|\alpha_2''(\varsigma)|}}\right)\mathrm{e}^{\mathrm{i}(\alpha_2(\varsigma)t_1\pm\pi/4)},
\label{eq:Psi1}
\end{equation}
where the `$+$' and `$-$' signs correspond to $\alpha_2''(\varsigma)>0$ and $\alpha_2''(\varsigma)<0$, and it is assumed that $\alpha_2''(\varsigma)\neq0$. If multiple stationary phase points exist for a given value of $x_1/t_1$, their contributions are obtained by summing the corresponding asymptotic expressions of the form \eqref{eq:Psi1}. Applying the same procedure to $\Psi_2^{(m)}$ yields an analogous expression, obtained simply by replacing $\Psi_1^{(m)}$ with $\Psi_2^{(m)}$ in \eqref{eq:Psi1}.

For the impact loading specified by \eqref{eq:impact}, we numerically verify that the dispersion curves are symmetric with respect to both $\xi$ and $s$. Therefore, the group velocity $c_g$ is an odd function of $\xi$. This implies that $\xi=-\varsigma$ must be a stationary phase point of $\alpha_1(\xi)$ provided that $\varsigma$ is a stationary phase point of $\alpha_2(\xi)$. Moreover, $\Psi_2^{(m)}$ is an even function of $\xi$ and an odd function of $s$, whereas $\Psi_1^{(m)}$ is odd in both variables. The integrals of $\Phi_1^{(m)}$ and $\Phi_2^{(m)}$ can be evaluated in the same manner, but evaluated at $\xi=-\varsigma$. Consequently, the asymptotic expressions for $\Psi_1^{(m)}$ and $\Psi_2^{(m)}$ coincide with those for $\Phi_1^{(m)}$ and $\Phi_2^{(m)}$, respectively. Summing the contributions from all stationary phase points, the asymptotic solutions of \eqref{eq:sol-u1u21} are expressed as
\begin{equation}
\begin{aligned}\label{eq:app-sol}
u_1&\approx\sum_{m=1}^{M}\left[\left(\Psi_1^{(m)}(s^{(m)},\varsigma,x_2)\sqrt{\frac{2}{\pi t_1|\alpha_2''(\varsigma)|}}\right){\cos{\left(t_1\alpha_2(\varsigma)\pm\pi/4\right)}}\right],\\
u_2&\approx\sum_{m=1}^{M}\left[\mathrm{i}\left(\Psi_2^{(m)}(s^{(m)},\varsigma,x_2)\sqrt{\frac{2}{\pi t_1|\alpha_2''(\varsigma)|}}\right){\sin{\left(t_1\alpha_2(\varsigma)\pm\pi/4\right)}}\right].
\end{aligned}
\end{equation}
From the above analysis we observe that both displacements are proportional to $(t_1|\alpha_2''(\varsigma)|)^{-1/2}$.

We note that the asymptotic solutions given in \eqref{eq:app-sol} break down at points where the group velocity is stationary, i.e., where $\alpha_2''(\varsigma)=0$. These points generally correspond to wavefronts. To obtain a uniformly valid approximation in the vicinity of these points, we employ a uniform asymptotic treatment based on the Airy and Scorer functions for degenerate stationary points. 

At a point $\xi=\varsigma$ where $\alpha_2''(\varsigma)=0$ but $\alpha_2'''(\varsigma)\neq 0$, we introduce the local coordinate $w=x_1-c_g^{(m)}(\varsigma)t_1$ in the vicinity of $\varsigma$. Retaining terms up to third order in the Taylor expansion of $\alpha_2(\xi)$ about $\varsigma$, we obtain
\begin{equation}
\alpha_2(\xi)\approx\alpha_2(\varsigma)+\frac{w}{t_1}(\xi-\varsigma)+\frac{1}{6}\alpha_2'''(\varsigma)(\xi-\varsigma)^3.
\end{equation}
Substituting this expansion into the integral involving $\Psi_1^{(m)}$ yields
\begin{equation}
\int_{-\infty}^{+\infty}\Psi_1^{(m)}\mathrm{e}^{\mathrm{i}\alpha_2(\xi)t_1}\,\mathrm{d}\xi
\approx e^{\mathrm{i}\alpha_2(\varsigma)t_1}
\int_{-\infty}^{+\infty}\Psi_1^{(m)}
\mathrm{e}^{\mathrm{i}t_1\left[w(\xi-\varsigma)/t_1+\alpha_2'''(\varsigma)(\xi-\varsigma)^3/6\right]}
\,\mathrm{d}\xi.
\end{equation}
The resulting integrals can be expressed in terms of the Airy function $\operatorname{Ai}$ and the Scorer function  $\operatorname{Gi}$ as
\begin{equation}
    \begin{aligned}
    &\int_0^{+\infty}\cos\,\left(t_1\left[\frac{w}{t_1}(\xi-\varsigma)+\frac{1}{6}\alpha_2'''(\varsigma)(\xi-\varsigma)^3\right]\right)\,\mathrm{d}\xi
=\frac{2^{1/3}\pi}{(t_1|\alpha_2'''(\varsigma)|)^{1/3}}\,\operatorname{Ai}\!\left(\pm\left|\varepsilon\right|\right),\\&
\int_0^{+\infty}\sin\,\left(t_1\left[\frac{w}{t_1}(\xi-\varsigma)+\frac{1}{6}\alpha_2'''(\varsigma)(\xi-\varsigma)^3\right]\right)\mathrm{d}\xi
=\frac{2^{1/3}\pi}{(t_1|\alpha_2'''(\varsigma)|)^{1/3}}\,\operatorname{Gi}\left(\pm\left|\varepsilon\right|\right), 
    \end{aligned}
\end{equation}
The positive and negative signs correspond to $\alpha_2'(\varsigma)\alpha_2'''(\varsigma)>0$ and $\alpha_2'(\varsigma)\alpha_2'''(\varsigma)<0$, respectively, and
\begin{equation}
    |\varepsilon|=\frac{2^{1/3}t_1^{2/3}|\alpha_2'(\varsigma)|}{(|\alpha_2'''(\varsigma)|)^{1/3}}.
\end{equation} 
Finally, exploiting the symmetry properties of $\Psi_1^{(m)}$ and $\Psi_2^{(m)}$, we obtain
\begin{equation}
    \begin{aligned}
       &u_1\approx\sum_{m=1}^{M}\Psi_1^{(m)}(s^{(m)},\varsigma,x_2)\,\mathrm{e}^{\mathrm{i}\alpha_2(\varsigma)t_1}\frac{2^{4/3}}{(t_1|\alpha_2'''(\varsigma)|)^{1/3}}\operatorname{Ai}(\pm|\varepsilon|), \\&
       u_2\approx\sum_{m=1}^{M}\Psi_2^{(m)}(s^{(m)},\varsigma,x_2)\,\mathrm{e}^{\mathrm{i}\alpha_2(\varsigma)t_1}\frac{2^{4/3} \mathrm{i}}{(t_1|\alpha_2'''(\varsigma)|)^{1/3}}\operatorname{Gi}(\pm|\varepsilon|).
    \end{aligned}\label{eq-asy-sol-for-high-sta}
\end{equation}
In contrast to \eqref{eq:app-sol}, displacements are now proportional to $(t_1|\alpha_2'''(\varsigma)|)^{-1/3}$.

\section{Finite element simulations}\label{Finite element simulations}

In this section, a finite element (FE) model is established in ABAQUS to perform a nonlinear analysis of wave propagation in pre-stressed compressible hyperelastic bilayer models. Since the compressible neo-Hookean strain energy function \eqref{eq:neo-Hookean} is not available as a built-in material model in ABAQUS, the user-defined subroutines UHYPER and VUMAT are developed in accordance with the ABAQUS user guidelines \citep{ABAQUS2013}.

An important aspect of the simulation setup is the coupling between the static and explicit analyses. Since an explicit dynamic step cannot be directly appended to a general static analysis step in ABAQUS, the simulation of elastic wave propagation is divided into two sequential stages. In the first stage, an initially undeformed bilayer model is constructed, and its uniform deformation is obtained using the UHYPER subroutine together with a general static analysis. In the second stage, a pre-deformed bilayer model corresponding to the uniformly deformed configuration obtained in the first stage is created. The VUMAT subroutine is then implemented in ABAQUS, and wave propagation in the impacted bilayer is analyzed using an explicit dynamic analysis.

\begin{figure}[H]
    \centering
    \includegraphics[width=0.6\linewidth]{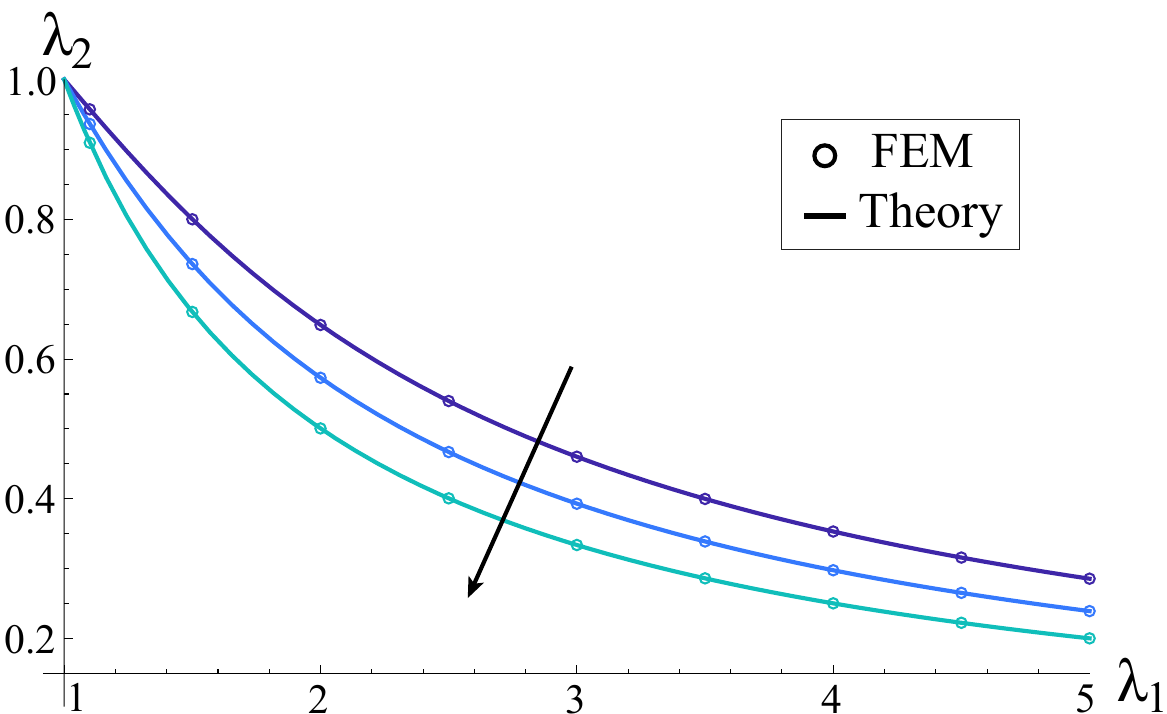}
    \caption{Comparison of the $\lambda_1$-$\lambda_2$ relationship obtained using the finite element method (FEM, open circles) and the theoretical predictions (solid lines). The arrow indicates the direction of increasing Poisson's ratio, with $\nu=0.3$, $0.4$, and $0.499$.}
    \label{fig:lambda2-lambda1-figure}
\end{figure}

We first validate the UHYPER subroutine by simulating the uniform stretching of a compressible bilayer. To represent an infinitely long bilayer, periodic boundary conditions are imposed on a representative cell, whose length-to-height ratio is set to 100 to ensure a sufficiently long wave-propagation duration. A uniform mesh is employed, and 4-node bilinear plane strain elements with reduced integration and hourglass control (CPE4R) are adopted in the general static analysis step. A uniform stretch along the horizontal direction (the $X_1$-direction in Figure \ref{fig:bilayer}(a)) is subsequently applied, resulting in a homogeneous deformation.

For the bilayer model, we impose traction-free boundary conditions on both the upper and lower surfaces, together with continuity conditions at the interface, by setting $r=2$ and $\zeta=1$. The relationship between $\lambda_1$ and $\lambda_2$ is determined numerically and compared with the analytical solution given by \eqref{eq:lambda2}, as shown in Figure \ref{fig:lambda2-lambda1-figure}. Poisson's ratio $\nu$ takes the values $\{0.3,\,0.4,\,0.499\}$. The excellent agreement in all three cases confirms the validity of the implemented UHYPER subroutine. In fact, for uniform stretching, the relationship between $\lambda_1$ and $\lambda_2$ is independent of the values of $r$ and $\zeta$.

As mentioned earlier, the user-defined subroutine VUMAT is implemented to simulate wave propagation in pre-stressed hyperelastic materials. The stresses obtained in the first step (uniform stretching) are taken as the initial stress state in the VUMAT subroutine. The Cauchy stress corresponding to the compressible neo-Hookean model \eqref{eq:neo-Hookean} is also implemented in the VUMAT subroutine and is given by \citep{pence2015compressible}
\begin{align}
\bm{\sigma}=\frac{\mu}{\det\mathbf{F}}\mathbf{B}+\left(\frac{2\mu\nu}{1-2\nu}(\det\mathbf{F}-1)-\frac{\mu}{\det\mathbf{F}}\right)\mathbf{I},
\label{eq:stress tensor}
\end{align}
where $\mathbf{I}$ denotes the second-order identity tensor.

\begin{figure}[H]
    \centering
    \includegraphics[width=0.6\linewidth]{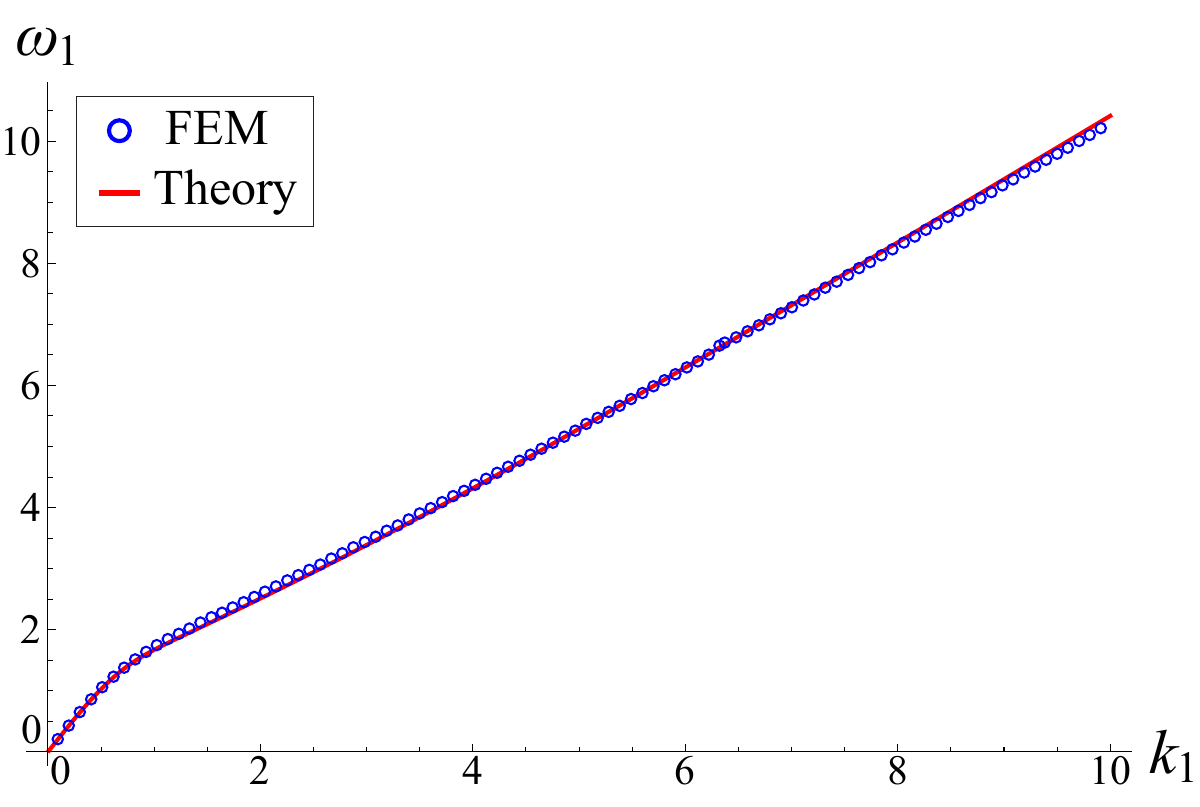}
    \caption{Comparison of the dispersion curve for the fundamental mode obtained from the FE simulation (blue circles) and the theoretical prediction (red solid line). The parameters are $\lambda_1=1.1$, $\nu=0.4$, $r=2$, and $\zeta=1$.}
    \label{fig:FEM-theory-omega-k-figure}
\end{figure}

To validate the developed VUMAT subroutine, we perform an FE simulation of impact wave propagation in a compressible bilayer with $\lambda_1=1.1$, $\nu=0.4$, $r=2$, and $\zeta=1$. To approximate the impulsive loading defined in \eqref{eq:impact}, the load is applied uniformly over a small spatial interval $[-1/2\,,1/2]$ and a short time interval $[0\,,1]$. The loading amplitude is chosen such that the total integral of the applied load is equal to unity. A two-dimensional Fast Fourier Transform (FFT) is then performed on the simulated displacement field to extract the frequency-wavenumber spectrum. In the FE simulation, a uniform mesh size of $0.05$ is adopted to ensure sufficient spatial resolution, providing an adequate number of elements per wavelength at least for $k_1=10$. The same CPE4R elements are used as well. The dispersion curve of the fundamental mode is extracted from the FFT results and compared with the theoretical prediction obtained from \eqref{eq:dispersion}, as shown in Figure \ref{fig:FEM-theory-omega-k-figure}. Excellent agreement is observed, providing a validation of the developed VUMAT subroutine.

Prior to a comprehensive comparison between the theoretical and FE solutions for impact waves, we first describe the numerical integration procedure adopted in the following examples. We exploit the odd and even symmetries implied by \eqref{eq:impact} (see Section \ref{Asymptotic solutions} for a detailed discussion), namely, $\Psi_2^{(m)}$ is an even function of $\xi$ and an odd function of $s$, whereas $\Psi_1^{(m)}$ is odd in both variables. According to these symmetry properties, the integrals in \eqref{eq:sol-u1u2} are simplified to
\begin{equation}
\begin{aligned}\label{eq:sym-u1u2}
u_1(x_1,x_2,t_1)&=-\frac{2}{\pi}\int_{0}^{+\infty}\sum^{\infty}_{m=1}\left[\Psi_1^{(m)}\sin\left(\frac{s^{(m)}}{\mathrm{i}}t_1\right)\sin(\xi x_1)\right]\mathrm{d}\xi,\\u_2(x_1,x_2,t_1)&=\frac{2\mathrm{i}}{\pi}\int_0^{+\infty}\sum^{\infty}_{m=1}\left[\Psi_2^{(m)}\sin\left(\frac{s^{(m)}}{\mathrm{i}}t_1\right)\cos(\xi x_1)\right]\mathrm{d}\xi.
\end{aligned}
\end{equation}
The integrals in \eqref{eq:sym-u1u2} are evaluated numerically over a truncated interval $[\xi_{\min},\,\xi_{\max}]$, rather than the original semi-infinite interval $(0,\,\infty)$. In the following, we demonstrate that this approximation provides high accuracy with a controllable error. To this end, we first examine the asymptotic behavior as $\xi\to0$. A comprehensive asymptotic analysis in both the long- and short-wavelength limits for a compressible bilayer was presented in \citet{liu2026wave}. Thus here we briefly summarize the main asymptotic results.

\begin{figure}[H]
		\centering
		{\subfigure[]{\includegraphics[width=0.49\textwidth]{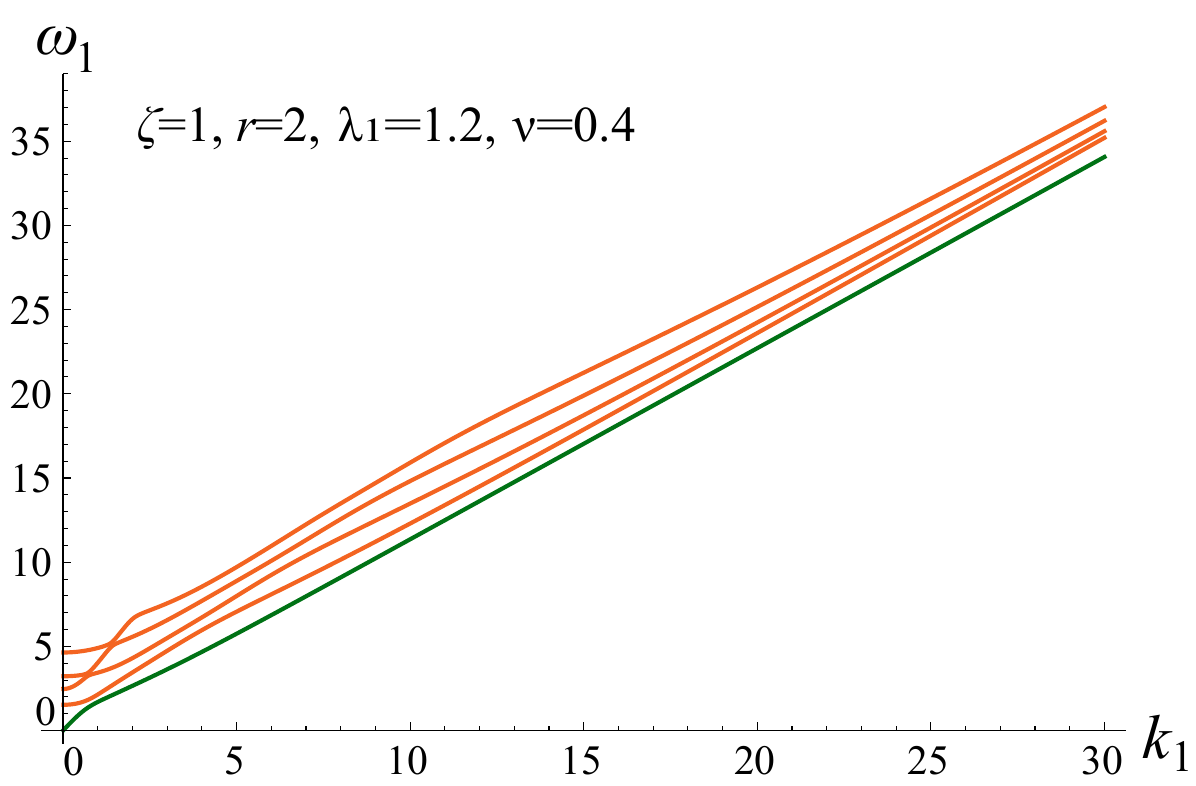}\label{fig:figure-omega1-k1-dispersion-relation-1}}~\subfigure[]{\includegraphics[width=0.49\textwidth]{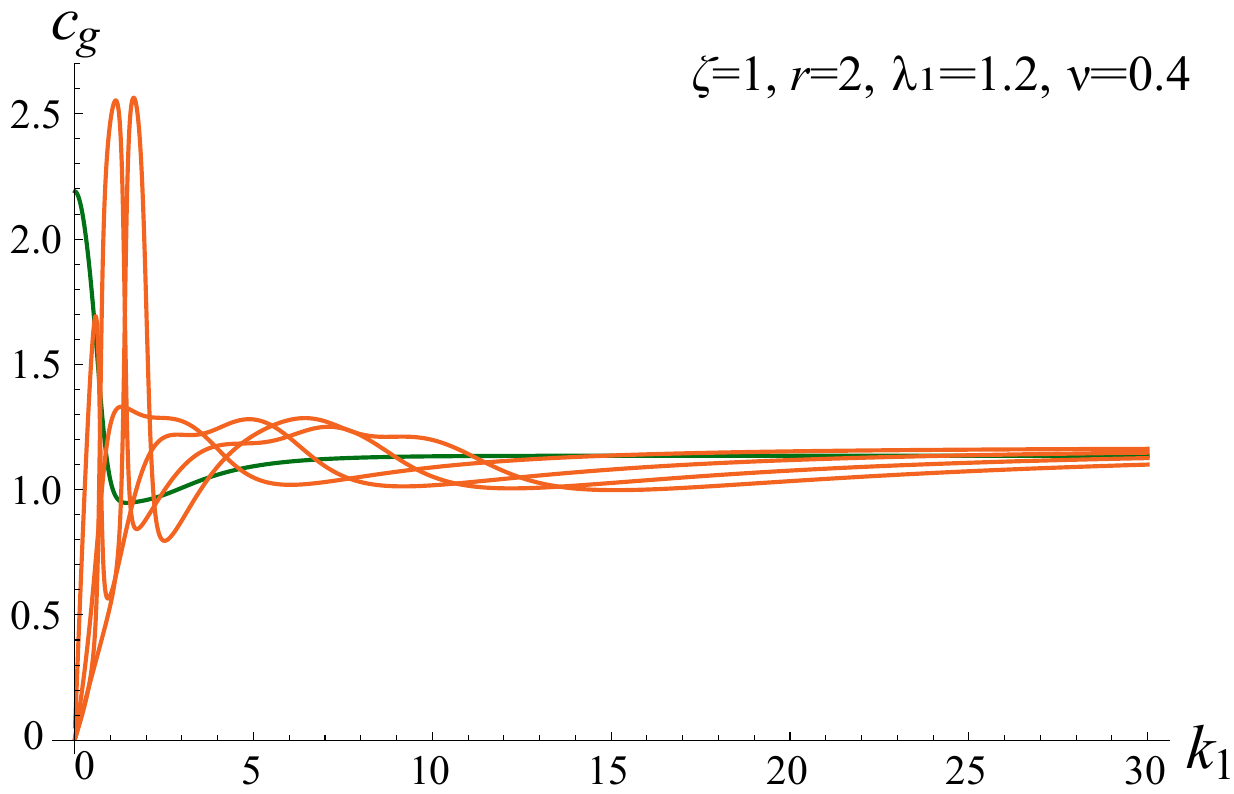}\label{fig:figure-cg-k1-relation-curve-1}}}
            \caption{Dependence of the (a) dimensionless frequency $\omega$ and (b) dimensionless group velocity $c_g$ on the dimensionless wave number $k_1$. The corresponding parameter values are indicated in the figure. The green curves denote the fundamental mode, whereas the orange curves represent the higher-order modes.}
            \label{fig:figure-dispersion-1}
\end{figure}

To facilitate interpretation, we specify $\lambda_1=1.2$, $\nu=0.4$, $r=2$, $\zeta=1$, and plot the dispersion curves together with the group velocity as functions of $k_1$ for the first five modes in Figure \ref{fig:figure-dispersion-1}. The curves corresponding to the fundamental mode are highlighted in green, whereas those for the higher-order modes are shown in orange. In the long-wavelength limit, the dimensionless frequency $\omega_1$ of the fundamental mode approaches zero, whereas it remains finite for the higher-order modes. Similarly, the group velocity of the fundamental mode approaches a finite value in the long-wavelength limit, while the group velocity curves of the higher-order modes originate from the origin. As detailed by \citet{rogerson1996small,wijeyewickrema2009wave,Kayestha2010time}, in the short-wavelength limit, the group velocity of the fundamental mode approaches the minimum of the Rayleigh wave velocity in a half-space composed of the upper layer, the Stoneley wave velocity at the interface between two half-spaces, and the shear and dilatational wave velocities of the two layers. However, for the present setup, \citet{liu2026wave} showed that interface waves are not permitted when $\lambda_1>1$. In contrast, the higher-order modes approach the lesser of the shear and dilatational wave velocities in each layer.

It should be emphasized that the superscript $m$ in \eqref{eq:sym-u1u2} denotes the quantities associated with the $m$th mode. For example, $m=1$ corresponds to the fundamental (first) mode (see the green curve in Figure \ref{fig:figure-omega1-k1-dispersion-relation-1}). For a given value of $\xi$, the pair $(\xi,s^{(m)})$ represents a point on the dispersion curve of the $m$th mode. As shown in Figure \ref{fig:figure-dispersion-1}, the fundamental mode has a zero cut-off frequency as $\xi\to 0$, while its corresponding group velocity approaches a finite value. In contrast, for the higher-order modes, the cut-off frequencies remain finite, whereas the corresponding group velocities approach zero. Following \citet{rogerson1996small, lashhab2015dispersion}, we can deduce that $s^{(1)}\sim\xi$, whereas $s^{(m)}\sim 1$, for $m>1$. A careful asymptotic analysis of the integrands in \eqref{eq:sym-u1u2} then yields, as $\xi\to 0$,
\begin{equation}
\left\{\begin{aligned}
     &\Psi_1^{(m)}\sin\left(\frac{s^{(m)}}{\mathrm{i}}t_1\right)\sin(\xi x_1)\sim \xi^2,\quad \Psi_2^{(m)}\sin\left(\frac{s^{(m)}}{\mathrm{i}}t_1\right)\cos(\xi x_1)\sim \xi^2,\quad m=1,\\
     &\Psi_1^{(m)}\sin\left(\frac{s^{(m)}}{\mathrm{i}}t_1\right)\sin(\xi x_1)\sim \xi^3,\quad \Psi_2^{(m)}\sin\left(\frac{s^{(m)}}{\mathrm{i}}t_1\right)\cos(\xi x_1)\sim \xi,\quad m=2,3,\ldots.
\end{aligned}\right.
\end{equation}
Consequently, provided that $\xi_{\min}$ is sufficiently small, the contributions from the interval $(0,\xi_{\min})$ to both integrals in \eqref{eq:sym-u1u2} are of order $o(\xi_{\min})$.

We next justify the choice of the upper truncation limit $\xi_{\max}$. In all subsequent examples, the value of $\xi_{\max}$ is selected on a case-by-case basis. The sole criterion is that further increasing $\xi_{\max}$ does not produce any noticeable change in the computed wave profiles. Moreover, a uniform discretization is adopted, with the interval $[\xi_{\min},\,\xi_{\max}]$ divided into $999$ equal subintervals. Our numerical calculations also indicate that the dominant contribution arises from the first few modes. In practice, retaining only the first five modes $(m=1,2,\ldots,5)$ in \eqref{eq:sym-u1u2} provides sufficient accuracy.

\begin{figure}[H]
		\centering
		{\subfigure[]{\includegraphics[width=0.9\textwidth]{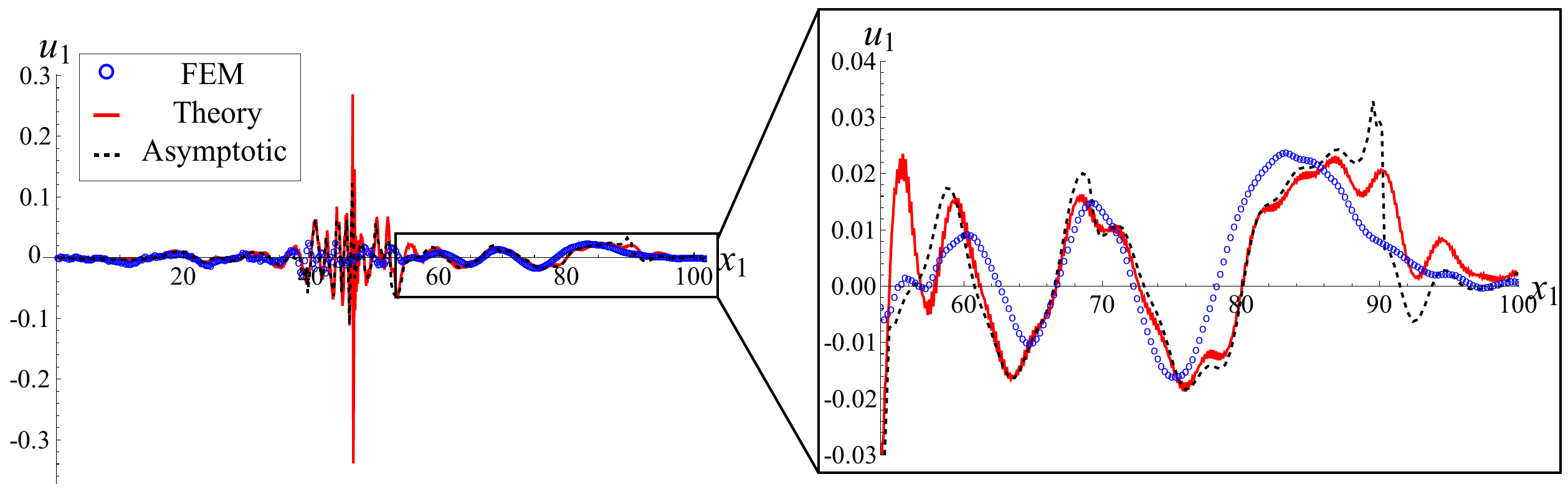}\label{fig:figure-u1-x1-verify-model-1}}\\\subfigure[]{\includegraphics[width=0.9\textwidth]{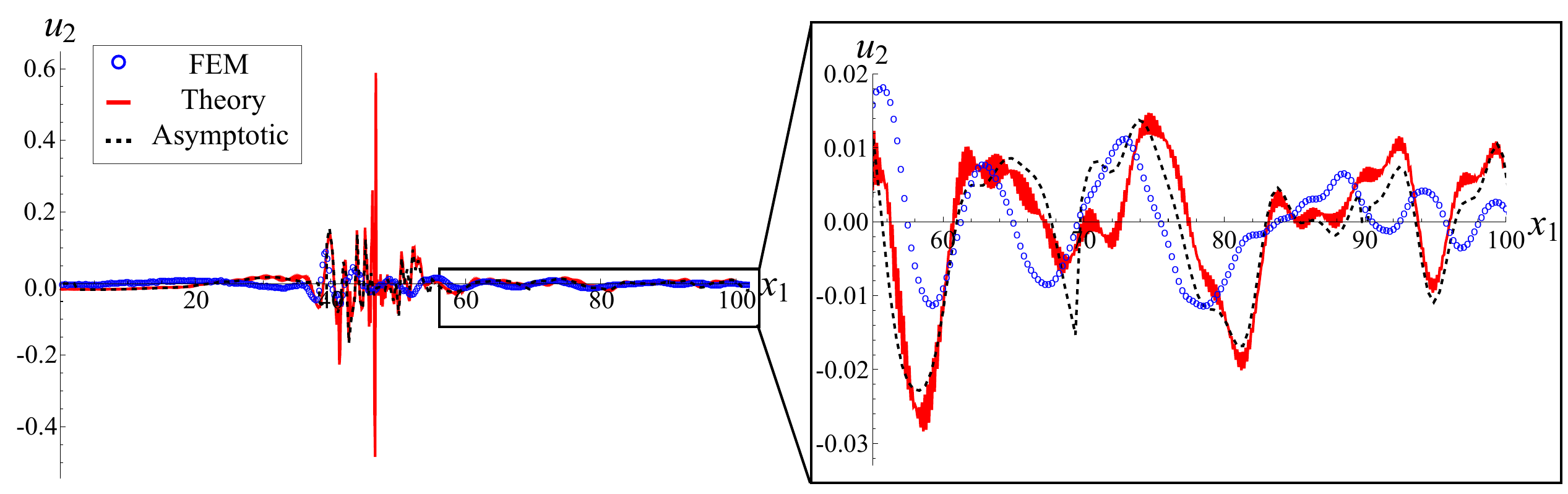}\label{fig:figure-u2-x1-verify-model-1}}}
            \caption{Distribution of the displacements $u_1$ and $u_2$ on the upper surface ($x_2=1$) at $t_1=42.17$. The parameters are $\lambda_1=1.2$, $\nu=0.4$, $r=2$, and $\zeta=1$. An enlarged view of the displacement profiles near the wavefront is also provided to highlight the agreement among the theoretical solution obtained by numerical integration (with the first five modes retained, i.e., $M=5$), the asymptotic solution, and the FE simulation.}
            \label{fig:figure-verify-model-1}
\end{figure}

Taking $\lambda_1=1.2$, $\nu=0.4$, $r=2$, and $\zeta=1$, we plot the horizontal displacement $u_1$ and the vertical displacement $u_2$ on the upper surface ($x_2=1$) at $t_1=42.17$ in Figure \ref{fig:figure-verify-model-1}. The numerical integration is performed over the interval $[0.03,30]$ with a uniform mesh size of $0.03$. For comparison, the theoretical solution obtained by numerical integration (red solid lines), the asymptotic solution (black dotted lines), and the FE result (blue open circles) are presented. In our FE simulations, the spatial mesh size and time step are set to $\Delta x_1=0.05$ and $\Delta t_1=0.005$, respectively. Periodic boundary conditions are imposed at both sides. The sensitivity of the FE results to the mesh size, time step, and boundary conditions is further examined in  \ref{Sensitivity}. In both panels, the sharp high-frequency feature near $x_1\approx47$ is located close to the arrival position predicted from the Rayleigh-wave velocity of the corresponding half-space \citep{weaver1982axisymmetric,erbay2000effects}. However, this feature is not fully resolved in the FE simulations. One factor contributing to this discrepancy is the difference in the applied excitations. The theoretical solution assumes an idealized point impulse, $g(x_1,t_1)=-\delta(x_1)\delta(t_1)$, whereas the FE simulation necessarily approximates this excitation using a spatially distributed load over $[-1/2,1/2]$ applied over a finite time interval $[0,1]$. Such spatial and temporal regularization suppresses the high-frequency components of the excitation, thereby partially accounting for the unresolved high-frequency feature.

Ahead of the wave peak corresponding to the velocity of Rayleigh surface wave of the half-space are relatively low-amplitude, low-frequency wave components. As shown in Figure \ref{fig:figure-verify-model-1}, good agreement is observed between the numerical integration and the asymptotic solution. Overall, despite some quantitative deviations, the theoretical predictions and the FE results agree closely when $x_1>55$ in both the wavelength and amplitude of the propagating disturbance. Therefore, the good agreement between the analytical and FE results over the resolvable wavelength range provides independent numerical support for the theoretical predictions. 

\begin{figure}[H]
		\centering
\includegraphics[width=0.49\textwidth]{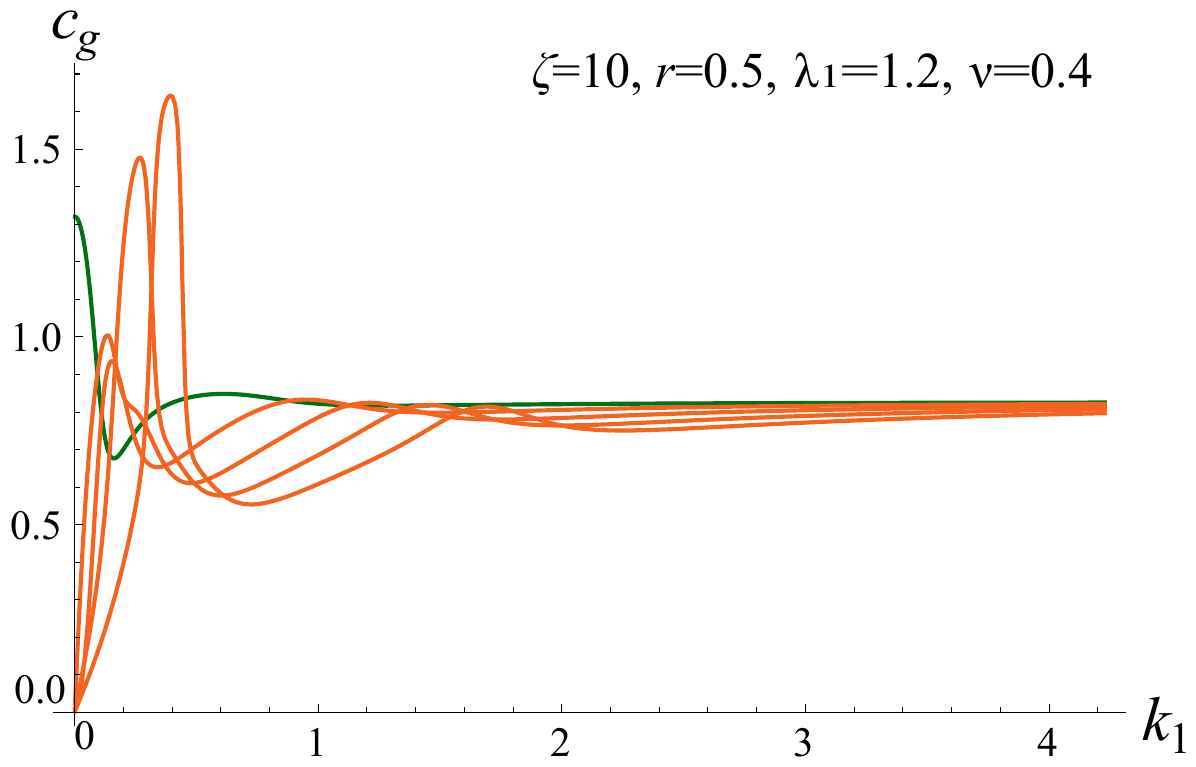}
            \caption{Dependence of the group velocity $c_g$ on $k_1$ for the first five modes. The parameter values are shown in the figure. The green curve corresponds to the fundamental mode, whereas all orange curves represent the higher-order modes.}
            \label{fig:figure-dispersion-2}
\end{figure}

\begin{figure}[H]
		\centering
		{\subfigure[]{\includegraphics[width=0.49\textwidth]{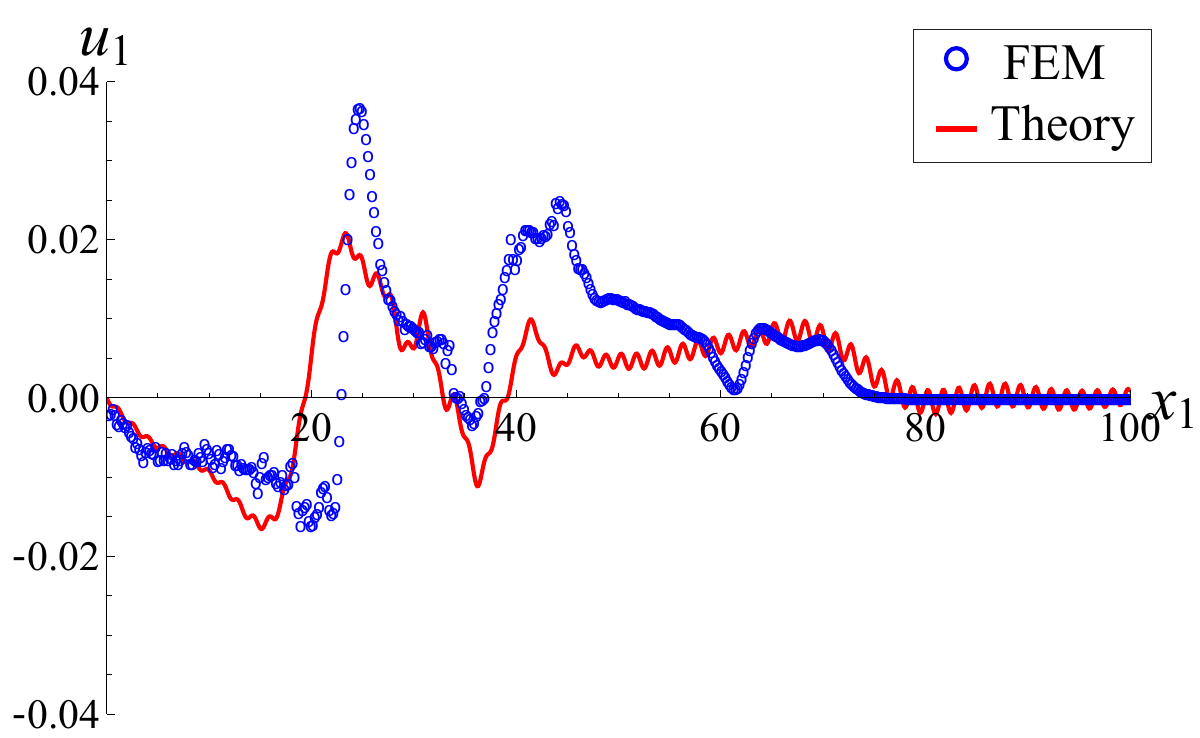}\label{fig:figure-u1-x1-verify-model-2}}~\subfigure[]{\includegraphics[width=0.49\textwidth]{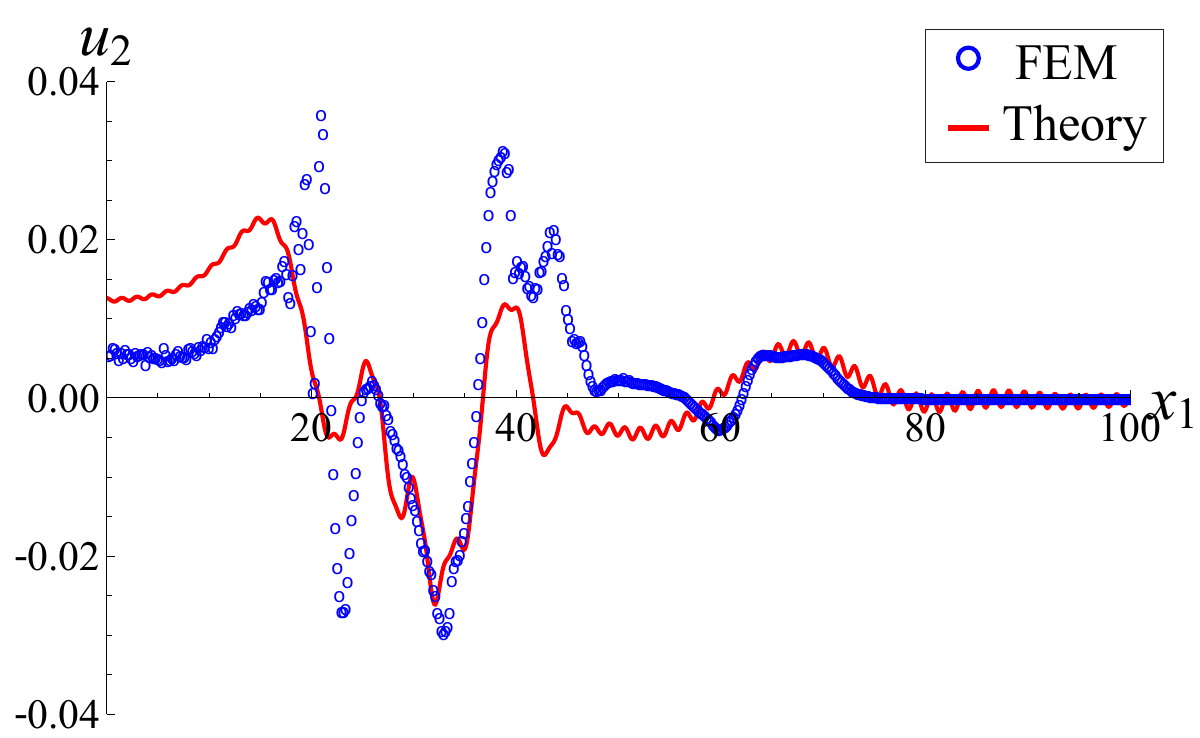}\label{fig:figure-u2-x1-verify-model-2}}}
            \caption{Comparison of the displacements $u_1$ and $u_2$ on the upper surface ($x_2=1$) at $t_1=42.17$ between the theoretical predictions (red solid curves) and FE results (open circles). The parameters are $\lambda_1=1.2$, $\nu=0.4$, $r=0.5$, and $\zeta=10$.}
            \label{fig:figure-verify-model-2}
\end{figure}

In the first example, the upper layer is softer than the bottom layer ($r=2$), and the two layers have the same thickness ($\zeta=1$). We next consider another scenario in which the bottom layer is ten times thicker than the top layer while becoming softer, with parameters $\lambda_1=1.2$, $\nu=0.4$, $r=0.5$, and $\zeta=10$. Similarly, the displacements $u_1$ and $u_2$ at $x_2=1$ and $t_1=42.17$ are plotted in Figure \ref{fig:figure-verify-model-2}. The FE results are represented by blue open circles, whereas the theoretical predictions are shown as red solid curves. The numerical integration is performed over the interval $[0.004,\,4]$ using a uniform mesh size of $0.004$. Unlike the previous case, the numerical integration converges much more rapidly, so $\xi_{\max}=4$ is sufficient to achieve the desired accuracy. To elucidate this difference, the group velocity $c_g$ is plotted as a function of $k_1$ in Figure \ref{fig:figure-dispersion-2}. Compared with Figure \ref{fig:figure-cg-k1-relation-curve-1}, it can be seen that the group velocities of all wave modes rapidly approach their short-wavelength limiting value.

In contrast to the wave profile displayed in Figure \ref{fig:figure-verify-model-1}, wave propagation in the case of a thicker bottom layer exhibits markedly different characteristics. From the short-wavelength limiting value of the group velocity $c_g$ for the fundamental mode (green curve in Figure \ref{fig:figure-dispersion-2}), the arrival of the wave peak is estimated to occur at approximately $x_1\approx34$ in Figure \ref{fig:figure-verify-model-2}. Seen from Figure \ref{fig:figure-u2-x1-verify-model-2}, this location corresponds to the deepest wave trough. In particular, the high-frequency oscillations surrounding the wave peak disappear, and the preceding wave components no longer exhibit pronounced harmonic oscillations. Therefore, the method of stationary phase is no longer applicable to this case. In addition, the trailing waves exhibit a slower decay than in the previous example. Similar to the previous case, the FE results show good qualitative agreement with the theoretical predictions. Overall, because the dominant wave signals on the upper surface are primarily low-frequency, they are expected to be more readily captured in experimental measurements.

To quantitatively assess the agreement between the theoretical and FE solutions, we introduce the normalized root mean square error (NRMSE) as a measure of the discrepancy between the two curves. Specifically, for two curves $f_1(z)$ and $f_2(z)$, the NRMSE is defined as \citep{willmott1985statistics,ma2025research,moghaddasi2025reinforcing},
\begin{equation}
\mathrm{NRMSE}(f_1(z),f_2(z))=
\frac{\sqrt{\frac{1}{K}\sum_{i=1}^{K}\left(f_1(z_i)-f_2(z_i)\right)^2}}
{\max(f_1(z))-\min(f_1(z))},
\quad z\in(z_1,z_2),
\label{eq:NRMSE-formulation}
\end{equation}
where $K$ denotes the number of discretization points within the interval $(z_1,z_2)$.

Taking the FE and theoretical results as $f_1$ and $f_2$, respectively, we evaluate the discrepancy between the FE and theoretical displacement distributions in Figure \ref{fig:figure-verify-model-1} over the wavefront region, i.e., $x_1\in(60,100)$, using $K=800$ discretization points. For Figure \ref{fig:figure-verify-model-2}, we use $K=2000$ points over the interval $x_1\in(0,100)$. The results are summarized in Table \ref{tab:NRMSE-of-FE}. All NRMSE values are below 5\%, indicating good agreement between the theoretical predictions and FE results within the considered regions.

\begin{table}[htbp]
\centering
\caption{NRMSEs between the FE and theoretical solutions.}\vspace{3mm}
\label{tab:NRMSE-of-FE}
\begin{tabular}{lcc}
\hline
Case & NRMSE of $u_1$ (\%) & NRMSE of $u_2$ (\%)\\
\hline
Figure \ref{fig:figure-verify-model-1} & 3.30 & 4.87\\
Figure \ref{fig:figure-verify-model-2} & 3.15 & 2.92\\
\hline
\end{tabular}
\end{table}

So far, we have primarily validated the proposed theoretical model by comparing its predictions with direct FE simulations of impact wave propagation in soft bilayers. While the two approaches exhibit good qualitative agreement, the theoretical framework offers several distinct advantages over direct FE simulations. In particular, it provides clear physical insight into the underlying wave phenomena, enables efficient exploration of a broad range of material and geometric parameters, and facilitates asymptotic analyses that reveal the mechanisms governing wave propagation. In the next section, we apply the proposed theoretical model to several representative biological tissues to investigate how subtle variations in material properties and geometry influence the characteristics of impact wave propagation.

\section{Applications to soft tissues}\label{Applications to soft tissues}

\begin{figure}[H]
		\centering      \includegraphics[width=1\linewidth]{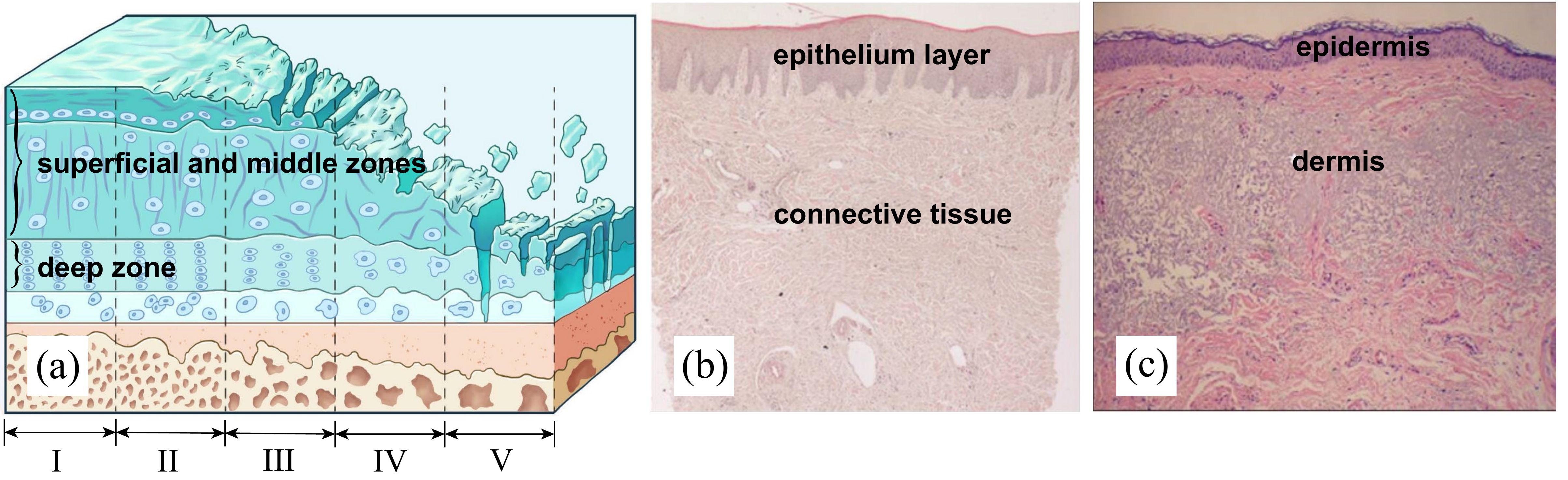}
        \caption{Layered structures of human articular cartilage, gingiva, and skin. (a) Schematic of the articular cartilage structure and the progression of osteoarthritis, with Stages I–V representing the transition from healthy cartilage to severe osteoarthritis. (b) Morphological architecture of the gingiva. (c) Stratified organization of the skin. Panel (a) is adapted from \citep{zhang2024research} (CC BY 4.0); panel (b) is adapted from \citep{francetti2019morphological} (CC BY 4.0); and panel (c) is adapted from \citep{barcaui2015study} (CC BY 4.0).}\label{fig:layered tissue}
\end{figure}

The semi-analytical and asymptotic solutions for the displacement fields have been derived in Sections \ref{Impact waves} and \ref{Asymptotic solutions}, respectively. In this section, we apply these solutions to several layered soft tissues to understand the propagation of impact waves in such media. Specifically, we consider three representative layered structures: human articular cartilage, gingiva, and skin. The layered configurations of these biological tissues are illustrated in Figure \ref{fig:layered tissue}. In modeling biological tissues, we assume that they are supported by a rigid and frictionless substrate, such that the theoretical framework developed in previous sections is applicable. Although this is an idealized approximation, it allows us to isolate the effects of bilayer geometry and material contrast on the impact response. Nevertheless, the results presented below are expected to provide useful insights into impact wave propagation in soft tissues under more realistic boundary conditions.

We note that optical coherence elastography (OCE) has recently been employed to experimentally characterize the mechanical properties of layered biological tissues \citep{singh2025optical}. For example, \citet{feng2023ultra} measured the shear moduli of bovine articular cartilage and human skin using OCE. Subsequently, the same technique was applied to the mechanical characterization of porcine gingiva by \citet{moon2024high}. More recently, \citet{li2025wireless} developed a wireless mechano-acoustic wave sensing technique for elastographic imaging of human skin, and \citet{ghaderi2026multilayered} proposed a material characterization method based on surface wave elastography
and laser Doppler vibrometry. These studies primarily relied on the Rayleigh wave velocity extracted from the dispersion relation of the fundamental mode. However, both experimentally measured and numerically simulated data directly correspond to displacement fields rather than dispersion curves or phase velocities. Moreover, neglecting the contributions of higher-order modes may be questionable in certain parameter regimes.

Motivated by these observations, rather than performing a comprehensive parametric study of impact wave propagation over a broad parameter space, which would be computationally expensive and of limited physiological relevance, we focus on how subtle changes in the modulus ratio $r$ and the thickness ratio $\zeta$ associated with specific diseases affect the characteristics of impact wave propagation. Our primary objective is to provide physical insight into the potential relevance to non-invasive characterization and detection of pathological changes by identifying distinctive wave signatures generated by external impact loading.

As demonstrated in Figures \ref{fig:figure-verify-model-1} and \ref{fig:figure-verify-model-2}, the wave profiles can exhibit highly intricate features. The asymptotic solutions of the displacement responses capture some features of the wave profiles. Since the required high-frequency oscillatory behavior of the integrand is uniquely exhibited by the human articular cartilage model among three models, asymptotic solutions are presented exclusively for this model. To facilitate the interpretation of the results presented in the following analysis, we utilize the corresponding group velocity curves, which are intrinsically related to the displacement response. In particular, we identify four characteristic velocities, denoted by $c_g^{\mathrm{l}}$, $c_g^{\mathrm{s}}$, $c_g^{\mathrm{min}}$, and $c_g^{\mathrm{max}}$. Here, $c_g^{\mathrm{l}}$ and $c_g^{\mathrm{s}}$ represent the long- and short-wavelength limiting group velocities, respectively, whereas $c_g^{\mathrm{min}}$ and $c_g^{\mathrm{max}}$ denote the local minimum and maximum of the group velocity curves. For completeness, the group velocity curves corresponding to all cases considered are provided in~\ref{Appendix-B}. Throughout the remainder of the paper, the wave components associated with these four characteristic velocities are referred to as wave types I–IV, respectively. For instance, wave type I wave type I is associated with the group velocity $c_g^{\mathrm{l}}$. It should be noted that the characteristic velocities and associated wave types do not necessarily correspond precisely to the peaks or troughs in the displacement responses. Instead, they serve as references for interpreting the wave-propagation characteristics and dominant oscillatory features.

\subsection{Human articular cartilage}\label{cartilage}

As illustrated in Figure \ref{fig:layered tissue}(a), human articular cartilage is typically composed of the superficial, middle, deep, and calcified zones, with the calcified zone lying beneath the deep zone and connecting the cartilage to the underlying bone. Osteoarthritis poses a threat to human health, as its onset is accompanied by progressive surface fibrillation, disruption of the collagen network, and loss of proteoglycans within articular cartilage \citep{tschaikowsky2022articular}. These degenerative changes are further manifested by pronounced cartilage thinning and deterioration of the layer-specific mechanical properties \citep{zhang2024research}. The five stages of osteoarthritis progression are highlighted in Figure \ref{fig:layered tissue}(a). 

To capture these pathological changes, we construct a series of simplified bilayer models representing healthy, moderate osteoarthritis, and severe osteoarthritis states. Unlike the three-layer model adopted by \citet{feng2023ultra}, in which the tissue is divided into non-calcified cartilage, calcified cartilage, and bone, we treat the calcified cartilage and the underlying bone as rigid because our primary interest lies in wave propagation within the deformable cartilage. Accordingly, the upper layer of the bilayer model represents the superficial and middle zones, whereas the lower layer represents the deep zone. The parameters of the healthy model ($\zeta=3/7$, $r=3$, $\nu=0.4$) are adopted from \citet{rayes2025high}. Since we focus on the change of material parameters, the pre-stretch is fixed at $\lambda_1=1.2$ for all examples. Consistent with the progression of osteoarthritis, the diseased models are characterized by an increased thickness ratio ($\zeta=2/3$ for moderate osteoarthritis and $\zeta=4/5$ for severe osteoarthritis), reflecting cartilage thinning, together with a reduced modulus ratio ($r=2$ for moderate osteoarthritis and $r=3/2$ for severe osteoarthritis), representing the diminished stiffness contrast between the superficial and deep regions. The corresponding group velocity curves for the fundamental mode are shown in Figure \ref{fig:cg-k1-model1}. Since the Poisson's ratio generally has a marginal influence on the dispersion curves, its value is fixed across the three models considered here \citep{liu2026wave}.

\begin{figure}[htbp]
     \centering
      {\subfigure[]{\includegraphics[width=0.49\textwidth]{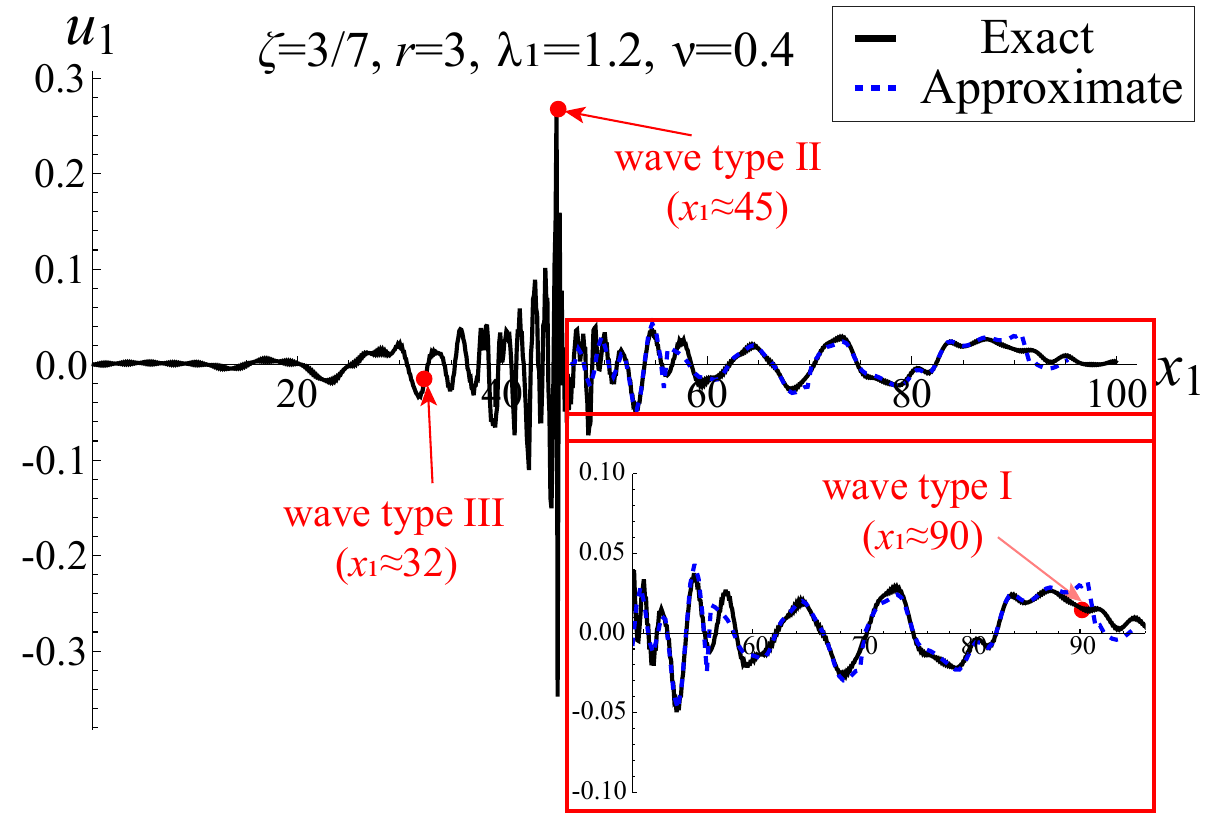}\label{fig:figure-u1-x1-model-group1-1}}~\subfigure[]{\includegraphics[width=0.49\textwidth]{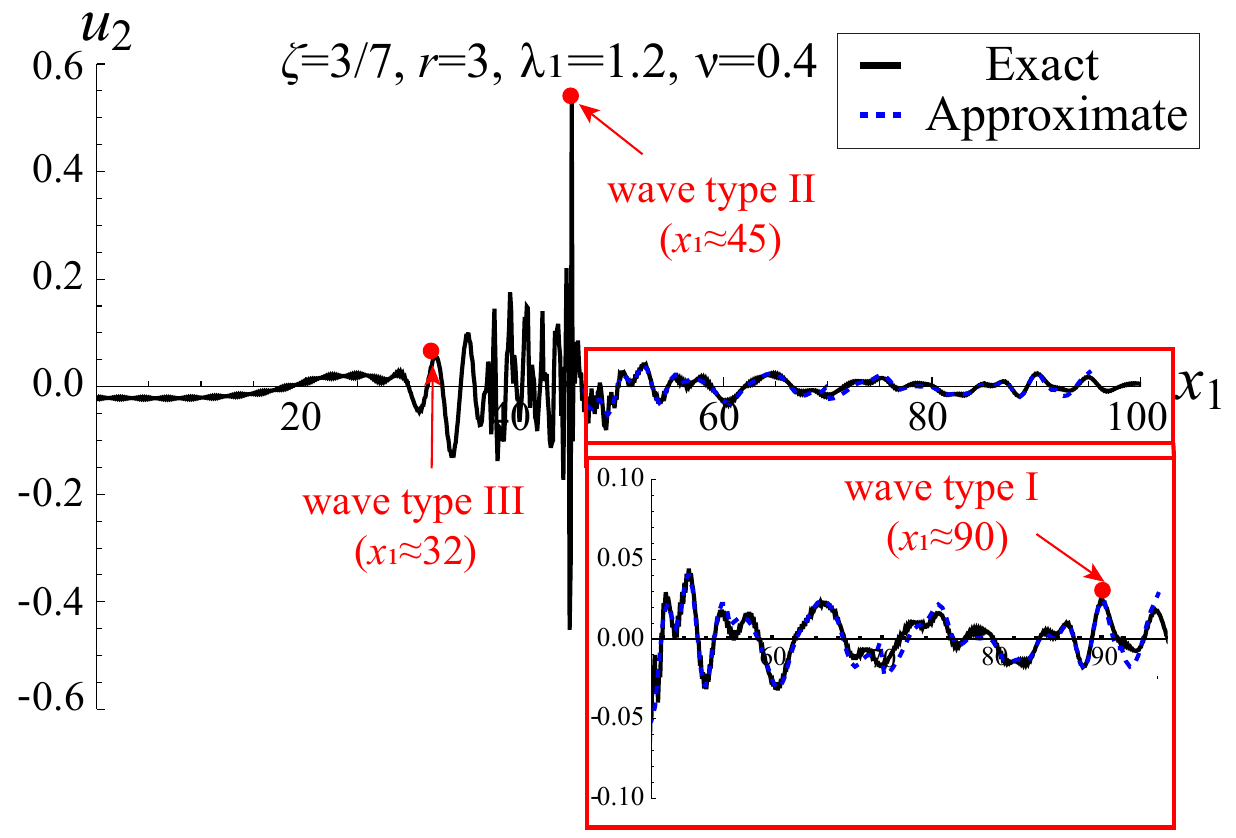}\label{fig:figure-u2-x1-model-group1-1}}\\\subfigure[]{\includegraphics[width=0.49\textwidth]{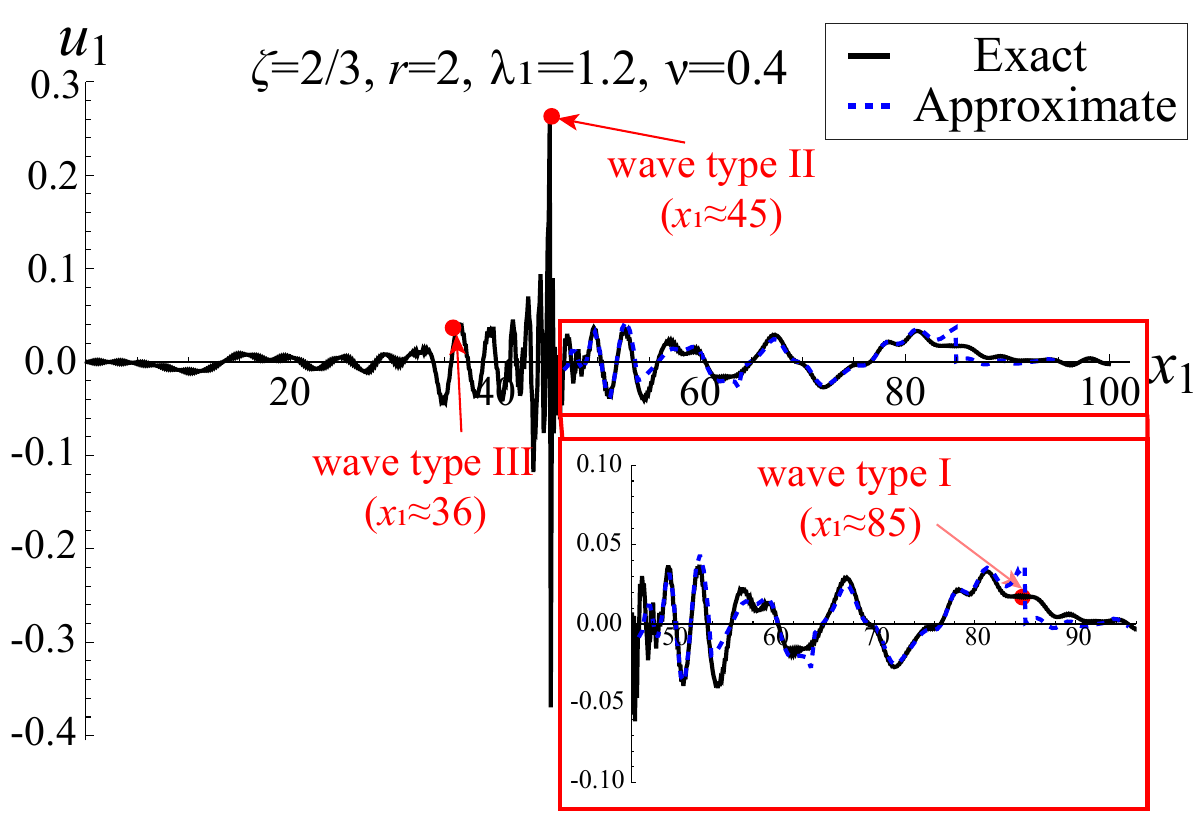}\label{fig:figure-u1-x1-model-group1-2}}~\subfigure[]{\includegraphics[width=0.49\textwidth]{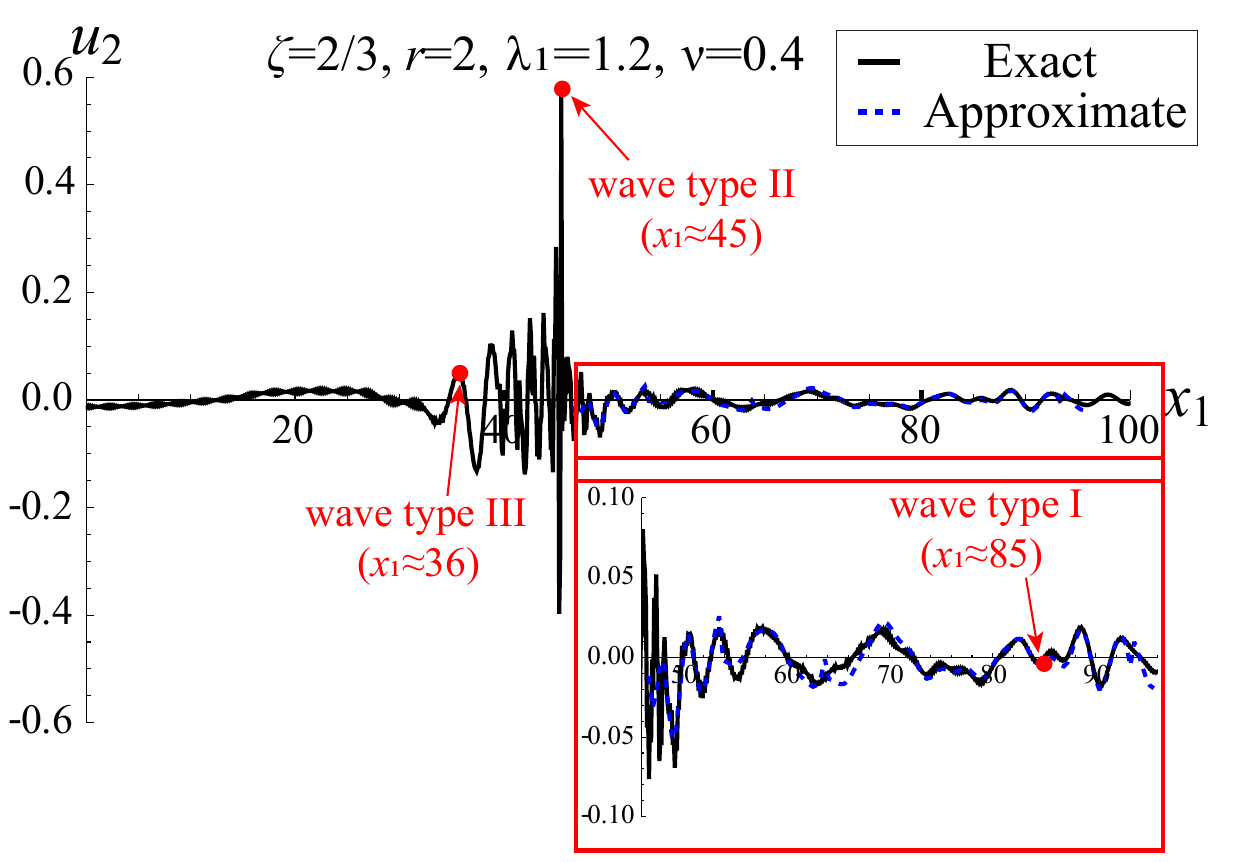}\label{fig:figure-u2-x1-model-group1-2}}\\\subfigure[]{\includegraphics[width=0.49\textwidth]{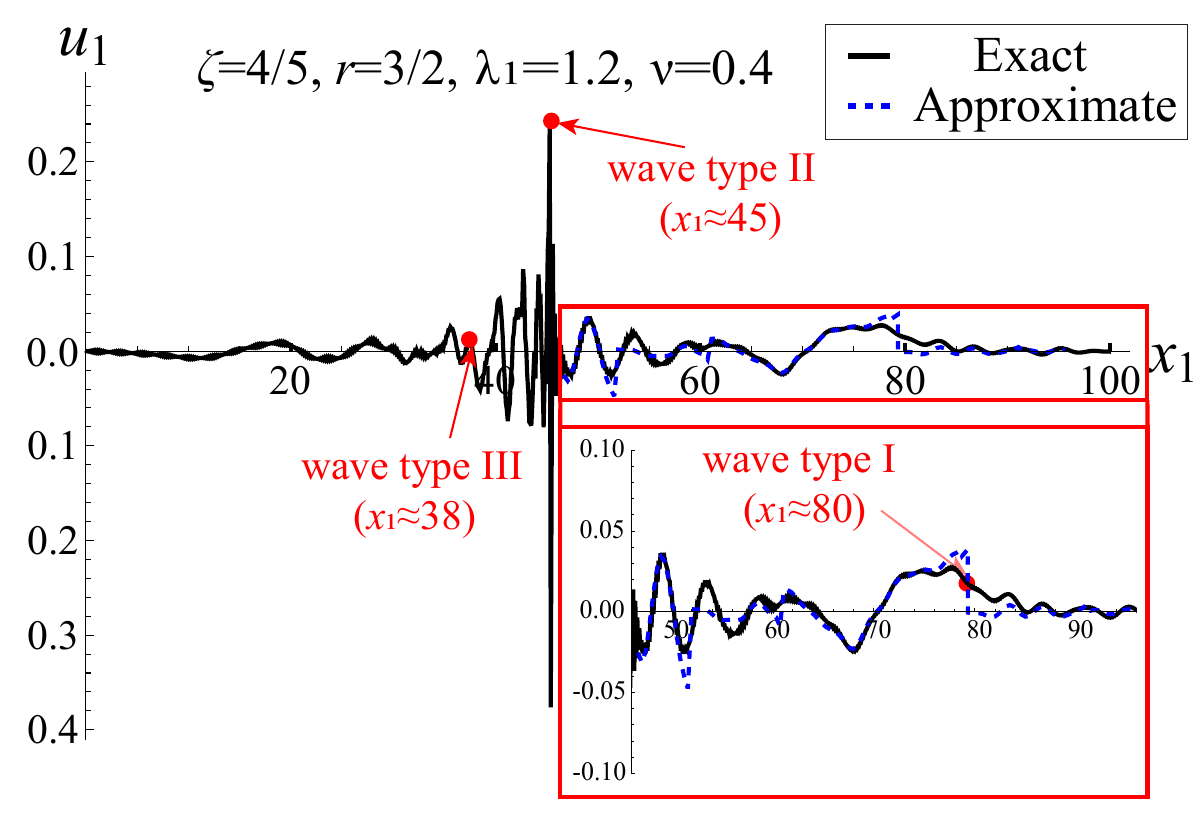}\label{fig:figure-u1-x1-model-group1-3}}~\subfigure[]{\includegraphics[width=0.49\textwidth]{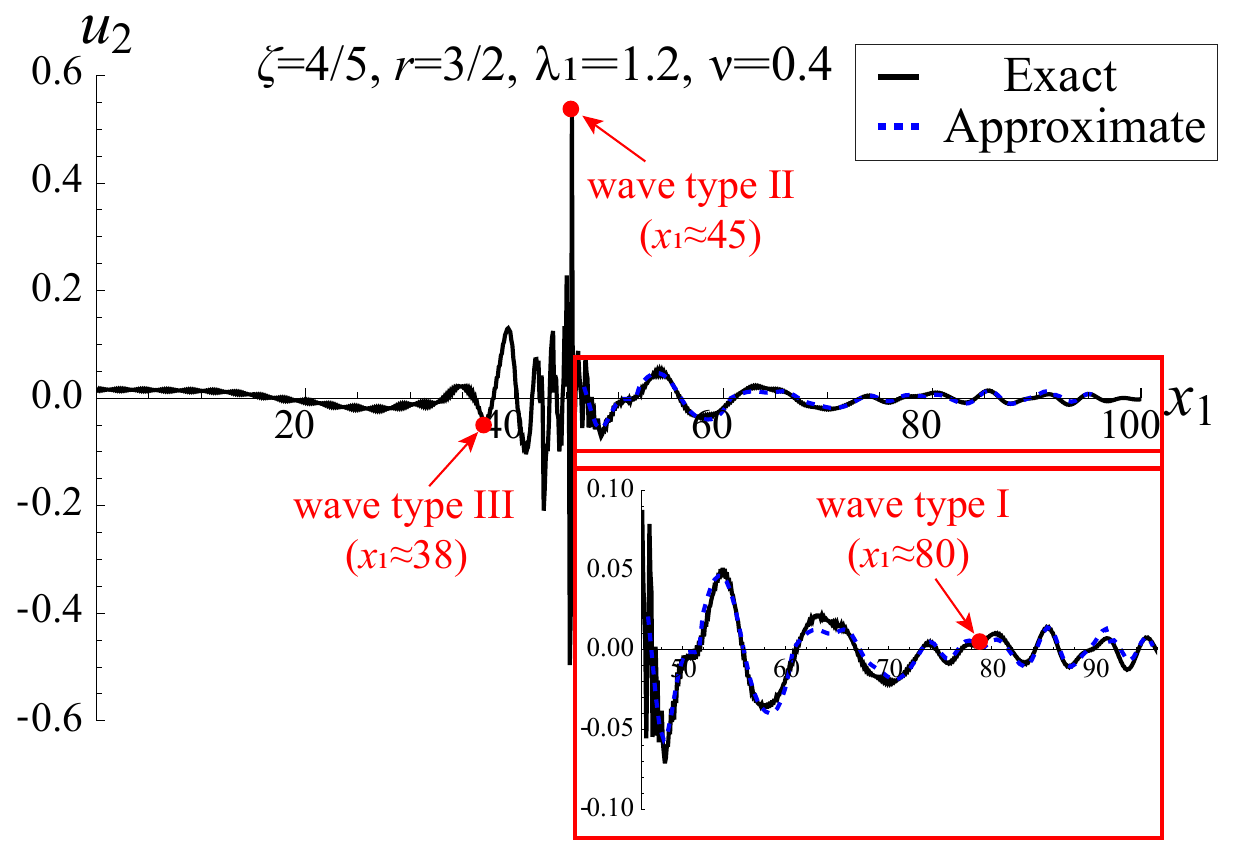}\label{fig:figure-u2-x1-model-group1-3}}}\caption{Dependence of the horizontal displacement $u_1$ and vertical displacement $u_2$ on $x_1$ for three articular cartilage models at $t_1=40$. The parameters are indicated in each panel, representing normal, moderate, and severe conditions in panels (a)–(b), (c)–(d), and (e)–(f), respectively. The semi-analytical solutions obtained by numerical integration (black solid curves) are compared with the asymptotic solutions (blue dashed curves), with detailed comparisons provided in the enlarged regions. The red dots indicate the locations of specific wave components.}\label{fig:figure-x1-model-group1}
\end{figure}

\begin{figure}[htbp]
     \centering
     {\subfigure[]{\includegraphics[width=0.49\textwidth]{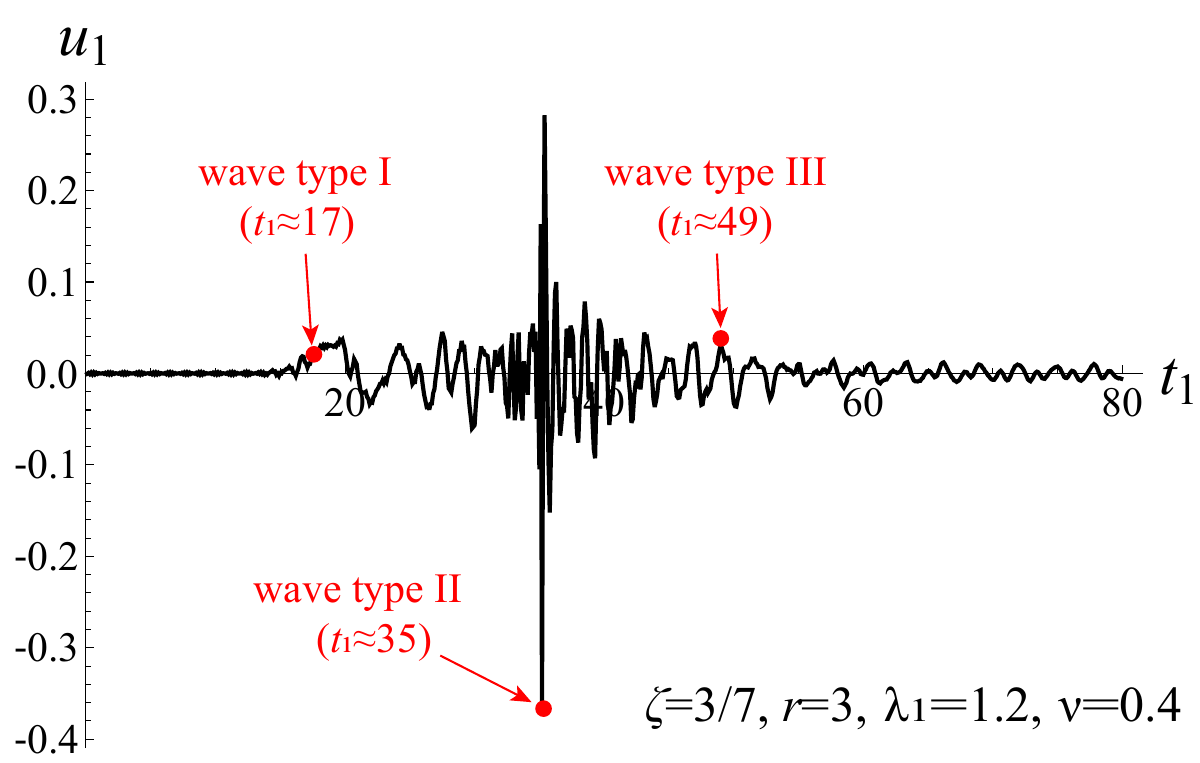}\label{fig:figure-u1-t-model-group1-1}}~\subfigure[]{\includegraphics[width=0.49\textwidth]{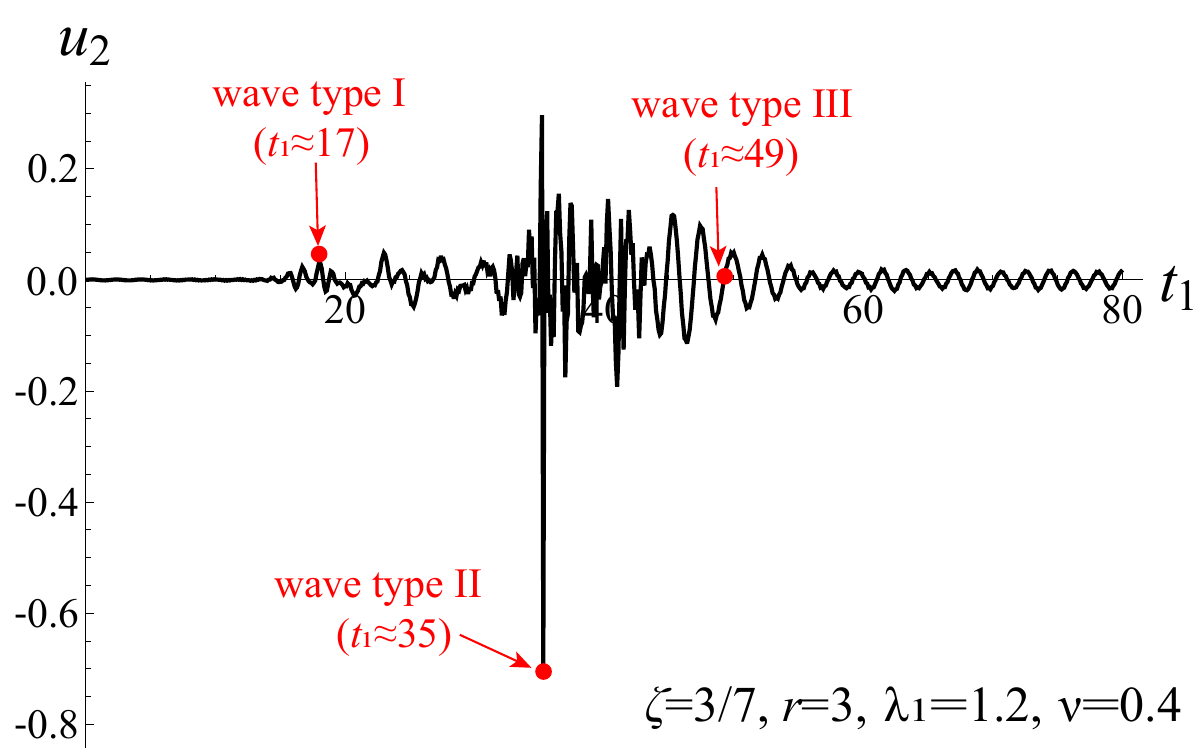}\label{fig:figure-u2-t-model-group1-1}}\\\subfigure[]{\includegraphics[width=0.49\textwidth]{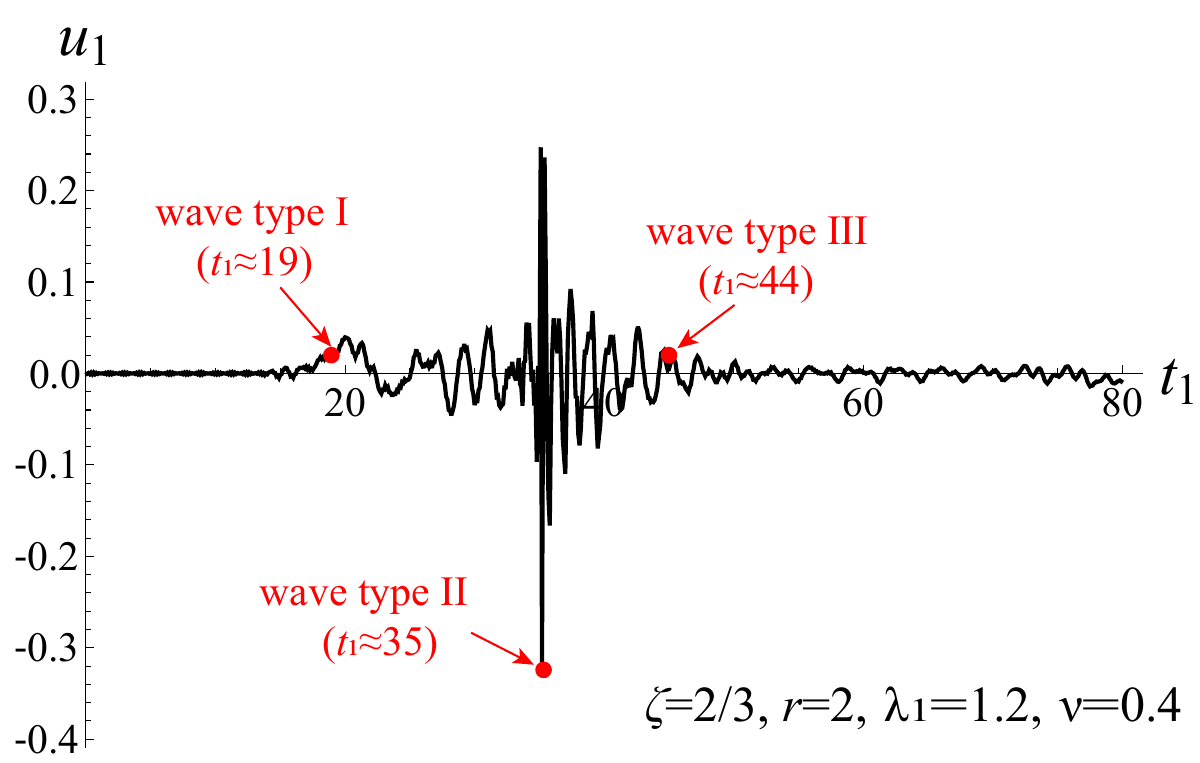}\label{fig:figure-u1-t-model-group1-2}}~\subfigure[]{\includegraphics[width=0.49\textwidth]{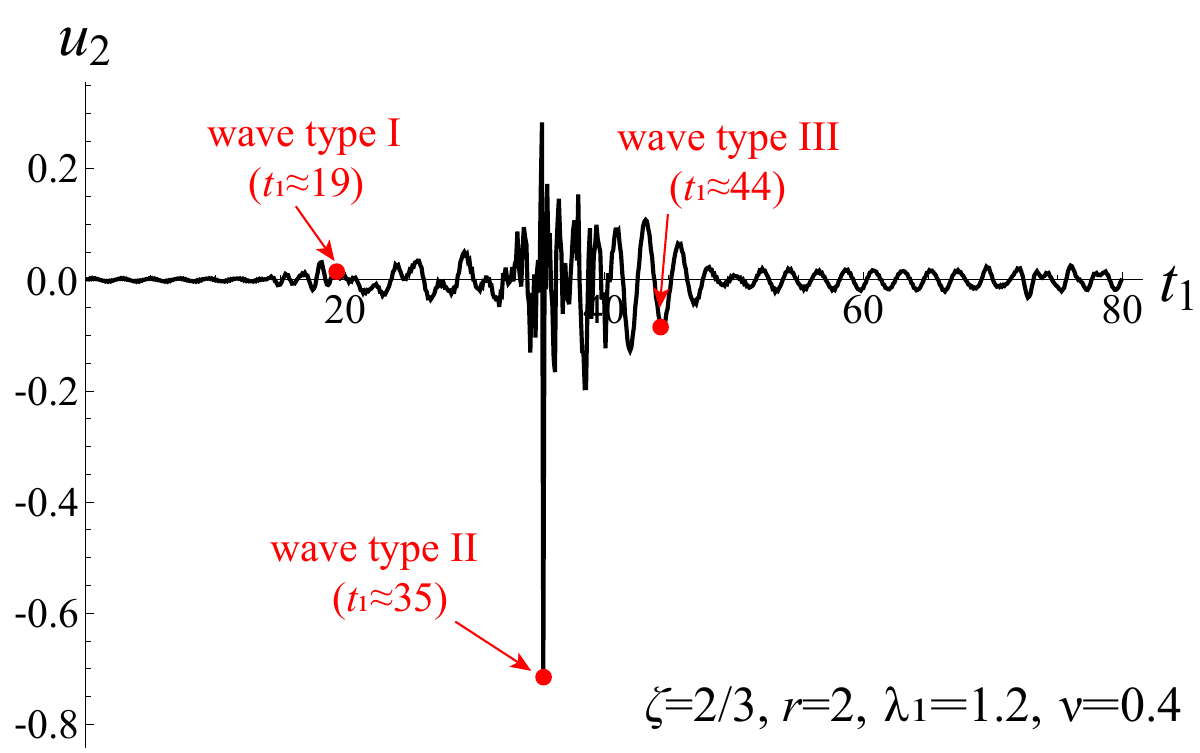}\label{fig:figure-u2-t-model-group1-2}}\\\subfigure[]{\includegraphics[width=0.49\textwidth]{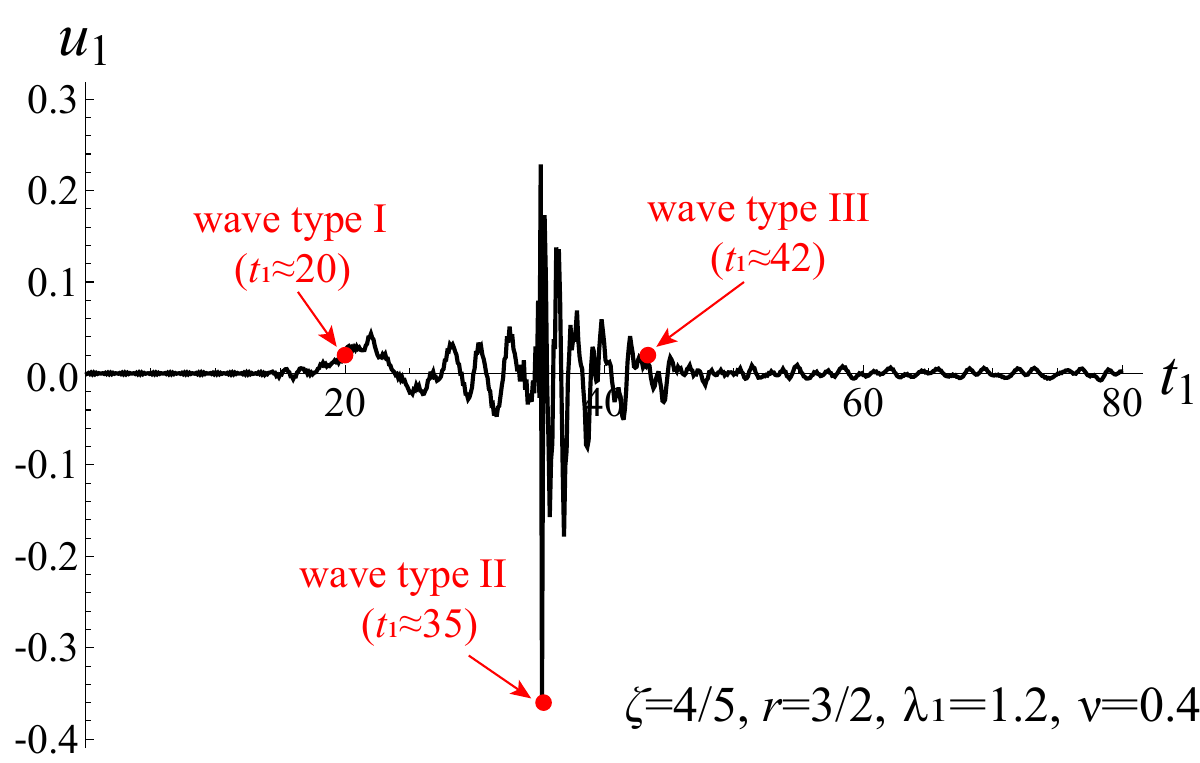}\label{fig:figure-u1-t-model-group1-3}}~\subfigure[]{\includegraphics[width=0.49\textwidth]{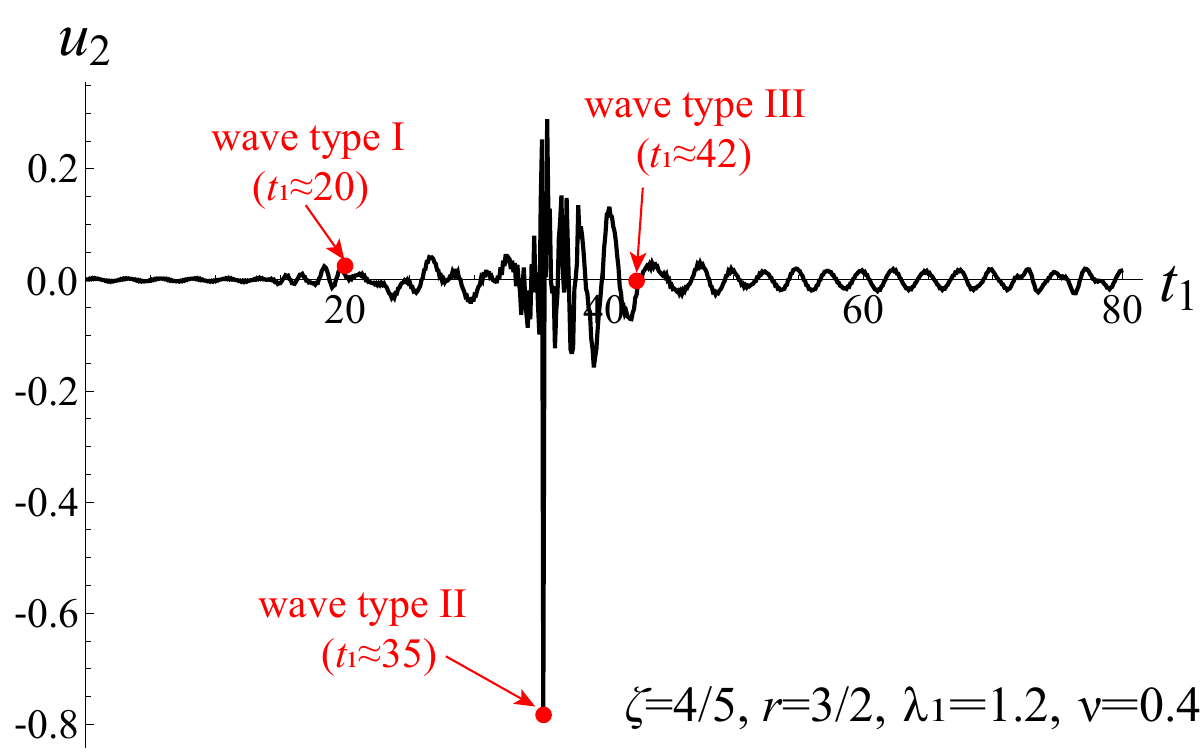}\label{fig:figure-u2-t-model-group1-3}}}\caption{Dependence of the horizontal displacement $u_1$ and vertical displacement $u_2$ on $t_1$ for three cartilage models at $x_1=40$. The parameters are presented in each panel, corresponding to normal, moderate, and severe conditions in (a)-(b), (c)-(d), and (e)-(f), respectively. The red dots indicate the locations of specific wave components.}\label{fig:figure-t-model-group1}
\end{figure}

We note that the numerical integration is performed over the interval $[0.03,\,30]$ using a uniform mesh size of $0.03$. Figure \ref{fig:figure-x1-model-group1} shows the displacement distributions at a fixed time $t_1=40$. Both the analytical solution based on \eqref{eq:sym-u1u2} (black solid curves) and the asymptotic solution given by \eqref{eq:app-sol} (blue dashed curves) are presented for comparison. In particular, the asymptotic solutions are only shown after the arrival of wave type II. Excellent agreement is observed in all three scenarios, indicating that the wave profile is inherently highly oscillatory. In this regime, destructive interference largely cancels out the rapid oscillations, and the macroscopic wave response is concentrated around the stationary points. We also mark the arrival positions of wave types I-III, corresponding to the characteristic group velocities $c_g^{\mathrm{l}}$, $c_g^{\mathrm{s}}$, and $c_g^{\mathrm{min}}$, respectively. 

It can be seen that wave type II, propagating at the characteristic speed $c_g^{\mathrm{s}}$ (corresponding to the Rayleigh wave velocity of a pre-stretched half-space composed of the upper layer with $\lambda_1=1.2$), consistently arrives at approximately $x_1\approx45$, regardless of variations in $\zeta$ and $r$. Referring to Figure \ref{fig:cg-k1-model1}, the short-wavelength limiting group velocity $c_g^{\mathrm{s}}$ remains constant at approximately $1.14$ for all three models. Moreover, wave type II exhibits the largest amplitude among all wave components. Wave type I, propagating at the larger characteristic velocity $c_g^{\mathrm{l}}$, arrives near the leading wavefront. In contrast, the arrival position of wave type III, associated with the characteristic velocity $c_g^{\mathrm{min}}$, defines the trailing boundary of the significantly disturbed region and remains within the first few wave peaks near the loading point. As osteoarthritis progresses, the arrival positions of wave types I and III move closer to one another (see Figures \ref{fig:figure-u2-x1-model-group1-1}, \ref{fig:figure-u2-x1-model-group1-2}, \ref{fig:figure-u2-x1-model-group1-3}), indicating that the region significantly affected by the impact wave becomes increasingly localized.

The vibrational responses of the three models are illustrated in Figure \ref{fig:figure-t-model-group1} for the particle located at $x_1=40$ and $x_2=1$ over the time interval $t_1\in[0,80]$. The red dots indicate the predicted arrival times of the three wave types. Consistent with the results in Figure \ref{fig:figure-x1-model-group1}, the maximum amplitude is associated with wave type II. The arrival times of wave type II are constantly observed at $t_1\approx35$, indicating that its propagation remains essentially unaffected by variations in parameters. The oscillations around the arrival of wave type II also exhibit a higher frequency compared with those in other time intervals, which is similar to the displacement responses. As osteoarthritis evolves, the arrival of wave type I is delayed, whereas that of wave type III is advanced. Therefore, the duration of the transient vibration is reduced, indicating that the time window for the displacement measurement becomes shorter.

We further demonstrate that the displacement responses shown in Figures \ref{fig:figure-x1-model-group1} and \ref{fig:figure-t-model-group1} are primarily governed by the fundamental mode---therefore only the group velocity curve of the fundamental mode is presented in Figure \ref{fig:cg-k1-model1}. This is also consistent with the results reported in Figure 2 by \citet{erbay2000effects} for an incompressible layer. These findings provide theoretical support for current elastography techniques, in which surface waves or shear waves associated with the fundamental mode are widely used to estimate elastic material properties across various fields \citep{feng2023ultra,georgas2024shear}. In the latter study \citep{georgas2024shear}, transient waves were generated by pulse excitation within a phantom surrounding articular cartilage. However, the articular cartilage was treated as a homogeneous medium, and the propagation velocity of the maximum displacement peak was identified as the shear wave velocity and used to inversely estimate the elastic modulus. Although the authors also reported several secondary peaks and troughs in the displacement response, this additional wavefield information was not fully exploited. Our findings reveal the transient wave distribution within layered media, demonstrating that the complete wavefield, including both the arrival times and phase characteristics of multiple wave components, contains substantially richer information than the peak velocity alone. This enables not only a more rigorous assessment of the validity of homogeneous-medium models but also the inverse identification of the material parameters of layered structures.

\subsection{Human gingiva}\label{gingiva}

An illustration of the bilayer geometry of human gingiva is shown in Figure \ref{fig:layered tissue}(b). Similar to the articular cartilage models, we consider that both the geometry and elastic properties of the gingival layers are altered during the progression and treatment of periodontitis, which disrupts the microarchitecture of gingival tissue. During disease progression, the gingival connective tissue becomes inflamed and edematous, accompanied by a marked reduction in elastic modulus \citep{shita2023increasing,jepsen2023biomechanical,makkar2026matrix}. Following therapeutic intervention, reactive gingival thickening occurs \citep{ayub2022non,kannan2025microneedling}, together with repair of the connective tissue matrix and partial restoration of the elastic modulus \citep{xue2023use}.

To characterize these pathological and reparative processes, three representative gingiva models are established: healthy gingiva, periodontitis-affected gingiva, and treated gingiva. The healthy gingiva is modeled with $\zeta=5$, corresponding to the measured layer thicknesses reported by \citet{russo2025anatomical}. We note that the elastic moduli of the individual layers of human gingiva have not been well established in the literature. Therefore, we refer to a recent study on the rheological properties of human gingiva \citep{dechelette2026rheological} and assume that the epithelium is stiffer than the underlying connective tissue. The remaining parameters are taken as $r=0.05$, $\lambda_1=1.2$, and $\nu=0.4$. In the periodontitis model, inflammatory swelling of the connective tissue is represented by increasing $\zeta$ to $8$ and reducing $r$ to $0.03$. In contrast, the treated gingiva model accounts for post-treatment tissue thickening by further increasing $\zeta$ to $10$ while restoring the modulus ratio to $r=0.05$. The corresponding group velocity curves for the three gingiva models are presented in Figure \ref{fig:cg-k1-model2}.

\begin{figure}[htbp]
\centering
{\subfigure[]{\includegraphics[width=0.48\textwidth]{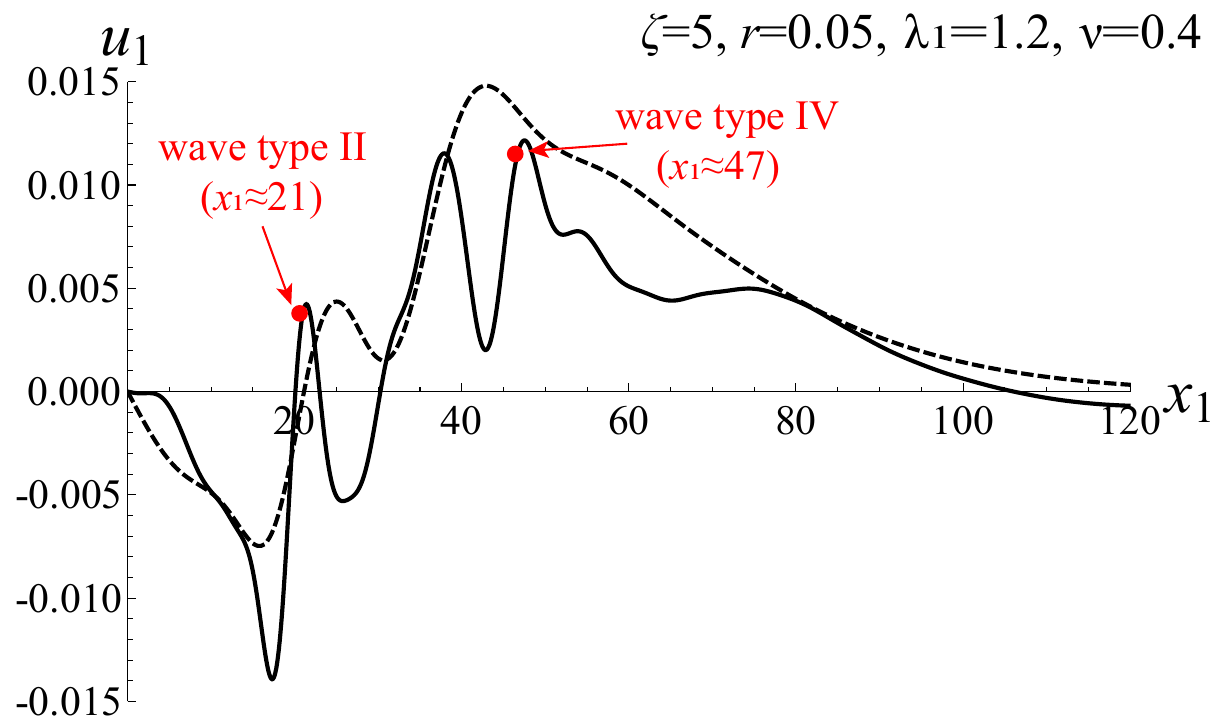}\label{fig:figure-u1-x1-model-group2-1}}\quad\subfigure[]{\includegraphics[width=0.48\textwidth]{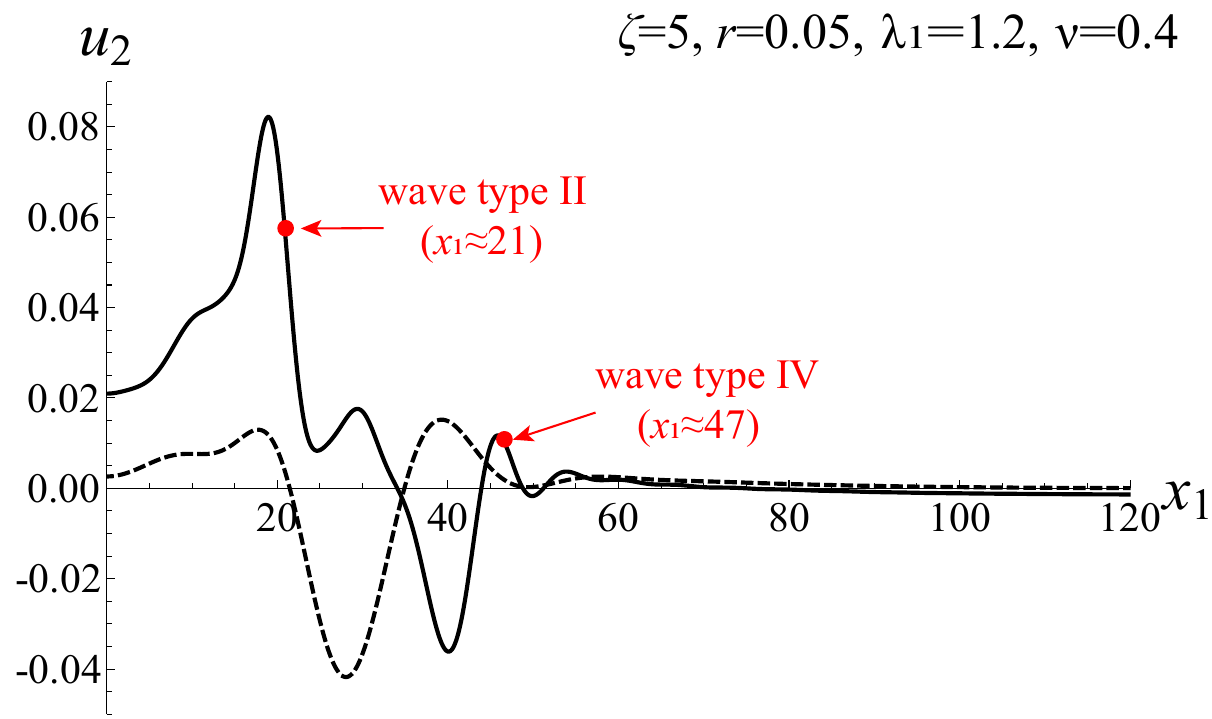}\label{fig:figure-u2-x1-model-group2-1}}\\\subfigure[]{\includegraphics[width=0.48\textwidth]{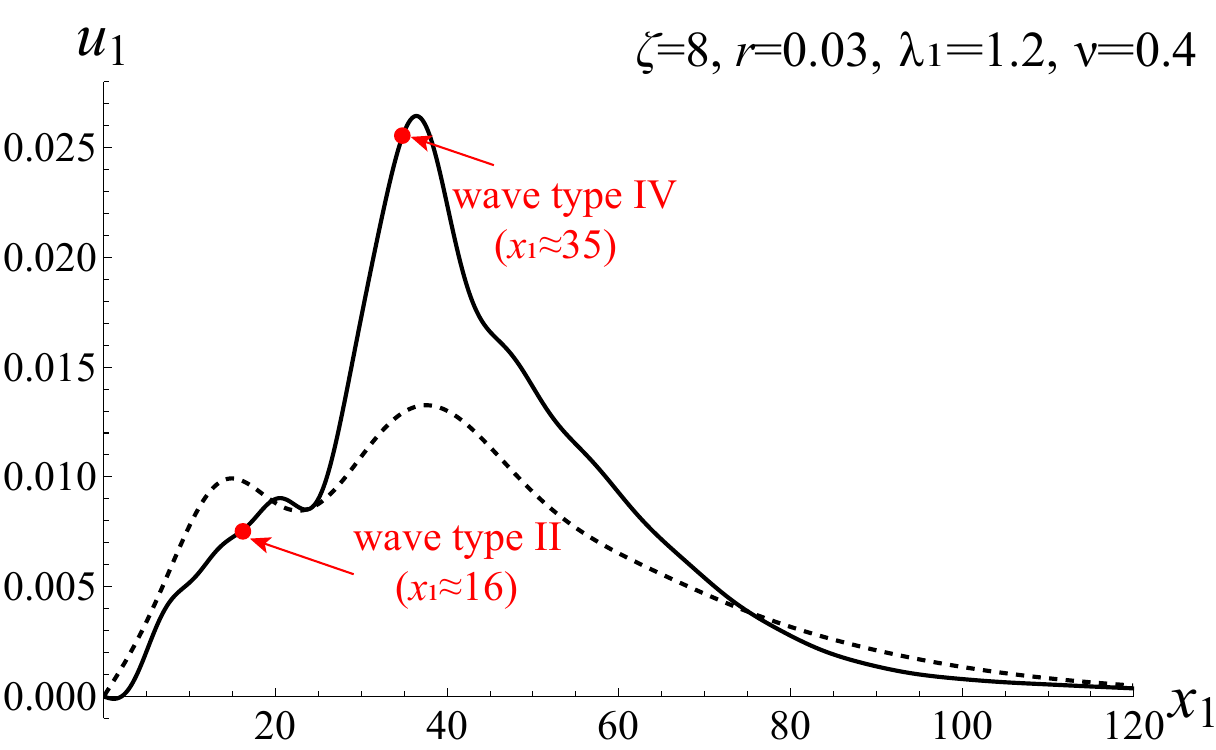}\label{fig:figure-u1-x1-model-group2-2}}\quad\subfigure[]{\includegraphics[width=0.48\textwidth]{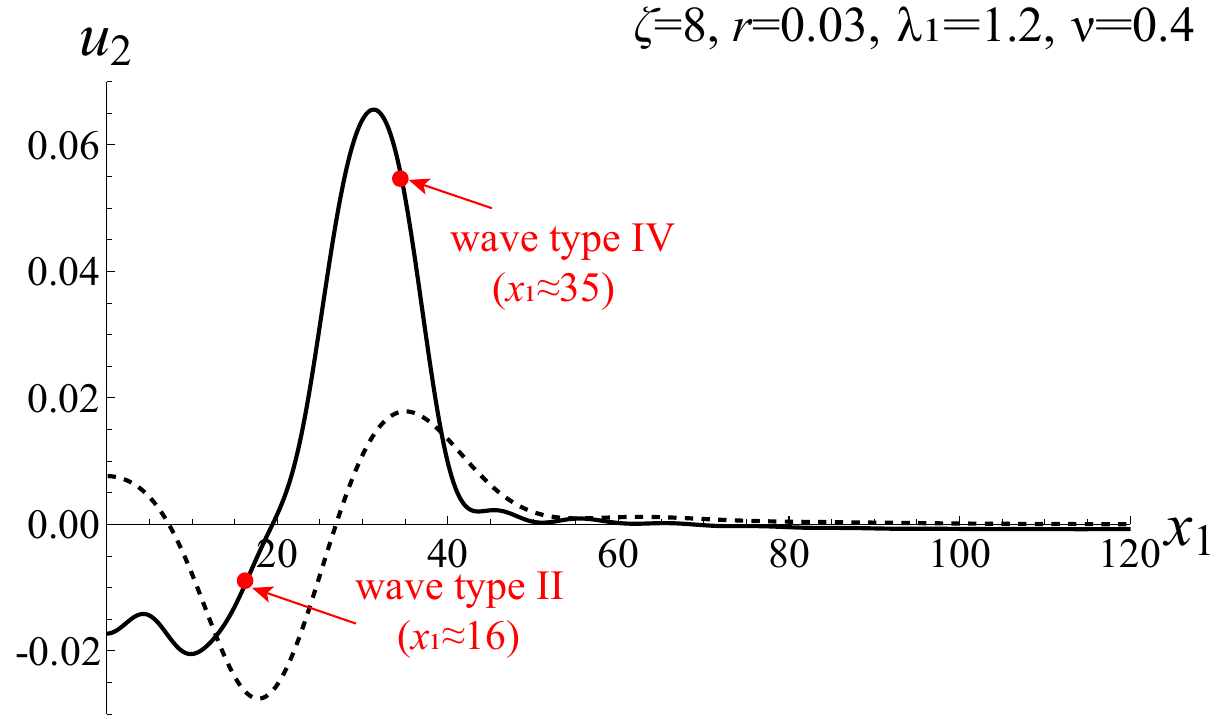}\label{fig:figure-u2-x1-model-group2-2}}\\\subfigure[]{\includegraphics[width=0.48\textwidth]{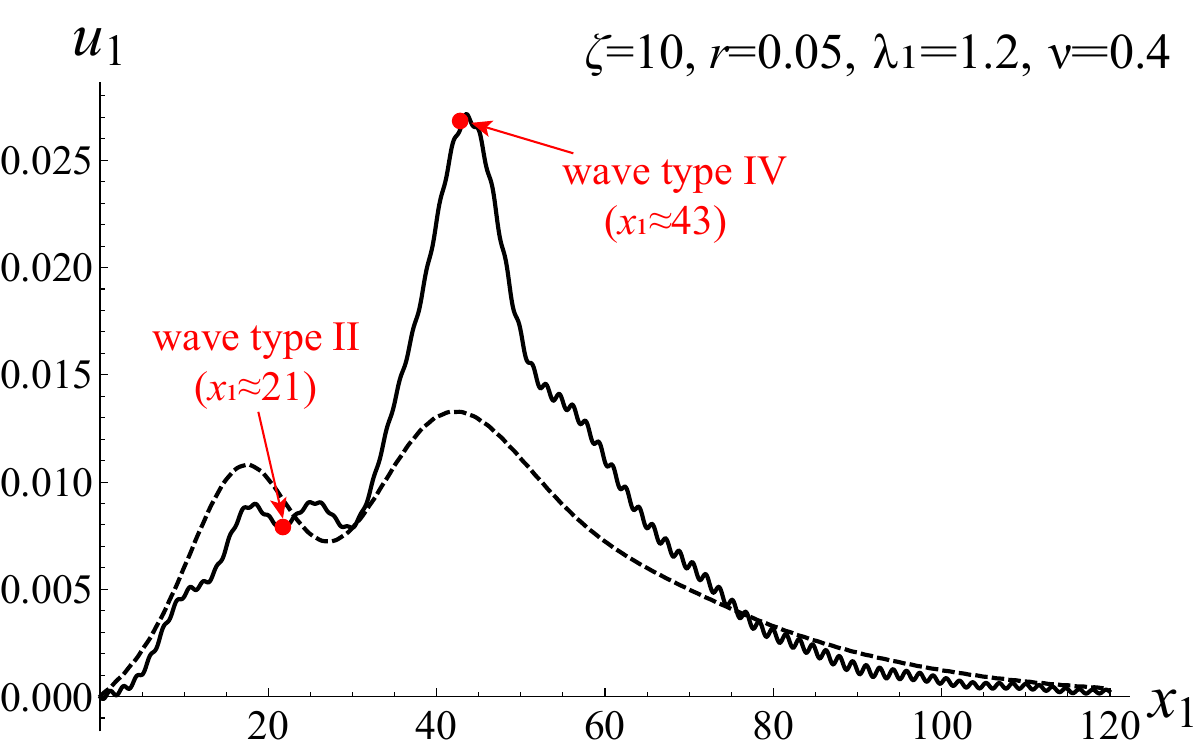}\label{fig:figure-u1-x1-model-group2-3}}\quad\subfigure[]{\includegraphics[width=0.48\textwidth]{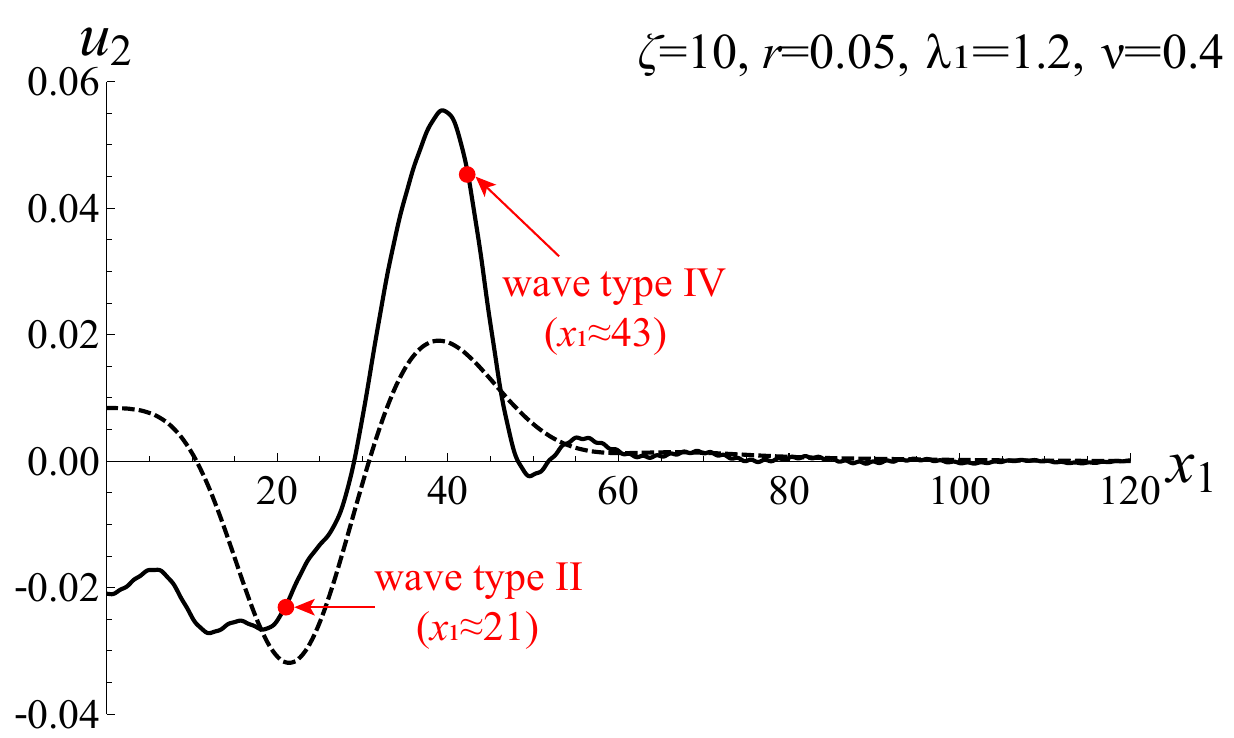}\label{fig:figure-u2-x1-model-group2-3}}}\caption{Dependence of the horizontal displacement $u_1$ and vertical displacement $u_2$ on $x_1$ for three gingiva models at $t_1=80$. The parameters are indicated in each panel, corresponding to the healthy, periodontitis-affected, and treated conditions in panels (a)–(b), (c)–(d), and (e)–(f), respectively. Dashed curves represent the displacement responses with only fundamental mode included, whereas the solid curves correspond to the displacement responses calculated by including the first five modes. Red dots are plotted to mark the arrival locations of the characteristic wave components.}\label{fig:figure-x1-model-group2}
\end{figure}

\begin{figure}[htbp]
\centering
{\subfigure[]{\includegraphics[width=0.48\textwidth]{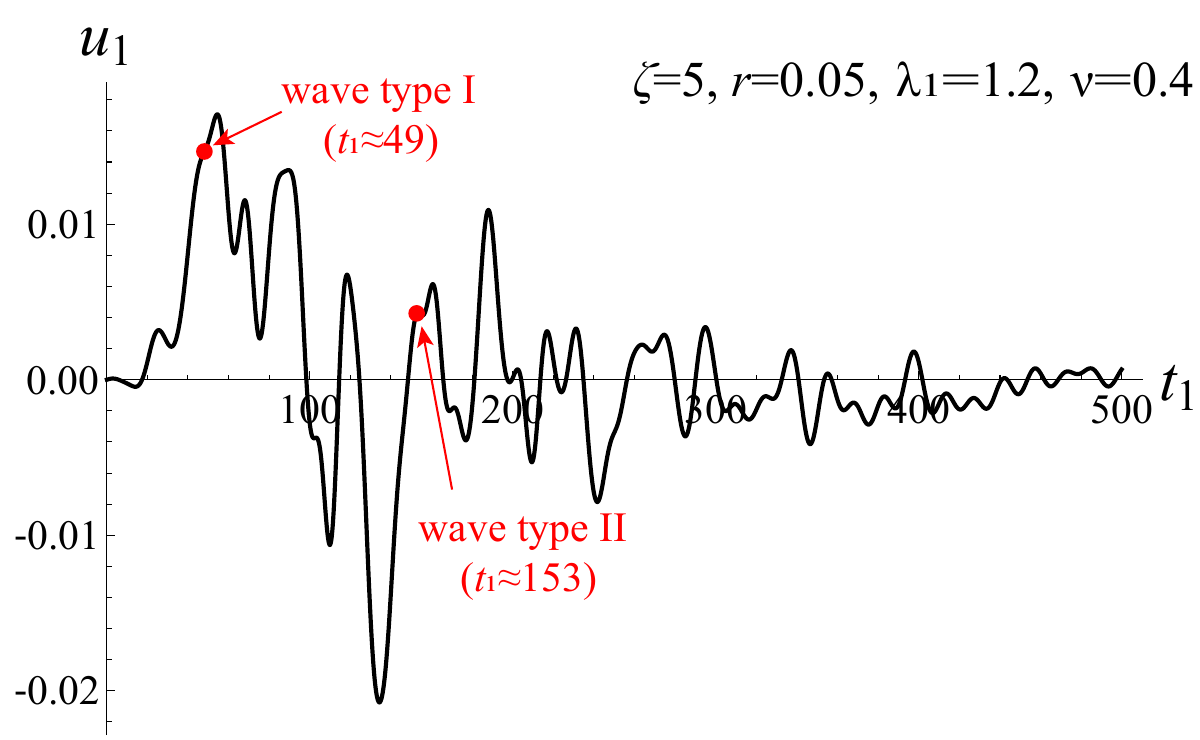}\label{fig:figure-u1-t-model-group2-1}}\quad\subfigure[]{\includegraphics[width=0.48\textwidth]{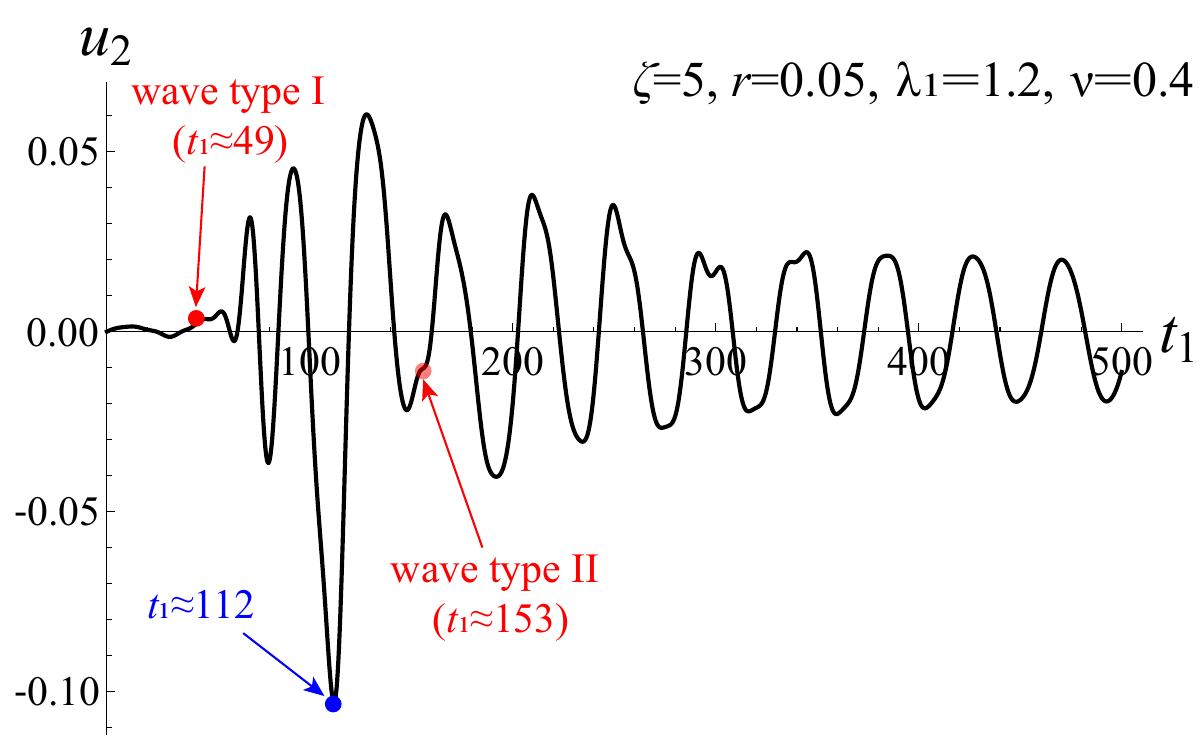}\label{fig:figure-u2-t-model-group2-1}}\\\subfigure[]{\includegraphics[width=0.48\textwidth]{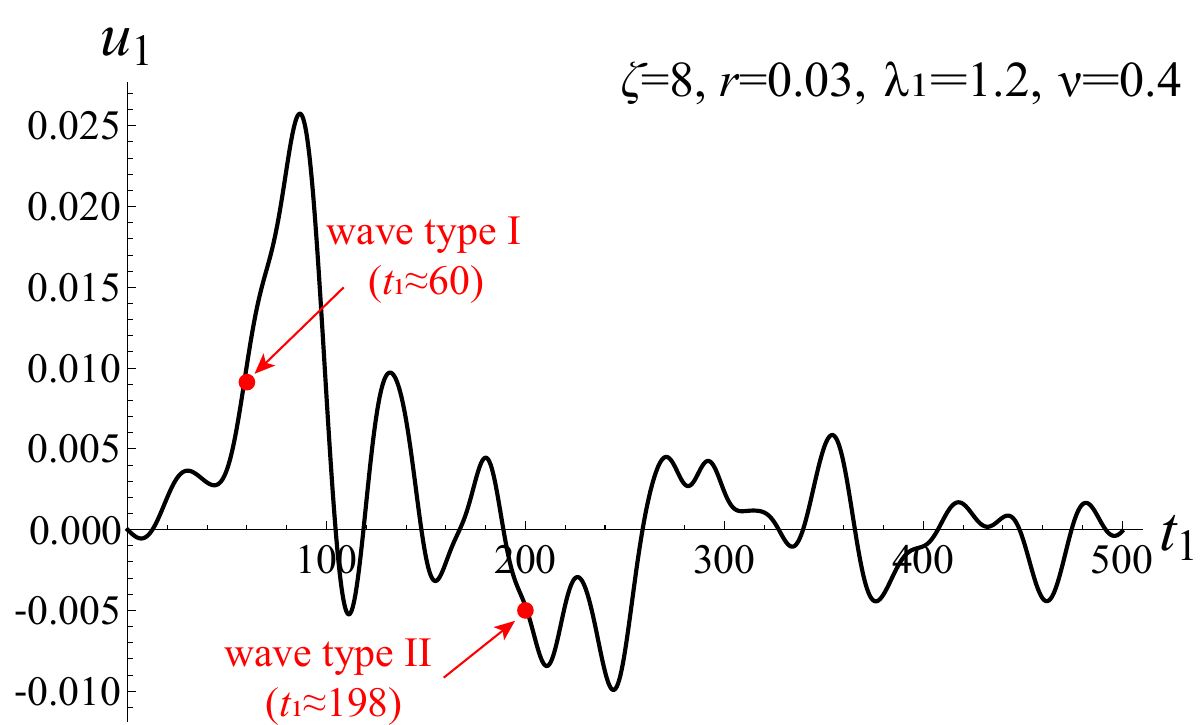}\label{fig:figure-u1-t-model-group2-2}}\quad\subfigure[]{\includegraphics[width=0.48\textwidth]{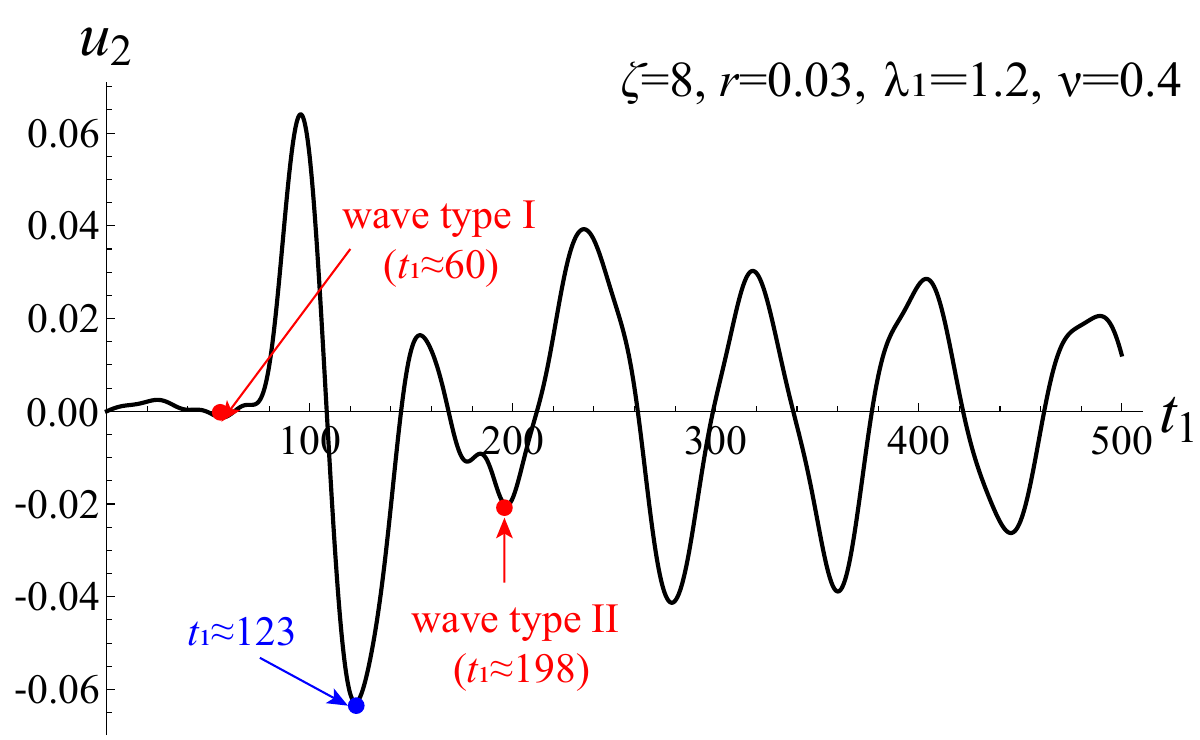}\label{fig:figure-u2-t-model-group2-2}}\\\subfigure[]{\includegraphics[width=0.48\textwidth]{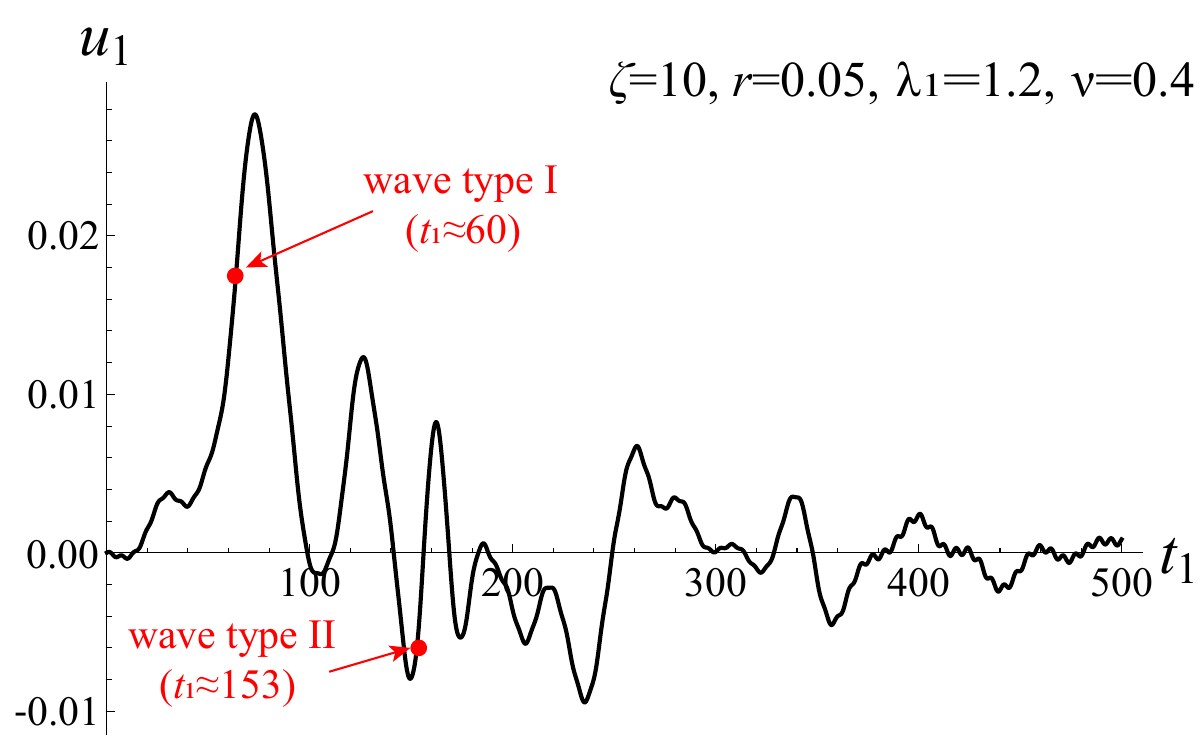}\label{fig:figure-u1-t-model-group2-3}}\quad\subfigure[]{\includegraphics[width=0.48\textwidth]{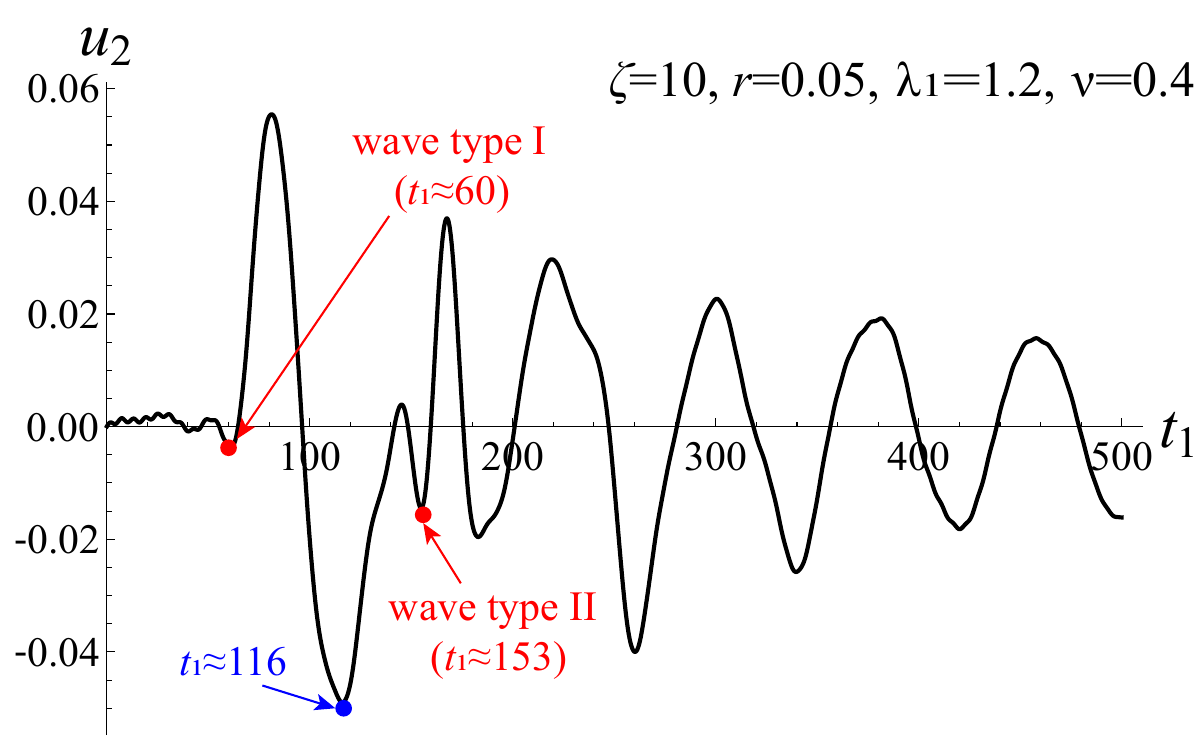}\label{fig:figure-u2-t-model-group2-3}}}\caption{Dependence of horizontal displacement $u_1$ and vertical displacement $u_2$ on $t_1$ for three gingiva models at $x_1=40$. The parameters are illustrated in each panel, representing the healthy, periodontitis-affected, and treated conditions in panels (a)–(b), (c)–(d), and (e)–(f), respectively. Red dots are plotted to mark the locations of specific wave components. The blue dots indicate the times of the deepest wave troughs.}\label{fig:figure-t-model-group2}
\end{figure}

In this subsection, numerical integration is conducted in the interval $[0.004,\,4]$ with a uniform mesh size of $0.004$ to ensure convergence while maintaining computational efficiency. The displacement responses for three representative models are plotted in Figure \ref{fig:figure-x1-model-group2} with $x_2=1$ and $t_1=80$. The dashed curves represent the responses obtained using only the fundamental mode, whereas the solid curves are computed by including the first five modes. In contrast to the examples presented in the previous subsection, when the surface layer becomes stiffer and thinner, the displacement responses are more strongly influenced by higher-order modes rather than the fundamental mode, particularly in the amplitudes of the peaks and troughs. We also confirm that the first five modes are enough to describe the displacement distribution.

To facilitate interpretation of the wave profiles, the arrival positions of wave types II and IV, corresponding to the group velocities $c_g^{\mathrm{s}}$ and $c_g^{\mathrm{max}}$, respectively, are marked by red dots, as they are associated with distinct features of the displacement curves. Because wave type I is not clearly identifiable in Figure \ref{fig:figure-x1-model-group2}, its arrival position is not marked. As shown in Figure \ref{fig:figure-u2-x1-model-group2-1}, the wave propagating at $c_g^{\mathrm{s}}$ occupies near the maximum displacement, whereas the wave propagating at $c_g^{\mathrm{max}}$ coincides with the last wave peaks in displacement curves. On the contrary, Figures \ref{fig:figure-u2-x1-model-group2-2} and \ref{fig:figure-u2-x1-model-group2-3} exhibit a different wave profile, in which the dominant wave peak is associated with wave type IV rather than wave type II. Besides, a localized wave packet emerges, which is generally advantageous for experimental measurements because its spatially concentrated waveform facilitates signal identification and improves measurement accuracy. Interestingly, a similar wave profile was reported by \citet{feng2022vivo} for skin, where the upper layer is likewise stiffer and thinner than the underlying substrate. 

We further display in Figure \ref{fig:figure-t-model-group2} the temporal evolution of the displacements as functions of $t_1$, with several characteristic features highlighted. The onset of the earliest arriving signal corresponds to wave type I, which propagates with the highest group velocity. For reference, the times corresponding to the principal wave troughs are marked by blue dots, which occur within the time interval $[112,123]$. A distinct change in the waveform marks the arrival of wave type II, after which the vibration amplitude gradually decays with increasing $t_1$. Since the contribution of wave type IV to the vibrational response is relatively weak, its arrival is not explicitly indicated. Notably, as periodontitis progresses, the arrival times of both wave types I and II increase, reflecting their decreased group velocities. Although the arrival time of wave type II partially recovers after treatment, wave type I continues to propagate slower than in the healthy model, owing to the persistent increase in gingival thickness caused by reactive tissue thickening.

It is worth noting that some existing elastography methods for human gingiva assume that shear waves propagate in an infinite homogeneous medium \citep{han2022preliminary,xue2023use}. A more recent study has accounted for the dispersive nature of surface waves by extracting dispersion curves using the FFT technique \citep{moon2024high}. In this study, the phase velocity at $16$ kHz was used to estimate the shear modulus. Our analysis of impact wave propagation characterizes the complete transient surface displacement field, which naturally incorporates the wave dynamics of the multilayer structure. Consequently, the elastic properties might be inferred directly from the displacement response.

\subsection{Human skin}\label{skin}

Human skin is a multilayered structure, as displayed in Figure \ref{fig:layered tissue}(c). To elucidate the progressive pathophysiology of skin sclerosis, the disease process is modeled through two distinct clinical stages: the oedematous phase, characterized by tissue swelling and initial stiffening, and the subsequent plateau phase, characterized by substantial collagen densification and markedly increased stiffness \citep{sobolewski2020applicability,herrick2022skin}. Based on experimental measurements of the elastic modulus and geometric dimensions of skin \citep{chavoshnejad2020surface}, the healthy skin model is defined by a thickness ratio of $\zeta = 50$ and a shear modulus ratio of $r = 0.06$. Similarly, the other two parameters take $\lambda_1 = 1.2$ and $\nu = 0.4$. 

In the oedematous phase, tissue swelling and initial stiffening are represented by increasing the thickness ratio and shear modulus ratio to $\zeta = 60$ and $r = 0.15$, respectively, while keeping the other parameters unchanged ($\lambda_1 = 1.2$, $\nu = 0.4$). In the plateau phase, the progressive collagen densification is captured by further increasing the shear modulus ratio to $r = 0.3$ while maintaining $\lambda_1 = 1.2$, $\nu = 0.4$, and $\zeta=60$. The corresponding group velocity curves for the three models are presented in Figure \ref{fig:cg-k1-model3}.

\begin{figure}[htbp]
\centering
{\subfigure[]{\includegraphics[width=0.48\textwidth]{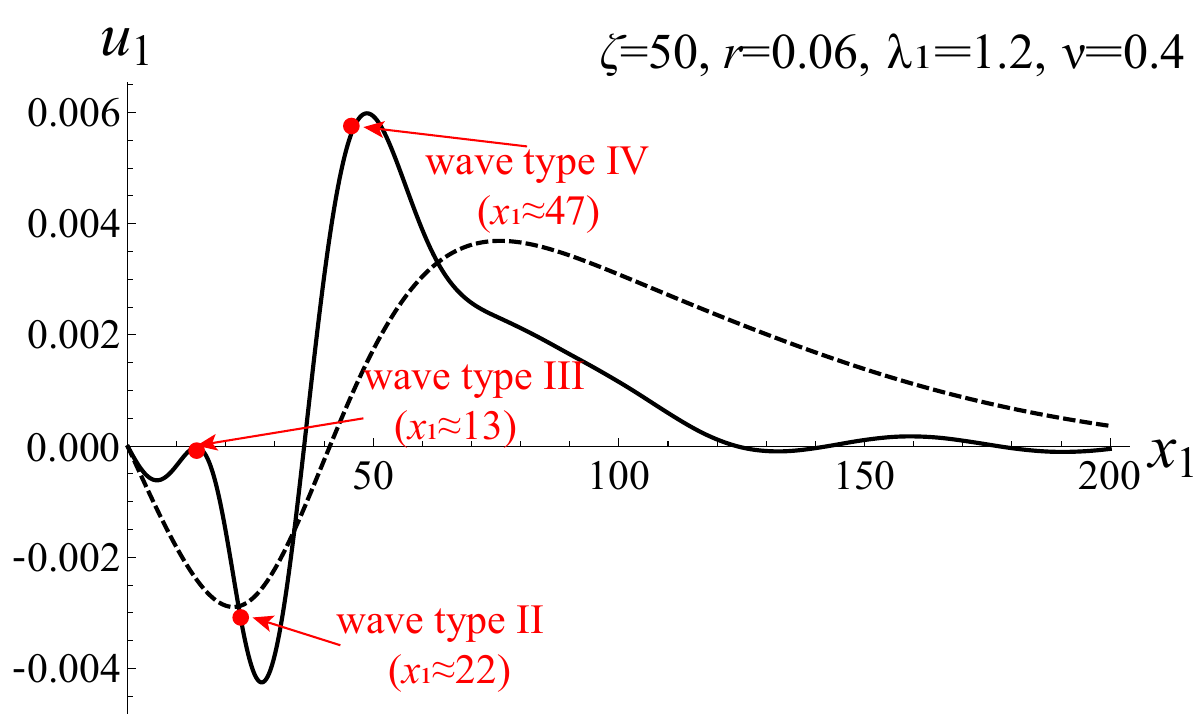}\label{fig:figure-u1-x1-model-group3-1}}\quad\subfigure[]{\includegraphics[width=0.48\textwidth]{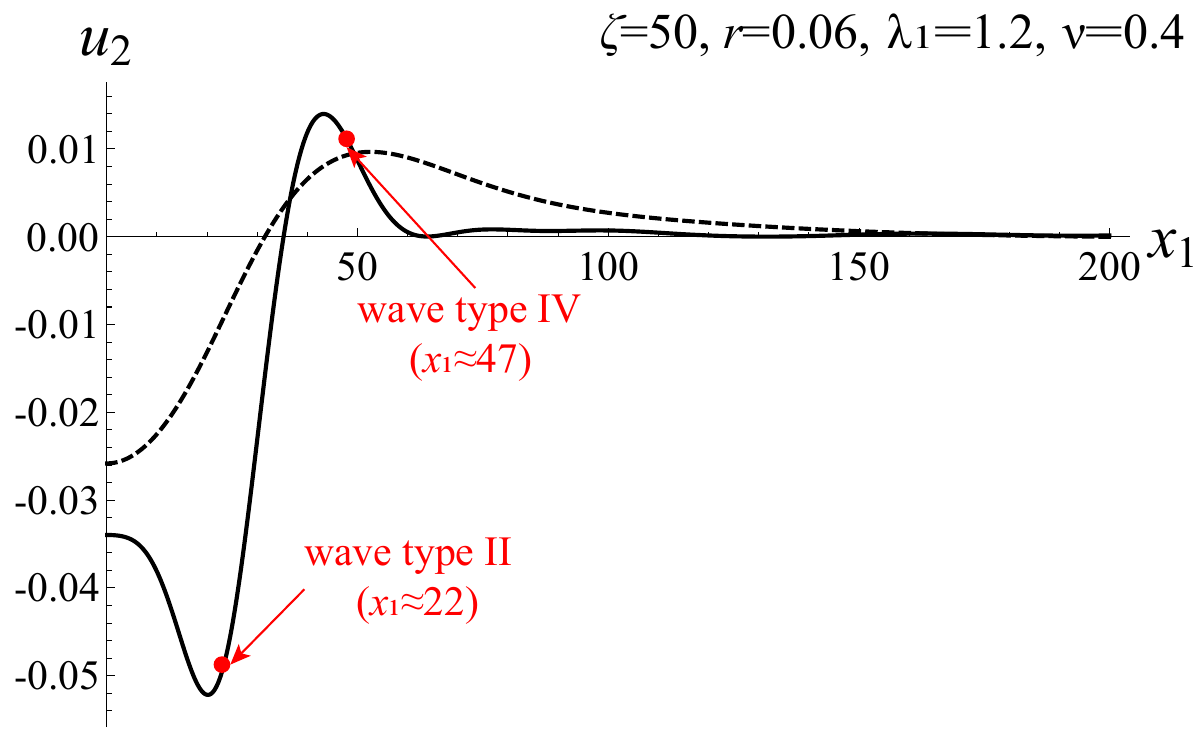}\label{fig:figure-u2-x1-model-group3-1}}\\\subfigure[]{\includegraphics[width=0.48\textwidth]{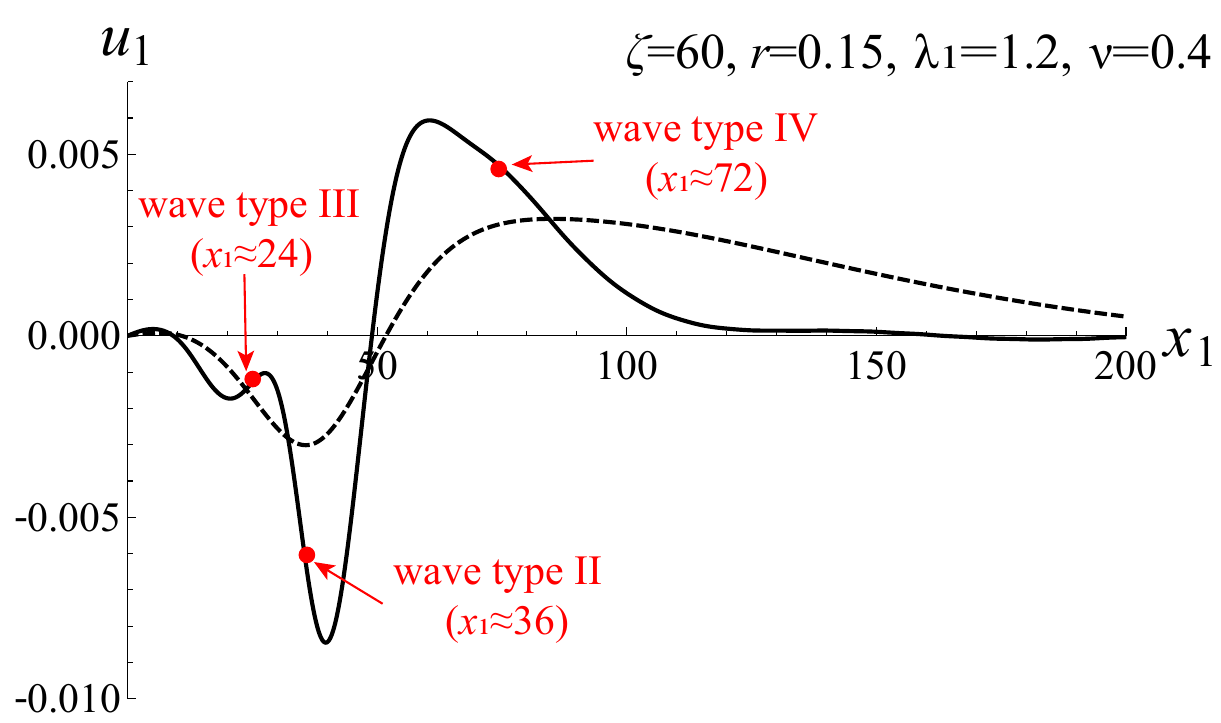}\label{fig:figure-u1-x1-model-group3-2}}\quad\subfigure[]{\includegraphics[width=0.48\textwidth]{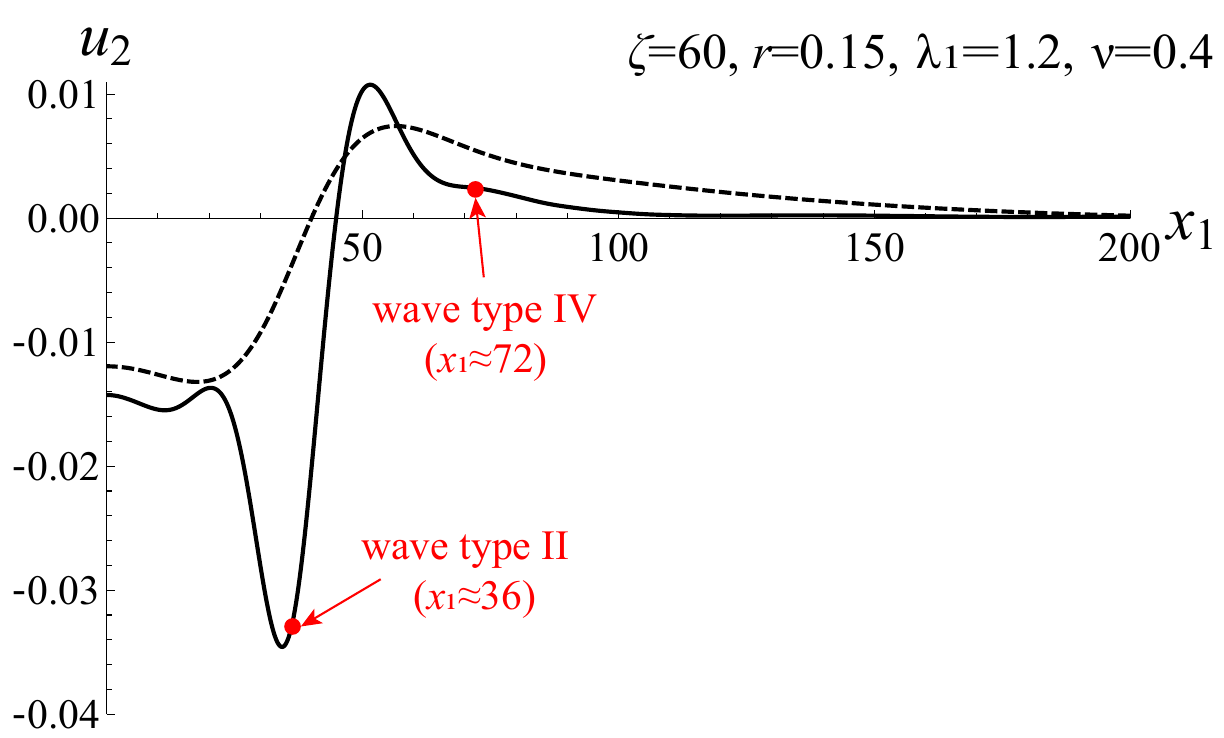}\label{fig:figure-u2-x1-model-group3-2}}\\\subfigure[]{\includegraphics[width=0.48\textwidth]{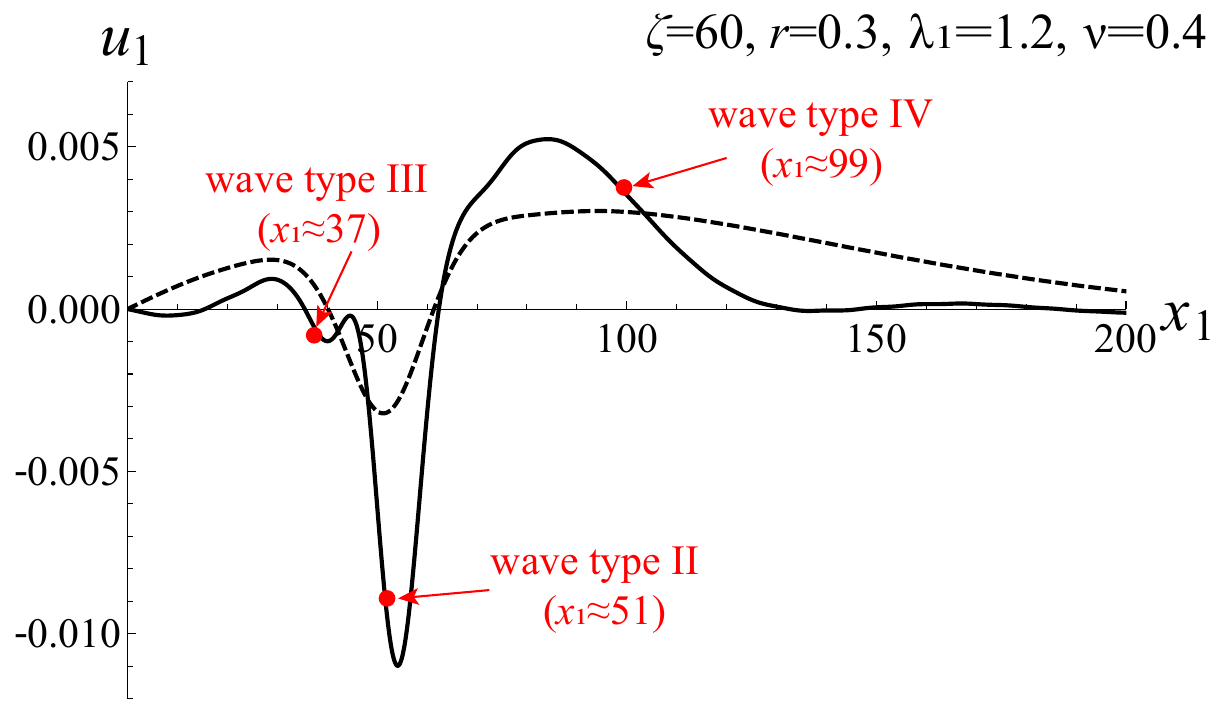}\label{fig:figure-u1-x1-model-group3-3}}\quad\subfigure[]{\includegraphics[width=0.48\textwidth]{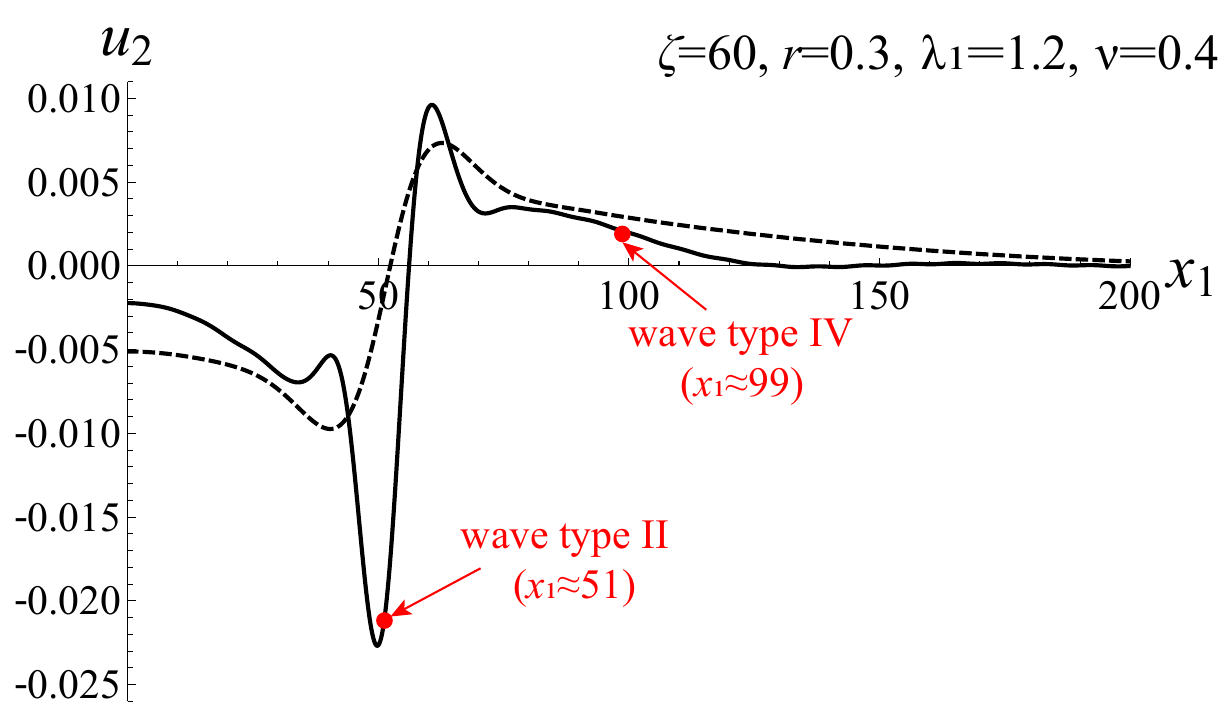}\label{fig:figure-u2-x1-model-group3-3}}}\caption{Dependence of the horizontal displacement $u_1$ and vertical displacement $u_2$ on $x_1$ for three skin models at $t_1=80$. The parameters are indicated in each panel, corresponding to healthy, oedematous, and plateau conditions in panels (a)-(b), (c)-(d) and (e)-(f), respectively. Dashed curves represent the displacement responses with only fundamental mode included, whereas the solid curves correspond to the displacement responses calculated by five modes. Red dots are plotted to mark the arrival locations of the relevant wave components.}\label{fig:figure-x1-model-group3}
\end{figure}

\begin{figure}[htbp]
\centering
{\subfigure[]{\includegraphics[width=0.48\textwidth]{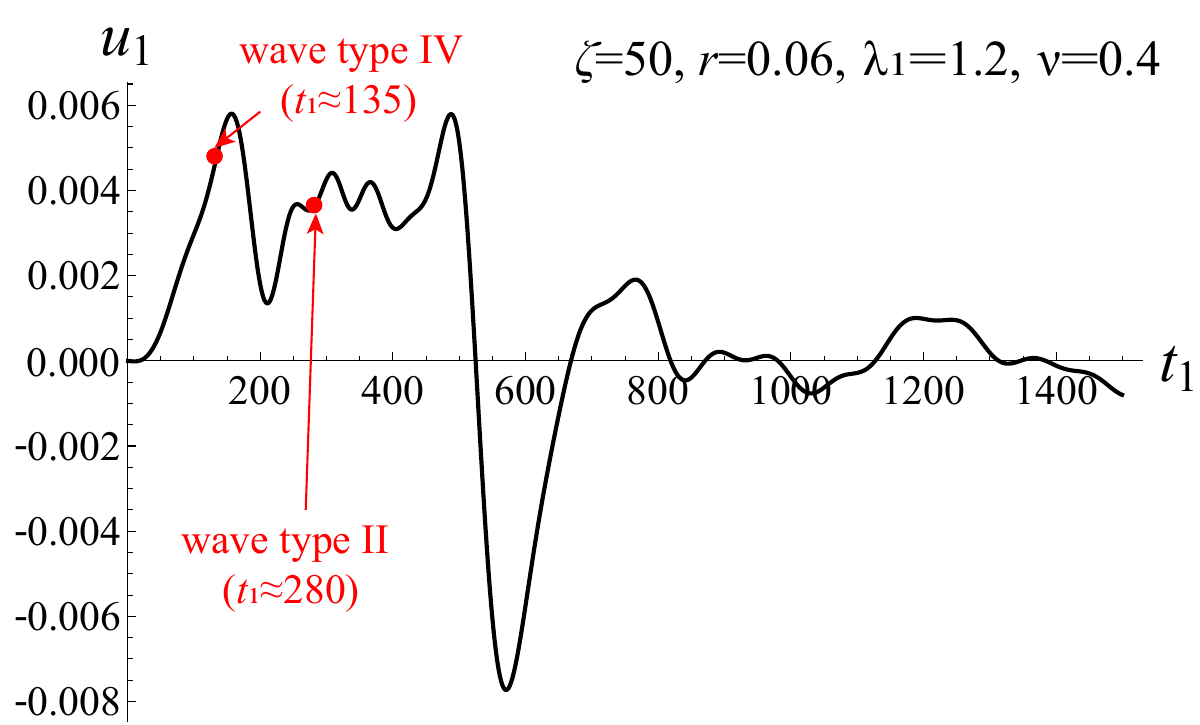}\label{fig:figure-u1-t-model-group3-1}}\quad\subfigure[]{\includegraphics[width=0.48\textwidth]{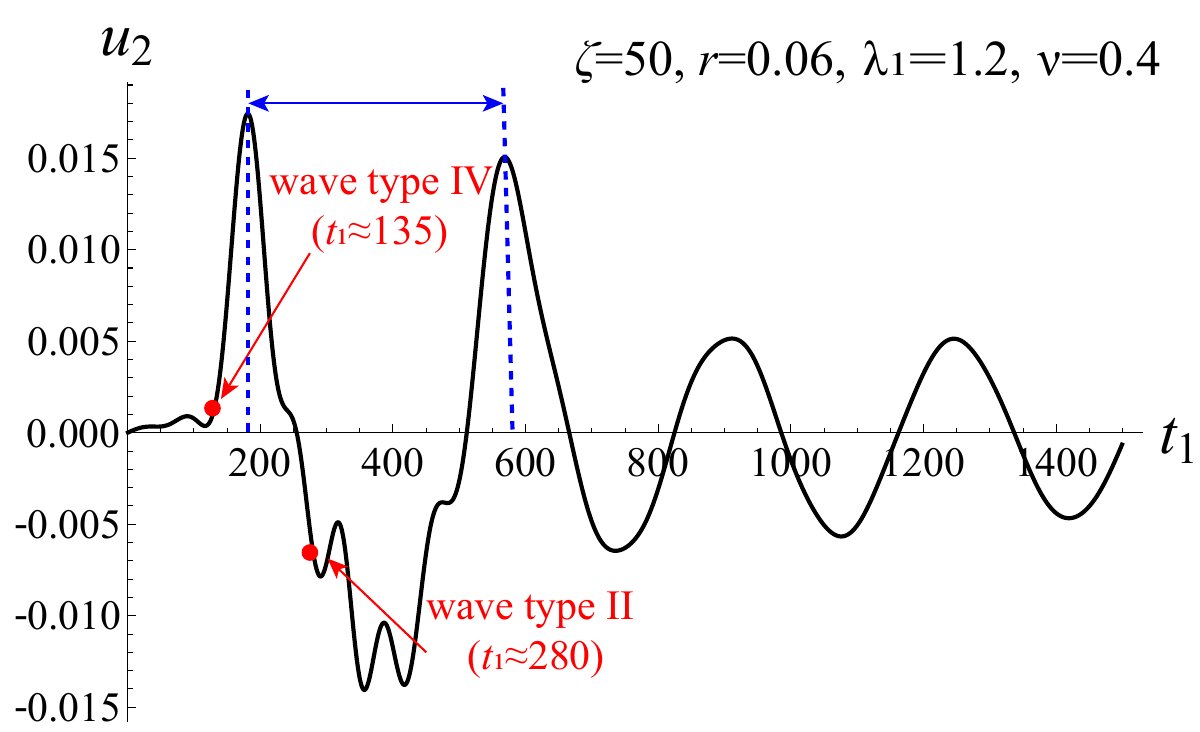}\label{fig:figure-u2-t-model-group3-1}}\\\subfigure[]{\includegraphics[width=0.48\textwidth]{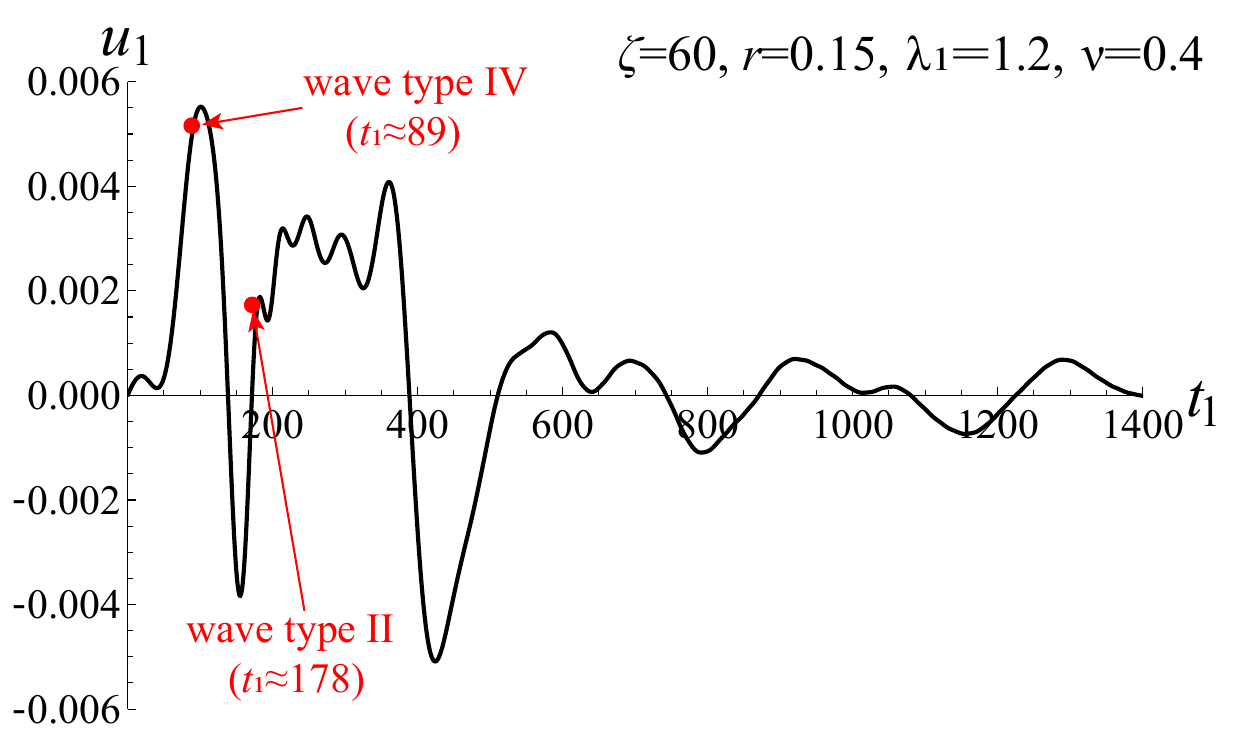}\label{fig:figure-u1-t-model-group3-2}}\quad\subfigure[]{\includegraphics[width=0.48\textwidth]{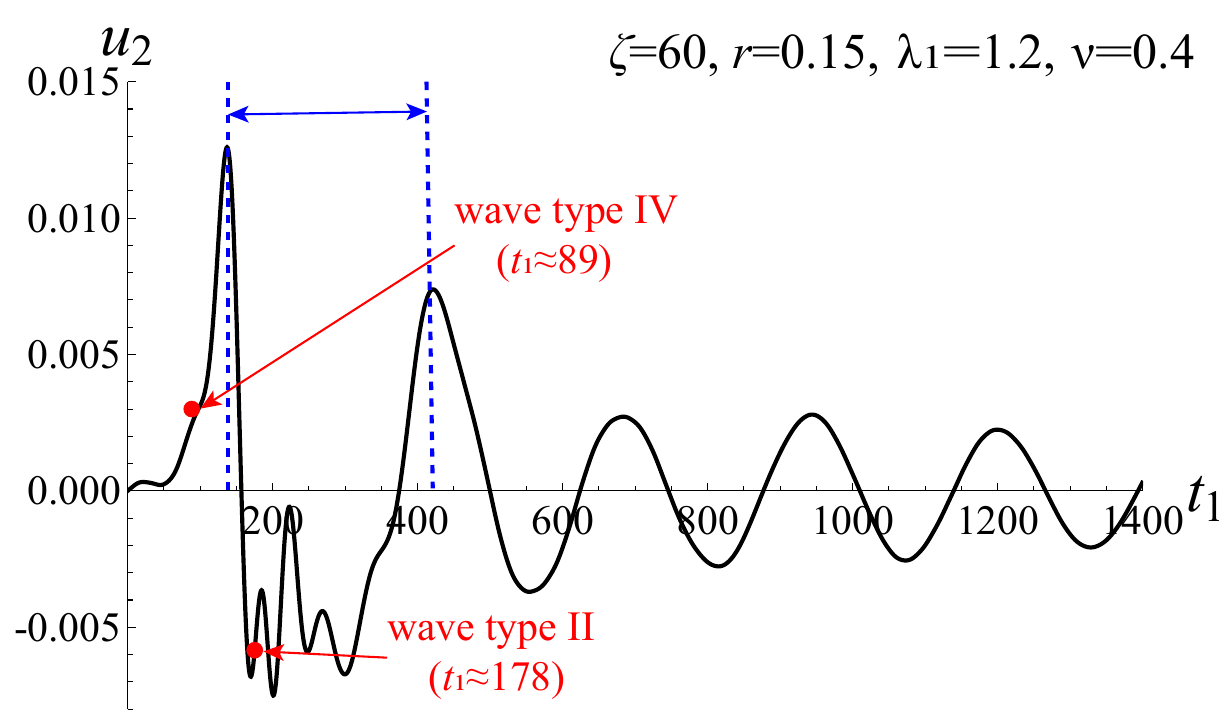}\label{fig:figure-u2-t-model-group3-2}}\\\subfigure[]{\includegraphics[width=0.48\textwidth]{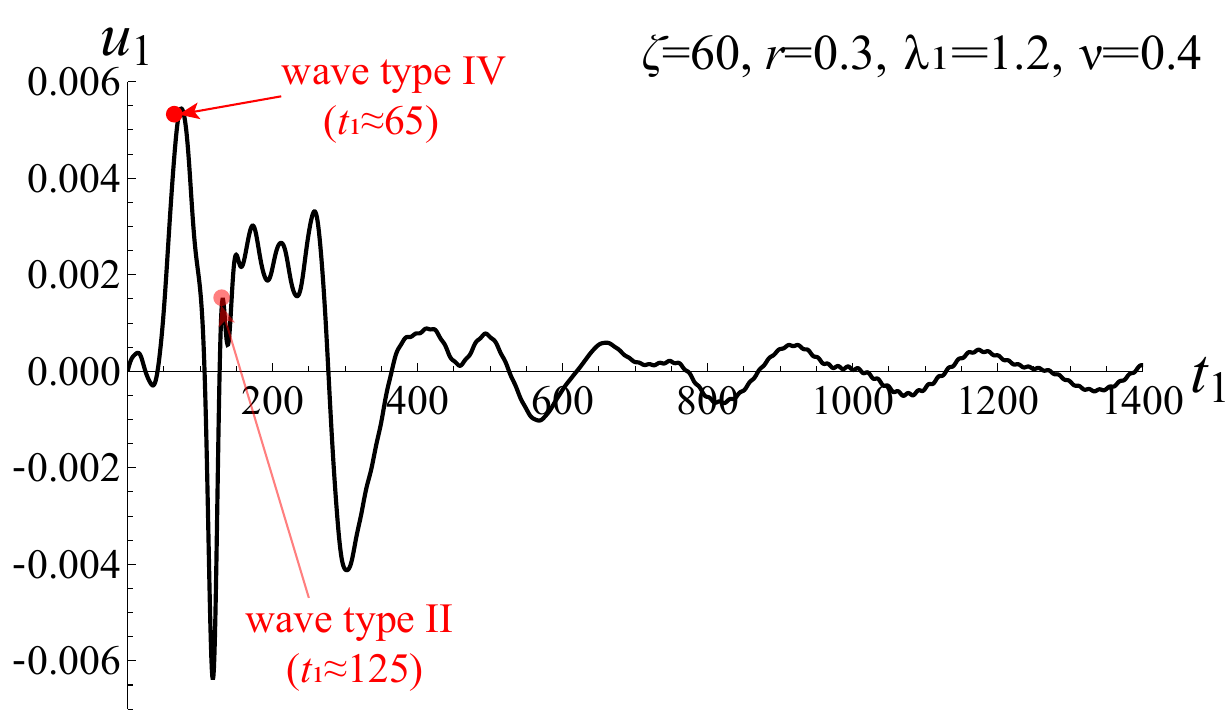}\label{fig:figure-u1-t-model-group3-3}}\quad\subfigure[]{\includegraphics[width=0.48\textwidth]{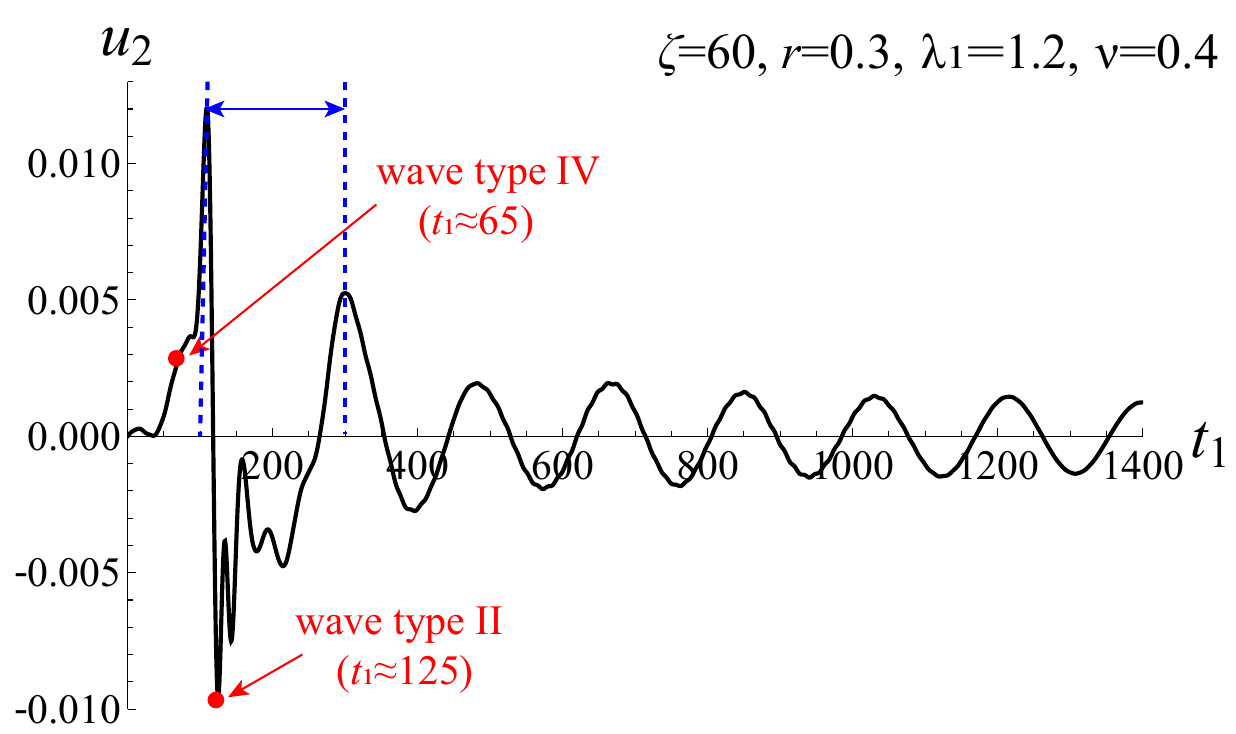}\label{fig:figure-u2-t-model-group3-3}}}\caption{Dependence of horizontal displacement $u_1$ and vertical displacement $u_2$ on $t_1$ for three skin models at $x_1=80$. The parameters are illustrated in each panel, representing the healthy , oedematous and plateau conditions in panels (a)-(b), (c)-(d) and (e)-(f), respectively. Red dots are plotted to mark the locations of the characteristic wave components. The blue dashed lines are plotted to mark the time difference between the first two wave peaks.}\label{fig:figure-t-model-group3}
\end{figure}

In this subsection, numerical integration is performed using a uniform step size approximately equal to one-thousandth of the domain interval $[0.0007,\,0.7]$, guaranteeing both convergence and computational efficiency. In these scenarios, as shown in Figure \ref{fig:cg-k1-model3}, the group velocity approaches its asymptotic value rapidly. The displacement distributions along $x_2=1$ at a fixed time $t_1=80$ are presented in Figure~\ref{fig:figure-x1-model-group3}. For comparison, the corresponding responses obtained using only the fundamental mode are shown as dashed curves, whereas those calculated by including the first five modes are represented by solid curves. Similar to the trends observed in Section \ref{gingiva}, using only the fundamental mode is insufficient to capture the actual displacements; accounting for higher-order modes is critical to accurately reflect the wave profile. The arrival positions of wave types II, III, and IV, corresponding to the group velocities $c_g^{\mathrm{s}}$, $c_g^{\mathrm{min}}$, and $c_g^{\mathrm{max}}$, respectively, are also indicated.

Because wave type III does not exhibit a clear characteristic in the vertical response, its arrival is not marked in Figures \ref{fig:figure-u2-x1-model-group3-1}, \ref{fig:figure-u2-x1-model-group3-2} and \ref{fig:figure-u2-x1-model-group3-3}. However, it is clearly discernible in the horizontal response ($u_1$), highlighted as a minor peak or trough in each panel. In addition, wave type II generally aligns with the wave troughs in Figures \ref{fig:figure-u2-x1-model-group3-1}, \ref{fig:figure-u2-x1-model-group3-2} and \ref{fig:figure-u2-x1-model-group3-3}. Conversely, the location associated with wave type IV shifts under parametric variation: for the skin model with the softest substrate (Figure \ref{fig:figure-u2-x1-model-group3-1}), wave type IV coincides nearly with the dominant peak, whereas as $\zeta$ and $r$ increase, its position migrates rightward. In general, the most prominent feature in the displacement response corresponds to the group velocity $c_g^{\mathrm{s}}$. Consequently, the dominant trough in Figure \ref{fig:figure-x1-model-group3} is better interpreted in conjunction with wave type II.

Figure \ref{fig:figure-t-model-group3} illustrates the displacement responses of the skin models at the particle located at $(x_1,x_2)=(80,1)$ over the time interval $t_1\in[0,1400]$. The red markers denote the arrival times of wave types II and IV. Wave type IV corresponds to the earliest arrival; however, for larger $\zeta$ and $r$, its arrival time deviates from the initial onset of vibration. As $\zeta$ and $r$ increase, the amplitude associated with wave type II increases and eventually becomes the most prominent trough. Since wave type III exhibits no clear signature in the time-domain vibration, its arrival is omitted from the plot. Other salient features, such as wave peaks in the vertical vibrational responses, are indicated by blue dashed lines. The intervals between successive peaks decrease from $387$ to $284$ and then to $191$ for the three skin models, respectively. Meanwhile, the amplitude of the second wave peak decreases, whereas that of the first wave peak remains nearly unchanged. These observations indicate that progressive subcutaneous sclerosis increases the characteristic frequency of the impact waves and accelerates their amplitude decrease over time $t_1$.

Recall that the first five modes are retained in all numerical integrations for the three tissue examples. The corresponding NRMSE values between the results obtained using four and five modes for the gingiva and skin models are also reported in Table \ref{tab:NRMSE-of-mode-number} in \ref{Convergence check}. Furthermore, although the upper integration limits for the gingiva and skin models are relatively small ($\xi_\mathrm{max}=4$ for the gingiva model and $\xi_\mathrm{max}=0.7$ for the skin model), we further examine the effects of these truncation limits in Figure \ref{figure-upper-limit-error-group}. The good agreement between the results obtained using different upper integration limits confirms the convergence of the corresponding numerical integrations.

We emphasize that elastography methods have been widely employed to characterize the stiffness of multilayered skin. For example, in OCE, a double-pulse excitation can be applied on the skin surface, and the first inflection point in the measured response was identified as the arrival of the surface wave \citep{liu2026nondestructive}. \citet{chawla2026assessment} further investigated the stiffness of the epidermis and dermis by applying excitations at different frequencies, with the resulting waves attributed to different skin layers based on the layer thicknesses. To account for the layered nature of skin, dispersion relations for skin-fat bilayers and skin-fat-muscle sandwich structures have been derived, and the fitted dispersion curves have been used for shear modulus inversion \citep{ma2024guided}. Simplified dispersion-based approaches have also been developed, where the average phase velocity at high frequencies is regarded as the Rayleigh wave velocity of an equivalent half-space, enabling estimation of the averaged Young's modulus of human skin \citep{li2025wireless}. More recently, inversion methods incorporating detailed layered structures and higher-frequency measurements have been developed to resolve the stiffness of individual skin layers \citep{feng2022vivo,feng2023ultra,ghaderi2026multilayered}. Since phase information is inherently available during the inversion process, deep-learning-based approaches have further been proposed to improve stiffness reconstruction from wave measurements \citep{feng2025deep,mieling2025scan,zhang2026depth}. Despite these advances, existing approaches primarily rely on phase velocity or dispersion information, while the physical interpretation of transient wave features remains less explored. Here, we analyze impact wave propagation in multilayer skin models and establish a direct relationship between characteristic features in the transient displacement responses and the corresponding group velocities. This connection provides physical insight into transient wave propagation and may further strengthen the theoretical foundation for future characterization of tissue properties.

\section{Conclusion}\label{Conclusion}

In this work, we investigated impact wave propagation induced by an impulsive surface load in a pre-stressed compressible hyperelastic bilayer resting on a smooth foundation. A general mechanics model was established within the framework of nonlinear elasticity \citep{ogden1997non,goriely2017mathematics}. By combining incremental elasticity theory with the Fourier-Laplace transform technique, a semi-analytical representation of the transient displacement field was derived. The asymptotic solutions under a Dirac delta excitation were further acquired using the stationary phase method and were shown to agree closely with numerical integration in the long-time regime. In particular, our framework can recover the results reported in \citet{erbay2000effects} in the incompressible limit when two layers are reduced to a single layer and the Heaviside step function is adopted as the impulsive loading. Finite element simulations based on ABAQUS, where user-defined UHYPER and VUMAT subroutines were implemented for the compressible neo-Hookean constitutive model, were performed to validate our theoretical predictions. The consistent agreement among the analytical solutions, asymptotic approximations, and finite element results establishes the reliability of the proposed framework for the low-frequency and dispersive wave content examined in pre-stressed bilayer materials.

To elucidate how variations in tissue properties influence transient wave signals, we applied the proposed theoretical framework to three representative layered biological tissues: human articular cartilage, gingiva, and skin. These tissues have been extensively investigated as targets for elastography owing to their pronounced layered architectures and stiffness contrasts \citep{feng2022vivo,han2022preliminary,feng2023ultra,xue2023use,georgas2024shear,moon2024high,feng2025deep}. Nevertheless, because wave propagation in pre-stressed layered soft tissues with both material and geometric nonlinearities remains poorly understood, most existing studies rely either on surface wave theories developed for linearly elastic media or on the implicit assumption that the fundamental mode alone governs the measured wave characteristics. Such assumptions inevitably overlook the contributions of higher-order modes, which, as demonstrated in the present study, may play a crucial role in shaping the transient wavefield.

For the human articular cartilage models, the spatial extent of the impact wave decreases as the thickness ratio $\zeta$ increases and the modulus ratio $r$ decreases. Notably, under the condition $\zeta<1$ and $r>1$, the group velocity $c_g^{\mathrm{s}}$ associated with wave type II remains essentially unchanged despite variations in $\zeta$ and $r$. This invariance indicates that the transient response is predominantly governed by the fundamental mode \citep{erbay2000effects}, thereby providing a theoretical justification for surface wave-based elastography methods that estimate elastic moduli from characteristic wave velocities \citep{feng2023ultra}. More importantly, our results show that the complete transient wavefield contains substantially richer information than the dominant wave velocity alone, suggesting the potential for inverse identification of multilayer cartilage properties from the full impact wave response.

For the human gingiva models, increasing the thickness ratio $\zeta$ increases the wavelength, whereas decreasing the modulus ratio $r$ reduces the propagation velocity of wave type IV. Moreover, wave type I is highly sensitive to $\zeta$ but exhibits only a weak dependence on $r$, while the arrival times of the characteristic waves increase progressively with periodontitis. These results demonstrate that the transient surface response is strongly influenced by the multilayer architecture of the tissue and therefore cannot be adequately described by the homogeneous half-space assumption commonly adopted in conventional elastography \citep{xue2023use}. Instead, the complete surface displacement field naturally captures the underlying multilayer wave dynamics, suggesting that transient displacement features may provide additional information for future inverse identification.

For the human skin models, increasing the thickness ratio $\zeta$ and modulus ratio $r$ enhances the contribution of wave type II while shifting the dominant wave feature associated with wave type IV. In particular, the transient waveform evolves from being dominated by wave type IV to wave type II as $r$ increases across the three skin models, demonstrating that higher-order modes can substantially influence the measured surface response and therefore cannot be neglected. 

Bearing in mind that the dimensionless time is set to $t_1=80$ in Sections \ref{gingiva} and \ref{skin}, we emphasize that the main features identified there persist as $t_1$ is increased. Although the wave becomes increasingly dispersed with time, the characteristic features shown in Figures \ref{fig:figure-x1-model-group2} and \ref{fig:figure-x1-model-group3} are preserved, and the fundamental mode alone is insufficient to capture the full wave characteristics. In particular, waves travelling at $c_g^\mathrm{s}$ do not necessarily coincide with the maximum wave peak or trough.

Overall, this study demonstrates that transient impact wave propagation in pre-stressed multilayer soft materials cannot, in general, be interpreted solely in terms of the fundamental mode. In addition, the use of measured dispersion curves to estimate soft-tissue elastic moduli has received considerable attention \citep{liou2019nondestructive}. For example, \citet{ma2024guided} used dispersion-based inversion with a linear elastic model to estimate the shear modulus of the skin layer. By showing how different modes contribute to the surface displacement under a prescribed excitation, the present forward model provides a basis for interpreting transient waveforms and selecting observables for future tissue characterization, including modal branches for dispersion-based inversion.

Finally, we emphasize that tissue characterization is fundamentally an inverse problem. Nevertheless, a clear and thorough understanding of the associated forward problem, such as the impact wave propagation investigated here, is essential. The proposed framework is expected to support the design of loading scenarios and the interpretation of experimentally observed wave signals in future studies. The present work is primarily theoretical and focuses on impact wave propagation in isotropic, purely elastic solids. Future work will also extend the framework to anisotropic and viscoelastic materials and investigate its applicability to more general transient loadings with prescribed frequency characteristics. Benchmark experiments will also be conducted to further assess the theoretical predictions and their relevance to transient-wave measurements in soft materials.

\bigskip
\section*
{Acknowledgment}\label{Acknowledgment} 
This work was supported by a grant from the National Natural Science Foundation of China (Project No. 12372072). 

\appendix
\renewcommand{\thefigure}{A.\arabic{figure}}
\setcounter{figure}{0}
\renewcommand{\thetable}{A.\arabic{table}}
\setcounter{table}{0}

\section{Dispersion relation}\label{Dispersion relation}

To determine the displacement fields of impact waves, accurate dispersion curves are required. Therefore, we derive in this appendix the dispersion relation governing wave propagation in a pre-stressed hyperelastic bilayer, where the lower layer is supported by a smooth rigid foundation. In our previous work, we presented a comprehensive analysis of wave propagation in compressible bilayers with free surfaces \citep{liu2026wave}. Here, we briefly summarize the derivation of the dispersion relation, highlighting only the key steps relevant to the present study. For consistency, we adopt the same notation as in the main text and outline only the general derivation applicable to both layers.

We look for a traveling wave solution of $\eqref{eq:gov-eq}_1$ as follows
\begin{equation}
\label{eq:tr-wave}
u_1=V_1(x_2)\sin{k(x_1-c_pt)},\quad
u_2=V_2(x_2)\cos{k(x_1-c_pt)},
\end{equation}
where $k$ denotes the wavenumber, $c_p$ represents the phase velocity, and $V_i(x_2)$ $(i=1,2)$ are unknown amplitude functions. These solutions describe waves propagating along the $x_1$ direction with identical wavenumber and phase velocity in both layers, thereby ensuring continuity conditions at the interface.

Substituting \eqref{eq:tr-wave} into \eqref{eq:gov-eq}$_1$ and retaining only the non-zero terms yields
\begin{equation}
\begin{aligned}\label{eq:ODE}
\left(\rho c_p^2-\mathcal{A}_{1111}\right)k^2V_1(x_2)+\left(\mathcal{A}_{1122}+\mathcal{A}_{2112}\right)kV_2'(x_2)+\mathcal{A}_{2121}V_1''(x_2)&=0,\\
\left(\rho c_p^2-\mathcal{A}_{1212}\right)k^2V_2(x_2)-\left(\mathcal{A}_{2211}+\mathcal{A}_{1221}\right)kV_1'(x_2)+\mathcal{A}_{2222}V_2''(x_2)&=0.
\end{aligned}
\end{equation}
Equation $\eqref{eq:ODE}_1$ is first used to express $V_2'(x_2)$ in terms of $V_1(x_2)$ and its derivatives. Differentiating $\eqref{eq:ODE}_2$ with respect to $x_2$ and inserting the resulting expression for $V_2'(x_2)$ provides the following fourth-order ordinary differential equation for $V_1(x_2)$:
\begin{align}\label{eq:ODE-4}
c_4V_1^{(4)}(x_2)+c_2k^2V_1''(x_2)+c_0k^4V_1(x_2)=0,
\end{align}
where the coefficients $c_4$, $c_2$, and $c_0$ are given by
\begin{equation}
\begin{aligned}\label{eq:c0-c4}
c_4&=\mathcal{A}_{2121}\mathcal{A}_{2222},\\
c_2&=\left(\mathcal{A}_{1122}+\mathcal{A}_{2112}\right)\left(\mathcal{A}_{2211}+\mathcal{A}_{1221}\right)+\mathcal{A}_{2121}\left(\rho c_p^2-\mathcal{A}_{1212}\right)+\mathcal{A}_{2222}\left(\rho c_p^2-\mathcal{A}_{1111}\right),\\
c_0&=\left(\rho c_p^2-\mathcal{A}_{1212}\right)\left(\rho c_p^2-\mathcal{A}_{1111}\right).
\end{aligned}
\end{equation}

Considering the general solution to \eqref{eq:ODE-4} in the form $V_1=e^{q k x_2}$, the corresponding characteristic equation yields four roots, given by $\pm q_1$ and $\pm q_2$, where
\begin{align}\label{eq:eigenvalues}
q_{1,2}=\sqrt{\frac{-c_2\pm\sqrt{c_2^2-4c_0c_4}}{2c_4}},
\end{align}
and `$+$' and `$-$' signs correspond to $q_1$ and $q_2$, respectively. The general solutions for $V_1(x_2)$ and $\hat{V}_1(x_2)$ can be expressed as
\begin{equation}
\begin{aligned}\label{eq:V1}
V_1(x_2)=C_1\cosh{(q_1kx_2)}+C_2\sinh{(q_1kx_2)}+C_3\cosh{(q_2kx_2)}+C_4\sinh{(q_2kx_2)},\\
\hat{V}_1(x_2)=C_5\cosh{(\hat{q}_1kx_2)}+C_6\sinh{(\hat{q}_1kx_2)}+C_7\cosh{(\hat{q}_2kx_2)}+C_8\sinh{(\hat{q}_2kx_2)},
\end{aligned}
\end{equation}
where $C_i$ $(i=1,\cdots,8)$ are arbitrary constants. The corresponding expressions for $V_2(x_2)$ and $\hat{V}_2(x_2)$ can be obtained from $\eqref{eq:ODE}_1$ and the counterpart equation for the lower layer, respectively.

We mention that the boundary conditions and continuity conditions are given by \eqref{eq:BC} by setting $g(x_1,t)=0$. These conditions lead to the following homogeneous system:
\begin{align}\label{eq:matrix}
\mathbf{H}\mathbf{C}=\mathbf{0},
\end{align}
where $\mathbf{C}=(C_1,C_2,C_3,C_4,C_5,C_6,C_7,C_8)^{\mathrm{T}}$ and the matrix $\mathbf{H}$ depends on the material and geometric parameters. For a non-trivial solution to exist, the coefficient matrix $\mathbf{H}$ must be singular. This condition yields
\begin{equation}\label{eq:dispersion}
\det\mathbf{H}=0,
\end{equation}
which defines the dispersion relation and determines the dispersion curves. In the following derivation, the frequency $\omega$ and group velocity $c_g$ are defined as
\begin{equation}\label{eq:omega}
\omega=kc_p,\quad c_g=\frac{\partial\omega}{\partial k}.
\end{equation}

The dispersion relation describes the relationship between the frequency $\omega$ and wavenumber $k$. Taking $\lambda_1=1.2$, $\nu=0.4$, $r=2$ and $\zeta=1$, we present the dispersion curves in Figure \ref{fig:figure-omega1-k1-dispersion-relation-1}, where all quantities are nondimensionalized.

\section{Sensitivity of the FE simulations to mesh size, time step, and boundary conditions}\label{Sensitivity}

\renewcommand{\thefigure}{B\arabic{figure}}
\setcounter{figure}{0}
\renewcommand{\thetable}{B\arabic{table}}
\setcounter{table}{0}

In this appendix, we examine the sensitivity of the FE results to the mesh size and time step, as well as the effects of the boundary conditions.

\begin{figure}[H]
     \centering
      {\subfigure[]{\includegraphics[width=0.49\textwidth]{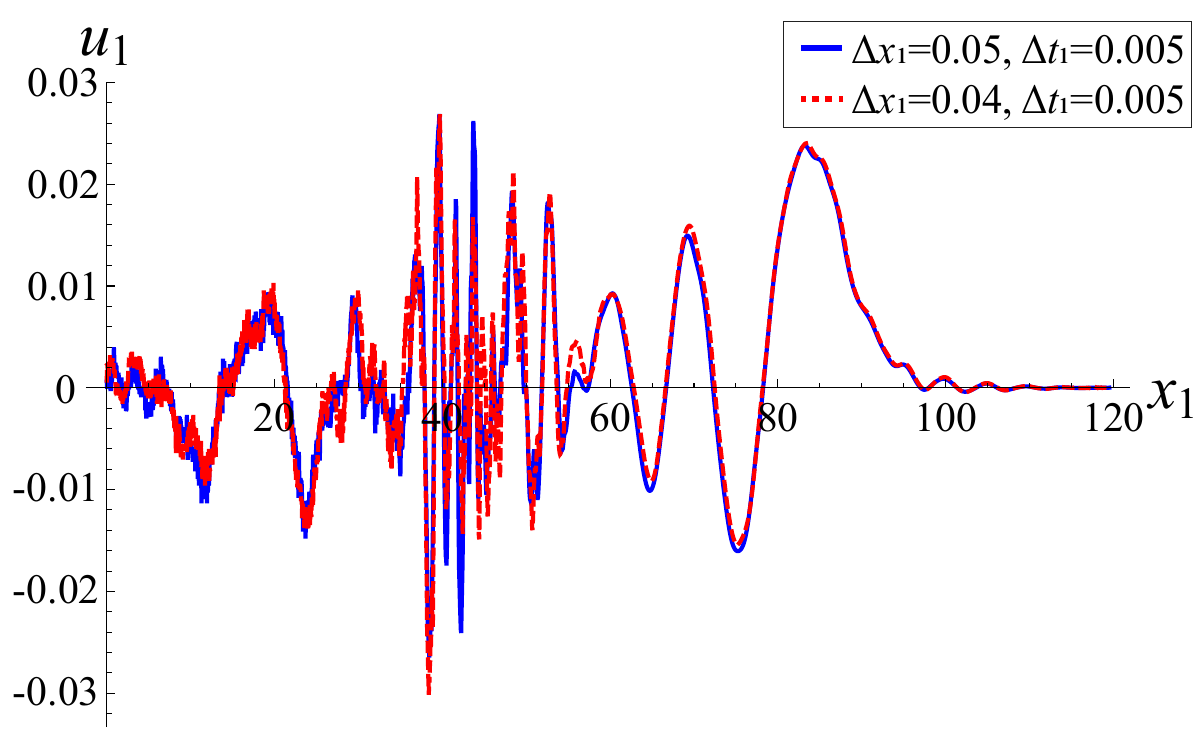}\label{figure-FEM-refined-mesh-u1}}\subfigure[]{\includegraphics[width=0.49\textwidth]{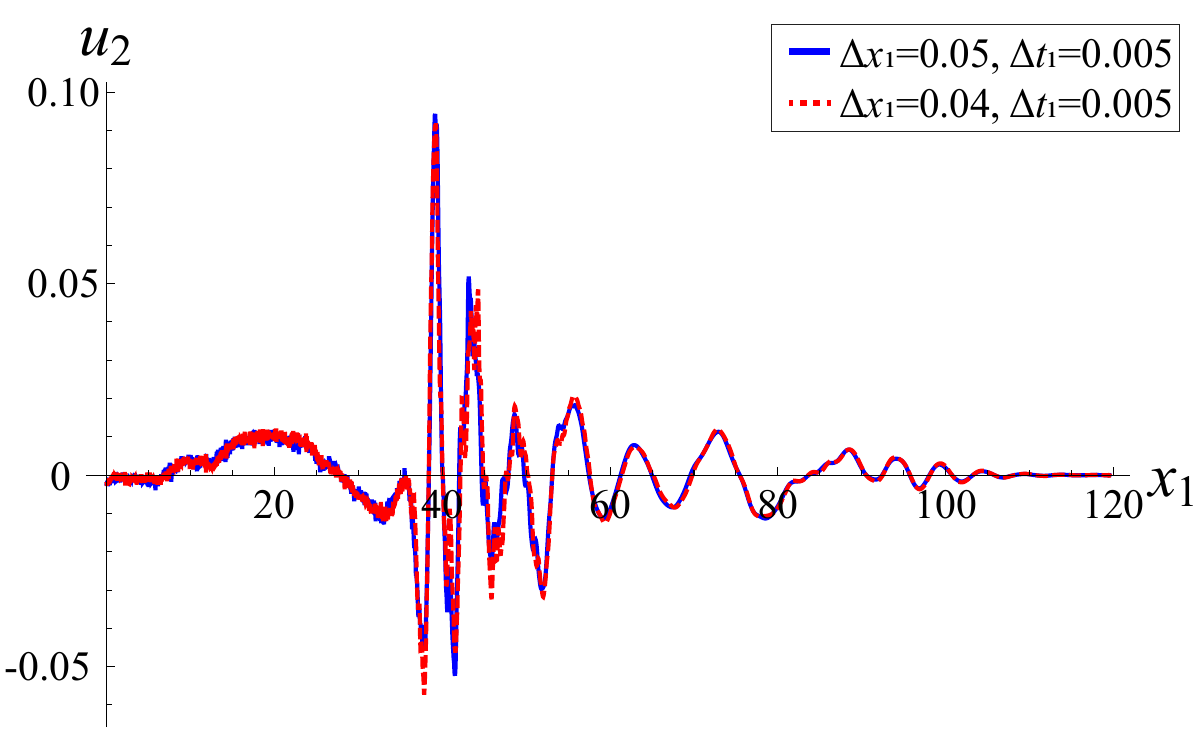}\label{figure-FEM-refined-mesh-u2}}}
      {\subfigure[]{\includegraphics[width=0.49\textwidth]{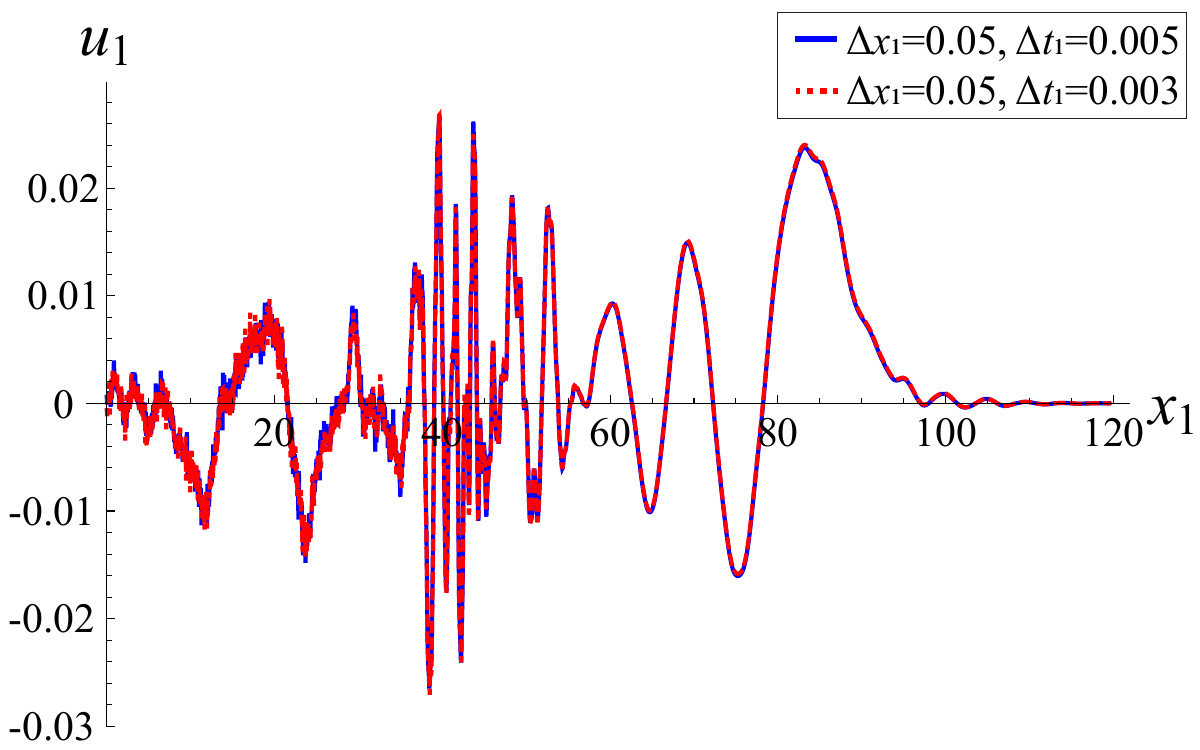}\label{figure-FEM-refined-step-u1}}\subfigure[]{\includegraphics[width=0.49\textwidth]{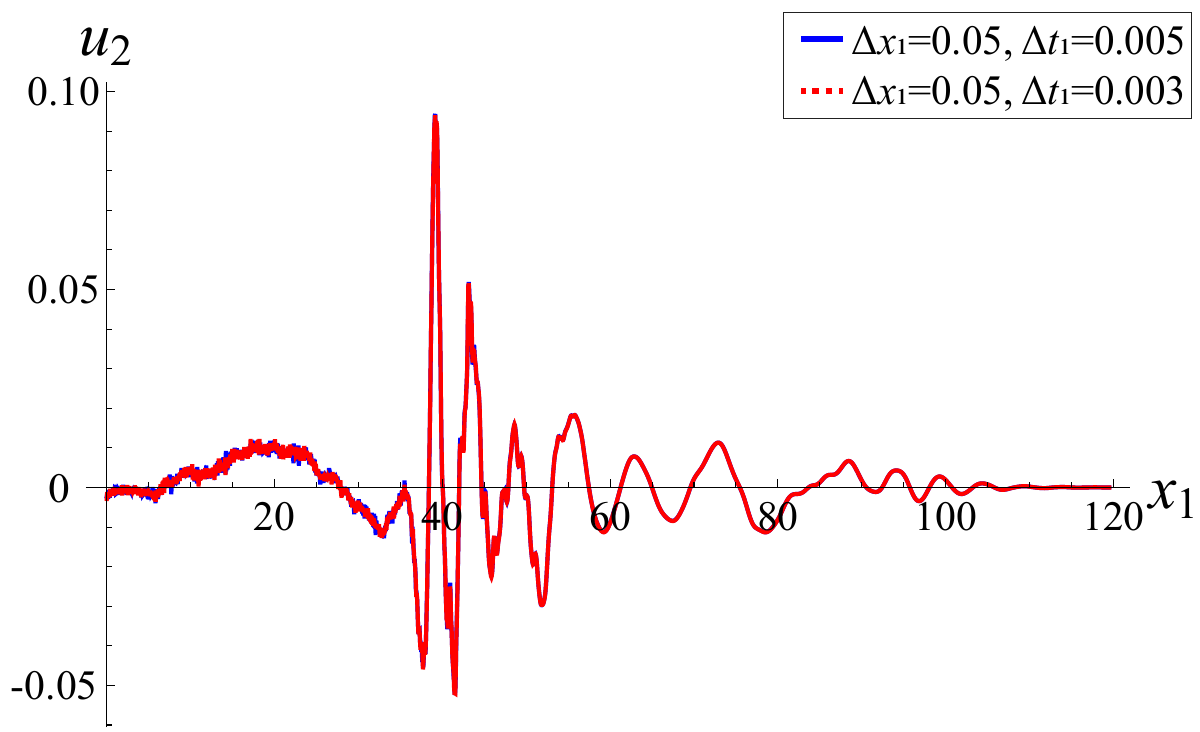}\label{figure-FEM-refined-step-u2}}}\caption{Comparison of the displacement distributions $u_1$ and $u_2$ on the upper surface obtained from FE simulation at $t_1=42.17$. The parameters are $\lambda_1=1.2$, $\nu=0.4$, $r=2$, and $\zeta=1$. Periodic boundary conditions are imposed in the simulations.}\label{fig:figure-FEM-refined-mesh}
\end{figure}

\begin{figure}[H]
     \centering
      {\subfigure[]{\includegraphics[width=0.49\textwidth]{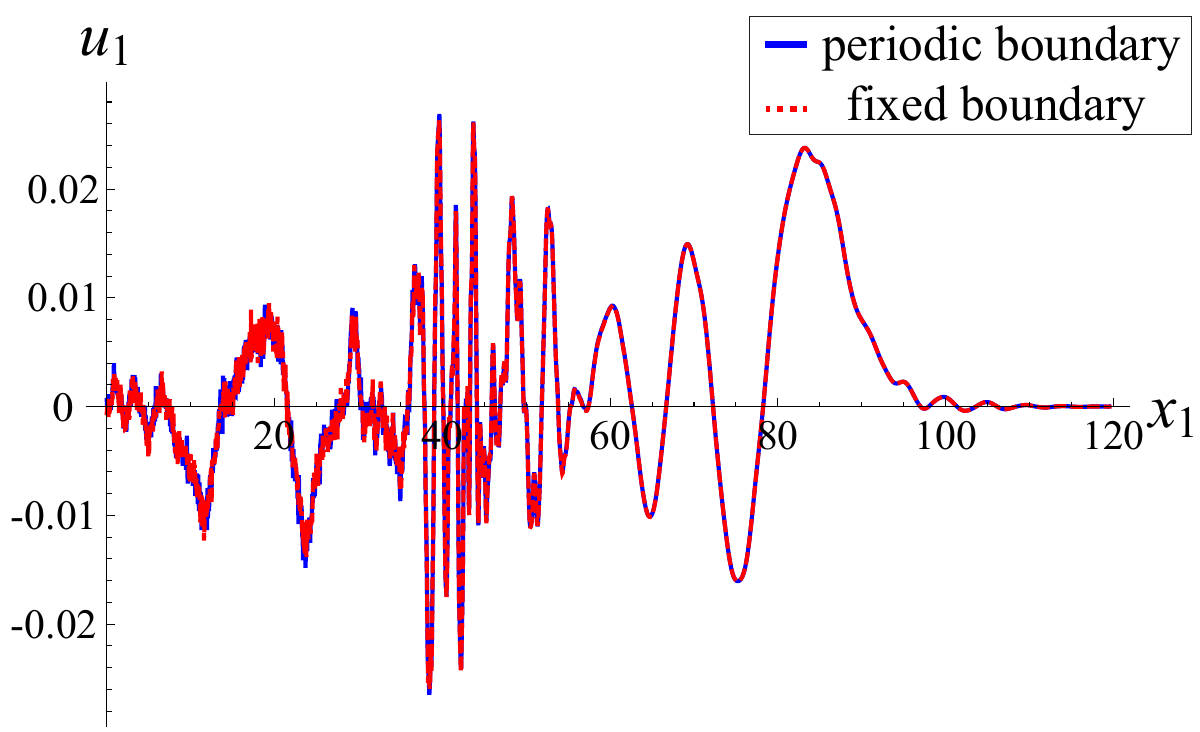}\label{figure-FEM-fixed-boundary-u1}}\subfigure[]{\includegraphics[width=0.49\textwidth]{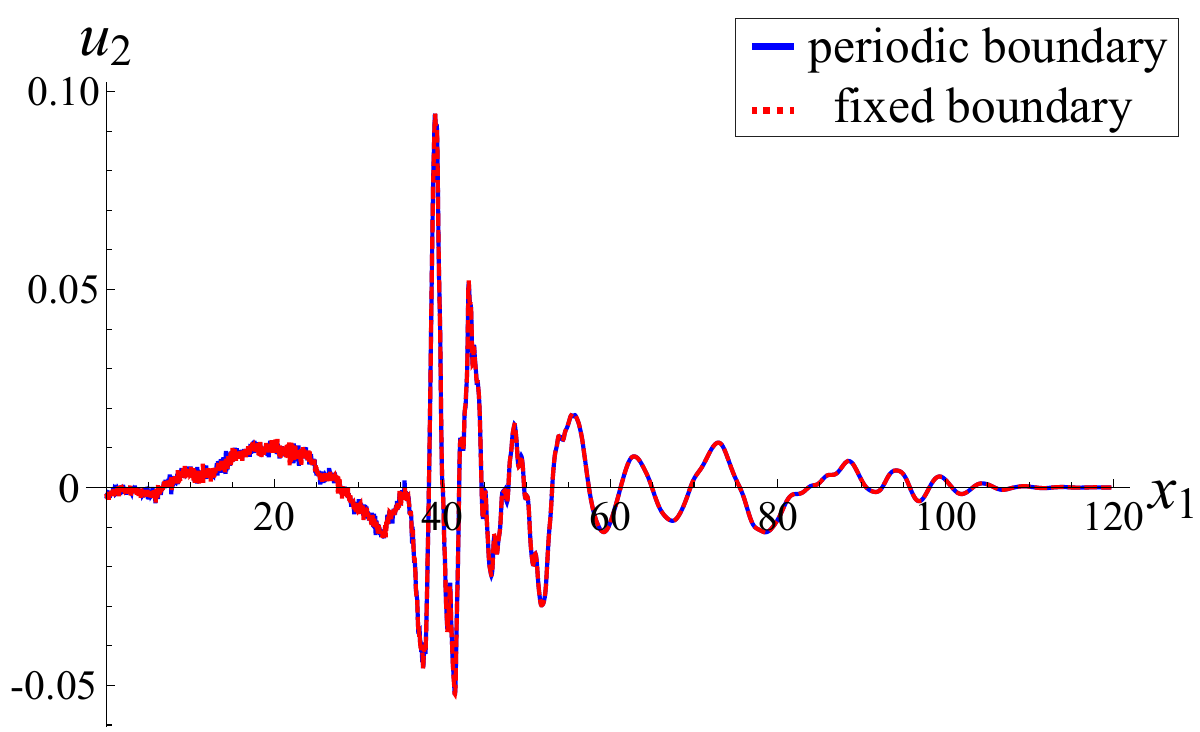}\label{figure-FEM-fixed-boundary-u2}}}\caption{Comparison of the displacement distributions $u_1$ and $u_2$ on the upper surface obtained from FE simulations at $t_1=42.17$. The parameters are $\lambda_1=1.2$, $\nu=0.4$, $r=2$, and $\zeta=1$. The simulations employ the same mesh size ($\Delta x=0.05$) and time step ($\Delta t=0.005$), but different boundary conditions: periodic boundary conditions (blue solid lines) and fixed boundary conditions (red dashed lines).}\label{fig:figure-FEM-fixed-boundary}
\end{figure}

\section{Group velocities used in Section \ref{Applications to soft tissues}}\label{Appendix-B}


\renewcommand{\thefigure}{C\arabic{figure}}
\setcounter{figure}{0}
\renewcommand{\thetable}{C\arabic{table}}

In this appendix, we illustrate the curves of the group velocity $c_g$ used in defining the four wave types in Section \ref{Applications to soft tissues}.

\begin{figure}[H]
        \centering
        \subfigure[]{\includegraphics[width=0.3\linewidth]{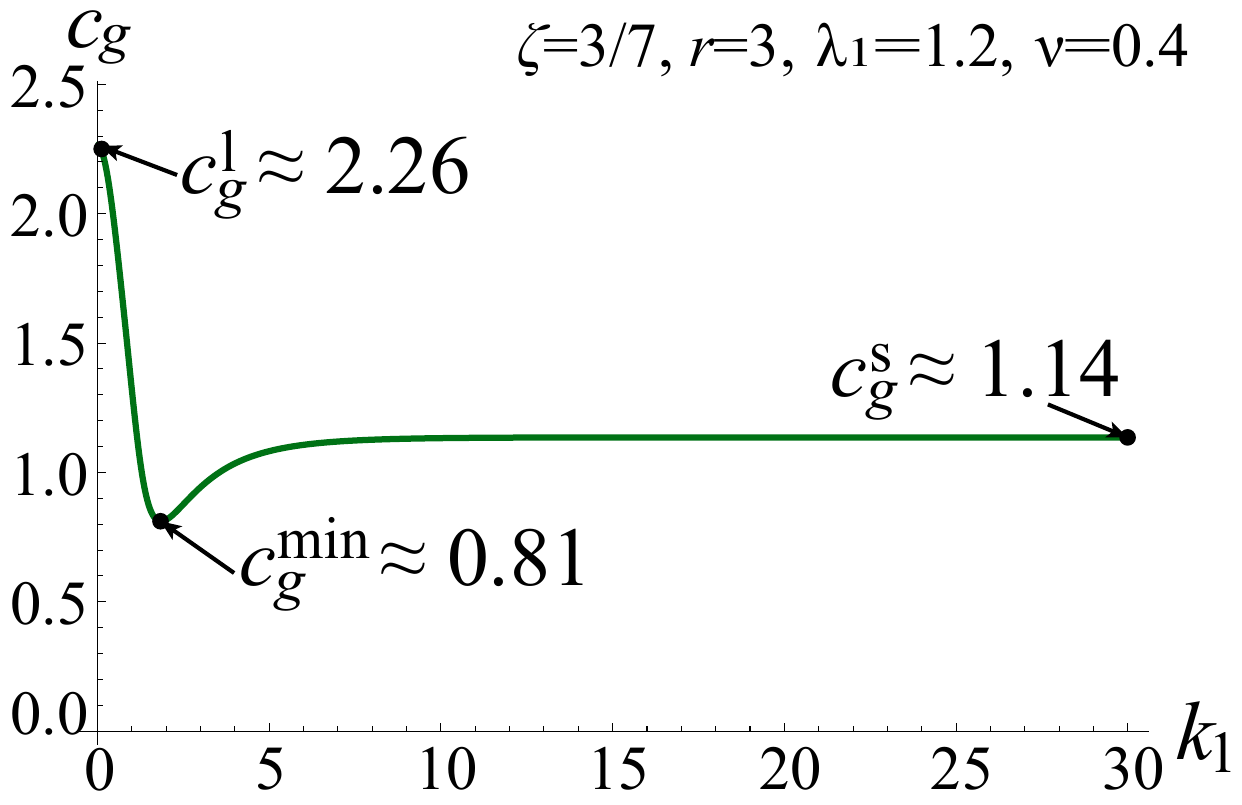}}\label{fig-cg-k1-model-group1-1}\quad\subfigure[]{\includegraphics[width=0.3\linewidth]{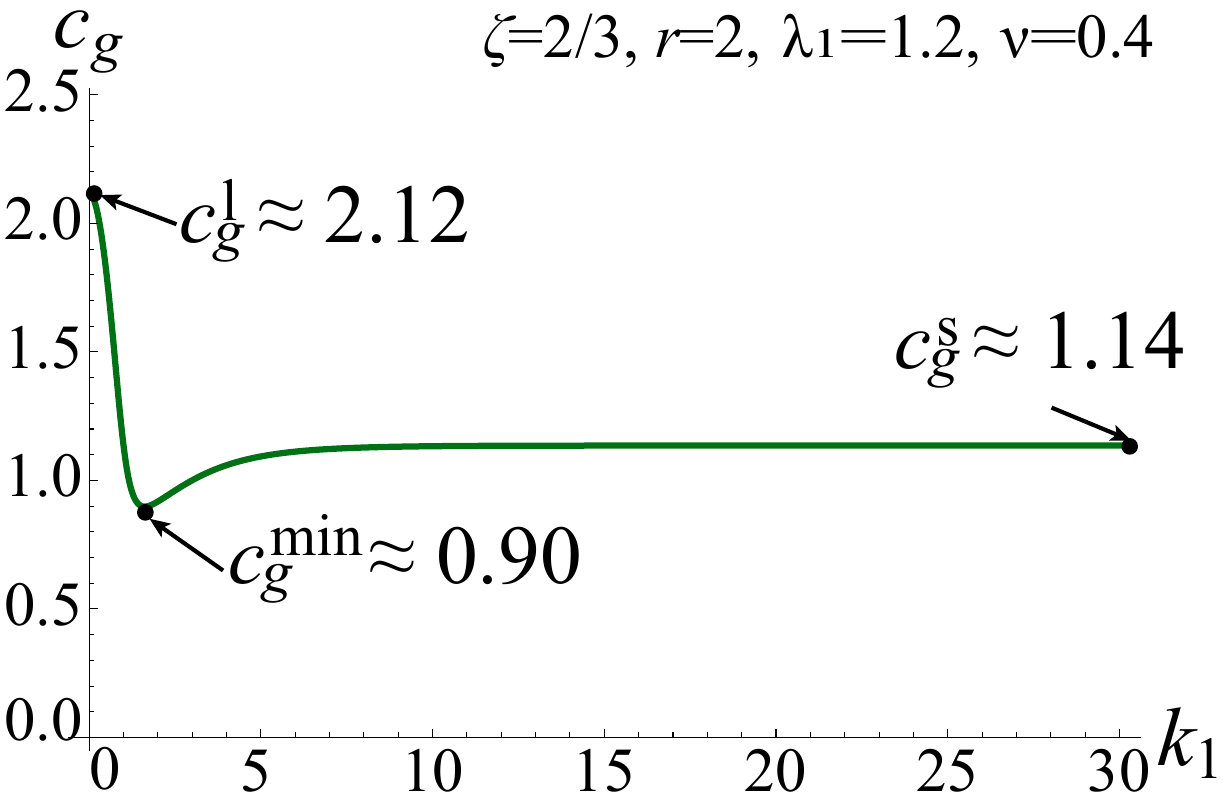}}\label{fig-cg-k1-model-group1-2}\quad\subfigure[]{\includegraphics[width=0.3\linewidth]{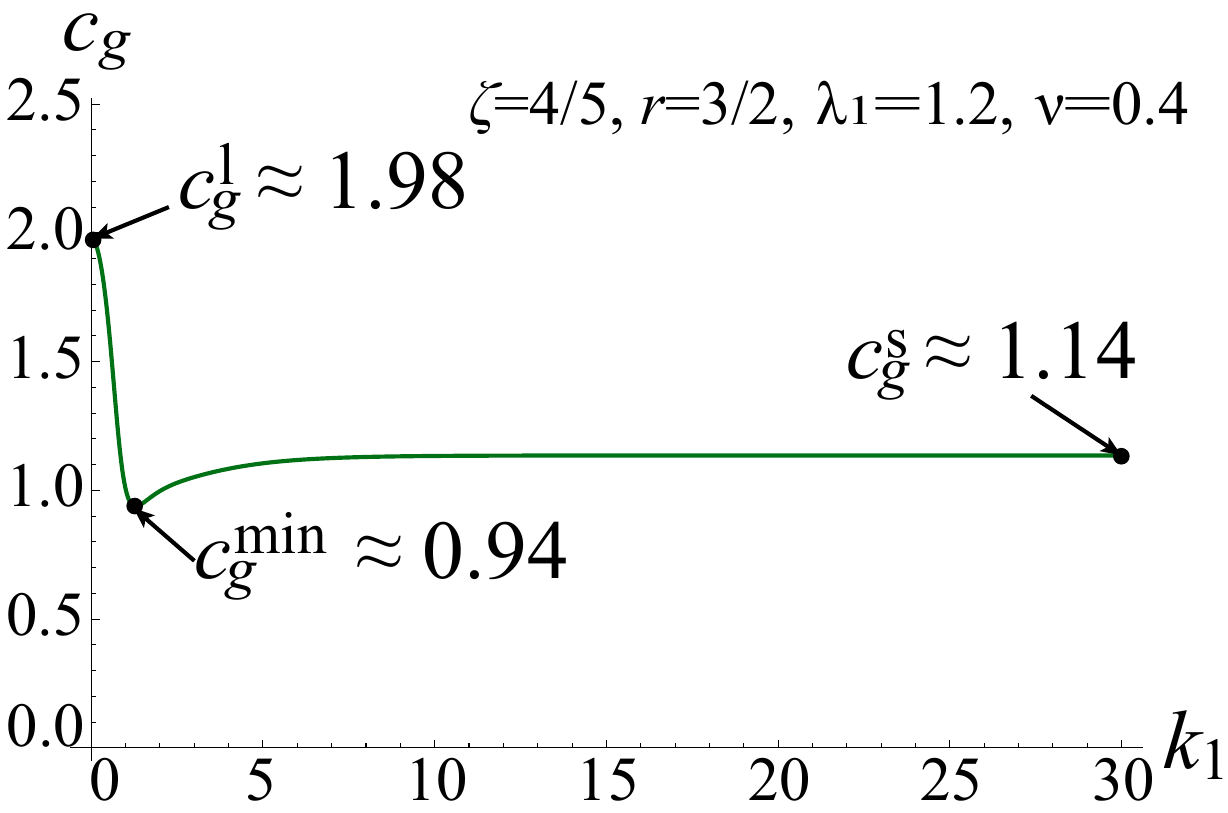}}\label{fig-cg-k1-model-group1-3}\caption{Group velocity curves of the fundamental mode for the three articular cartilage models, with the main parameters and characteristic group velocities annotated. In each panel, the group velocities $c_g^{\mathrm{l}}$, $c_g^{\mathrm{s}}$, and $c_g^{\mathrm{min}}$ correspond to the long-wavelength limit, short-wavelength limit, and minimum group velocity, respectively.}\label{fig:cg-k1-model1}
\end{figure}

\begin{figure}[H]
        \centering
        \subfigure[]{\includegraphics[width=0.3\linewidth]{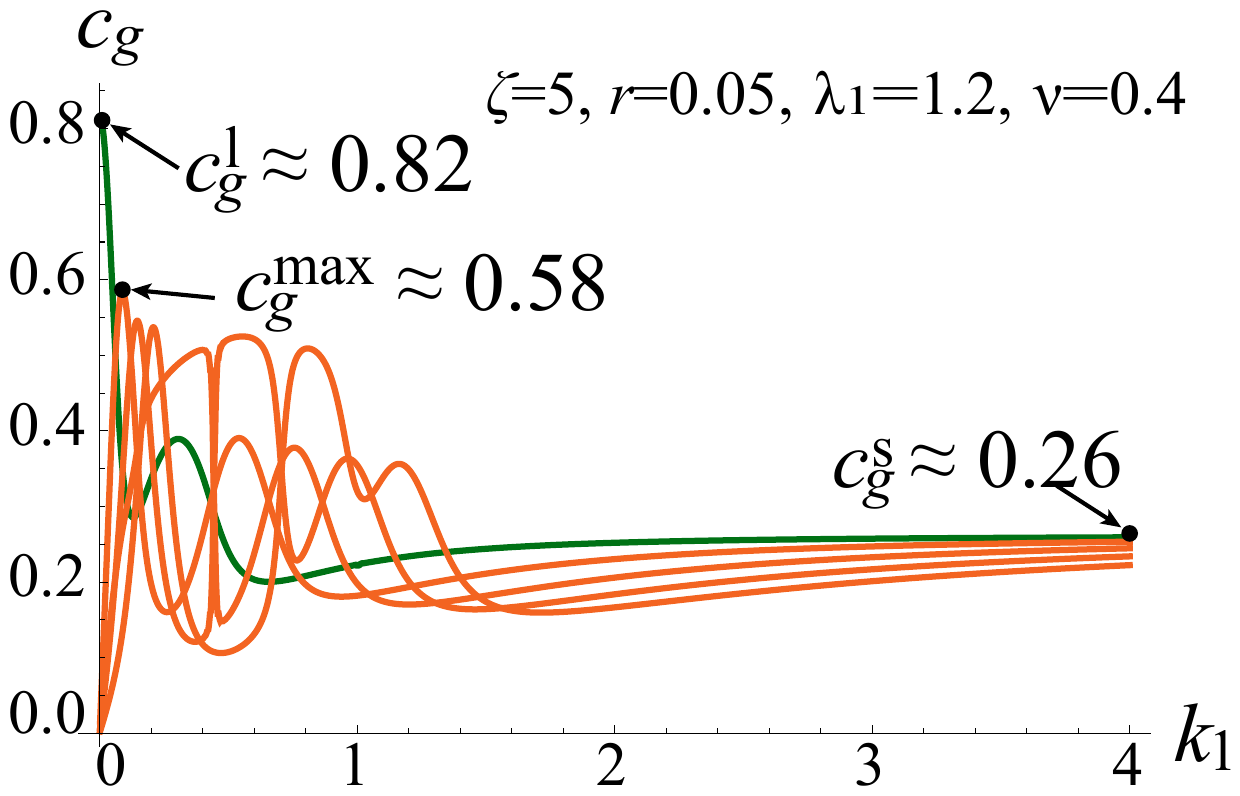}}\label{fig-cg-k1-model-group2-1}\quad\subfigure[]{\includegraphics[width=0.3\linewidth]{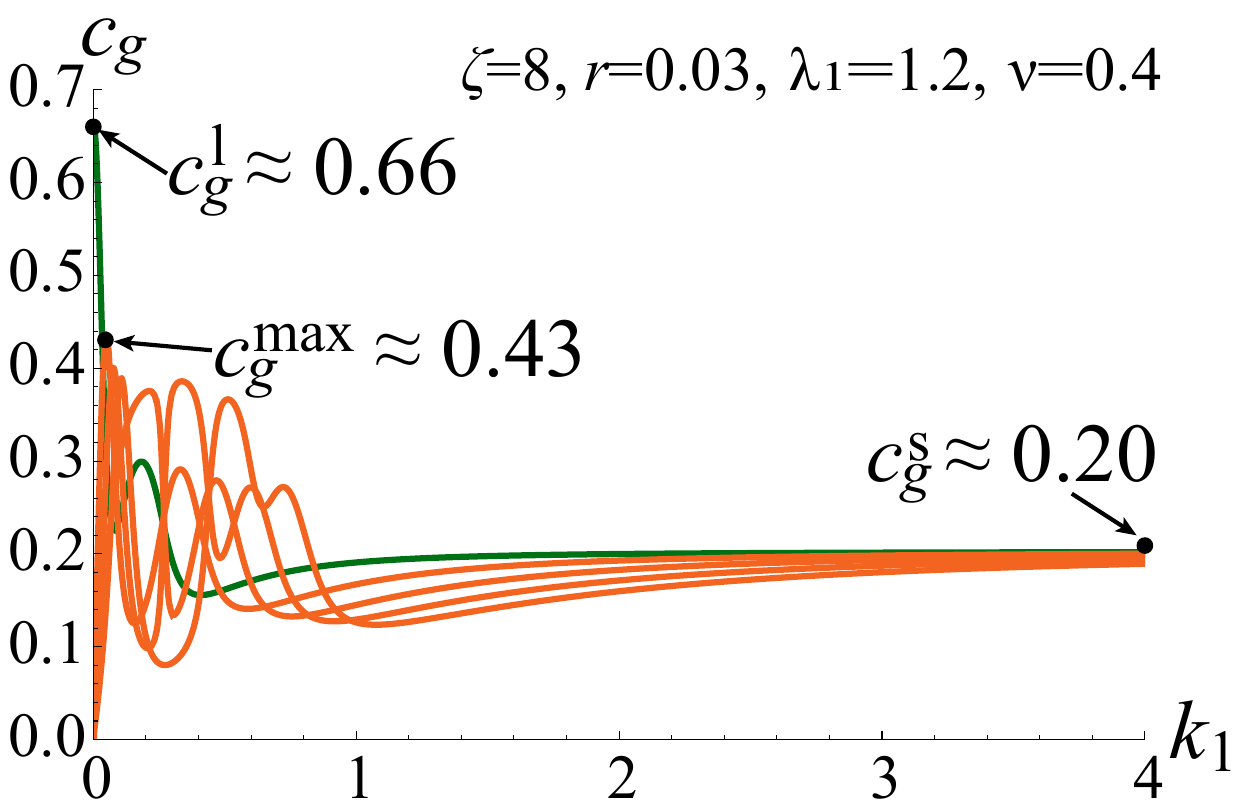}}\label{fig-cg-k1-model-group2-2}\quad\subfigure[]{\includegraphics[width=0.3\linewidth]{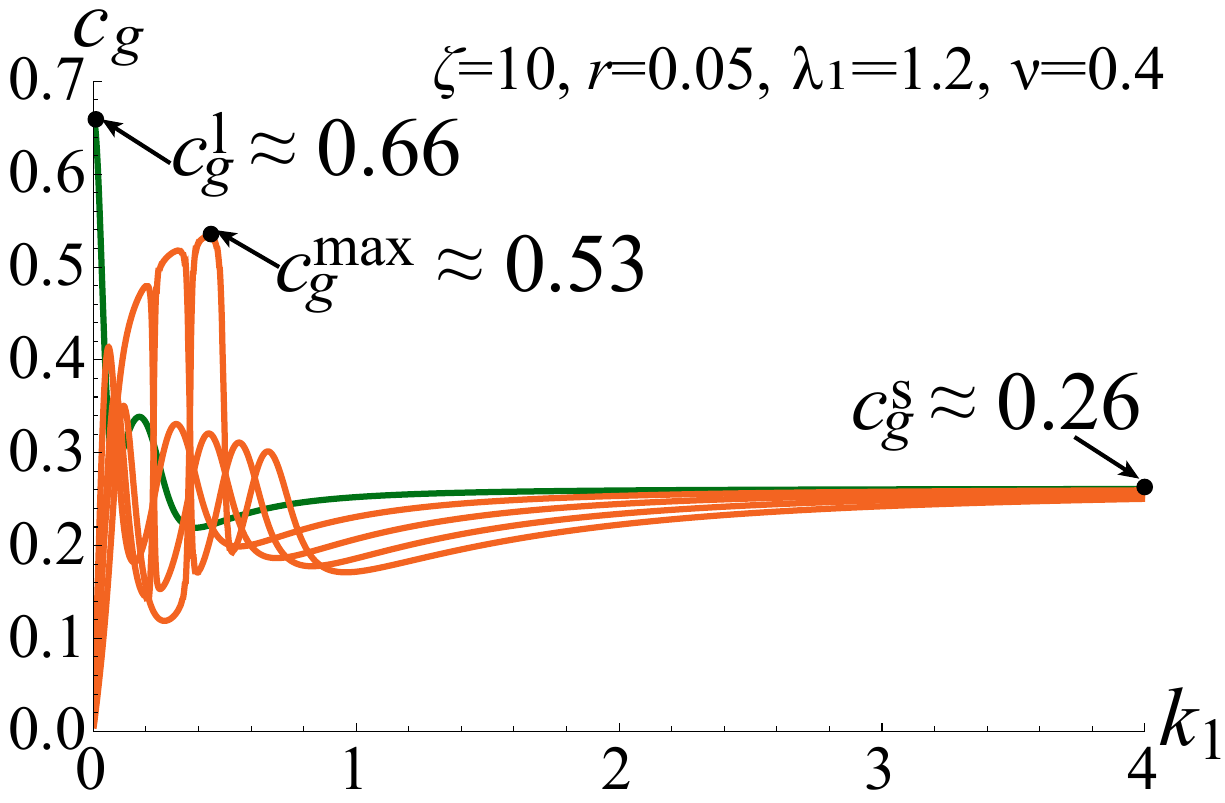}}\label{fig-cg-k1-model-group2-3}\caption{Group velocity curves of the first five modes for the three gingiva models, with the main parameters and characteristic group velocities annotated. In each panel, the group velocities $c_g^{\mathrm{l}}$, $c_g^{\mathrm{s}}$, and $c_g^{\mathrm{max}}$ correspond to the long-wavelength limit, short-wavelength limit, and local maximum group velocity, respectively.}\label{fig:cg-k1-model2}
\end{figure}

\begin{figure}[H]
        \centering
        \subfigure[]{\includegraphics[width=0.3\linewidth]{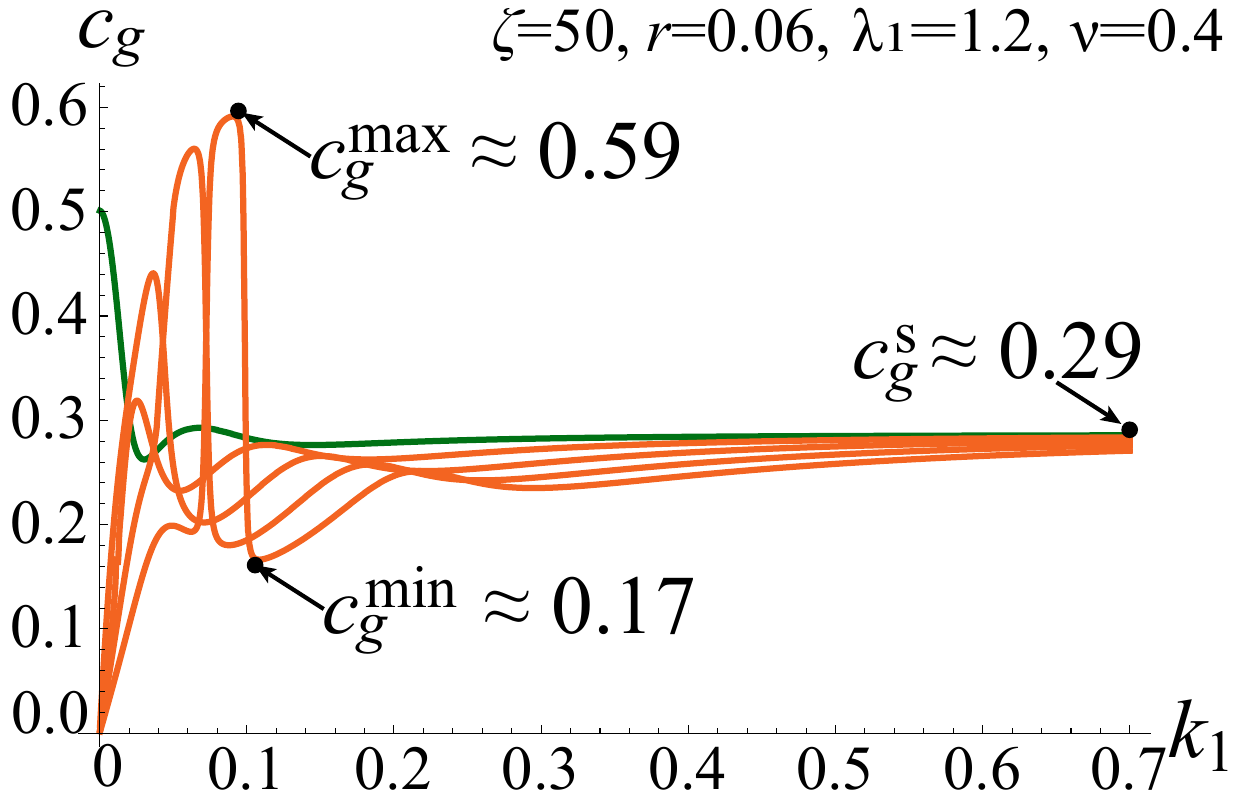}}\label{fig-cg-k1-model-group3-1}\quad\subfigure[]{\includegraphics[width=0.3\linewidth]{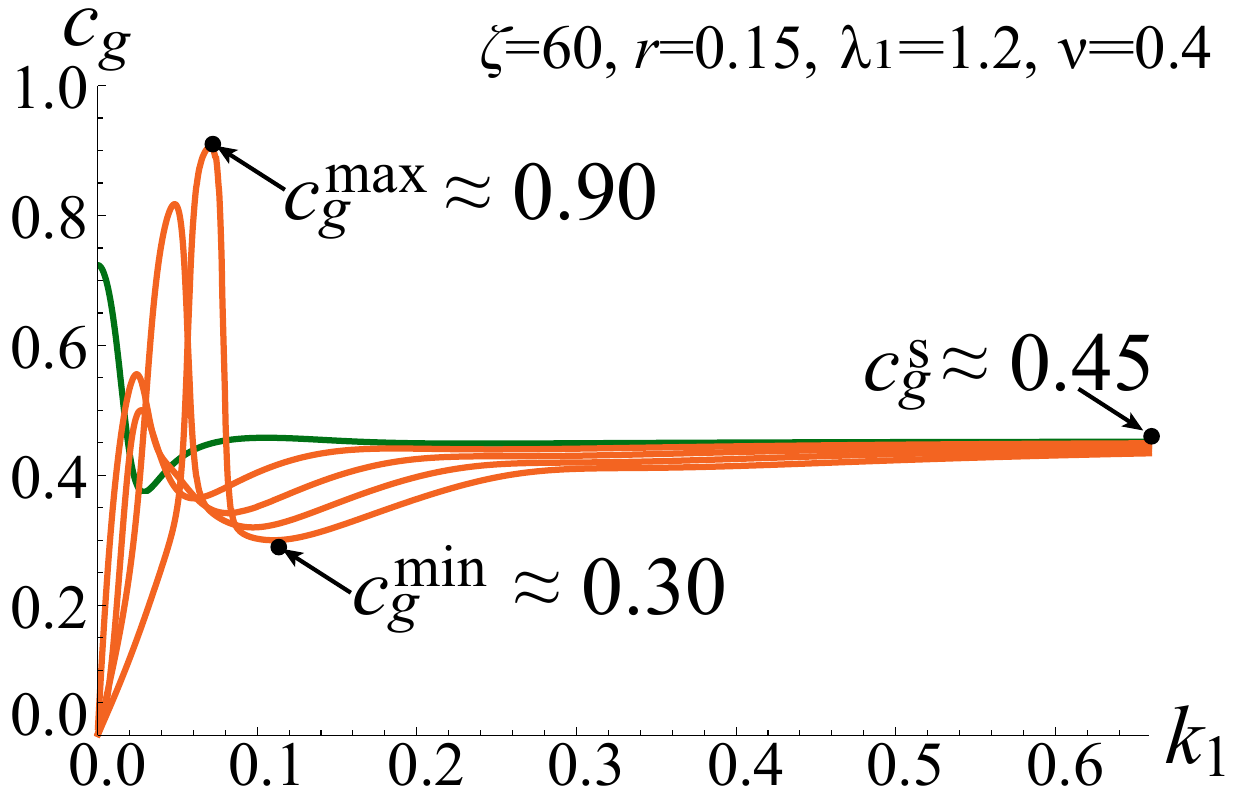}}\label{fig-cg-k1-model-group3-2}\quad\subfigure[]{\includegraphics[width=0.3\linewidth]{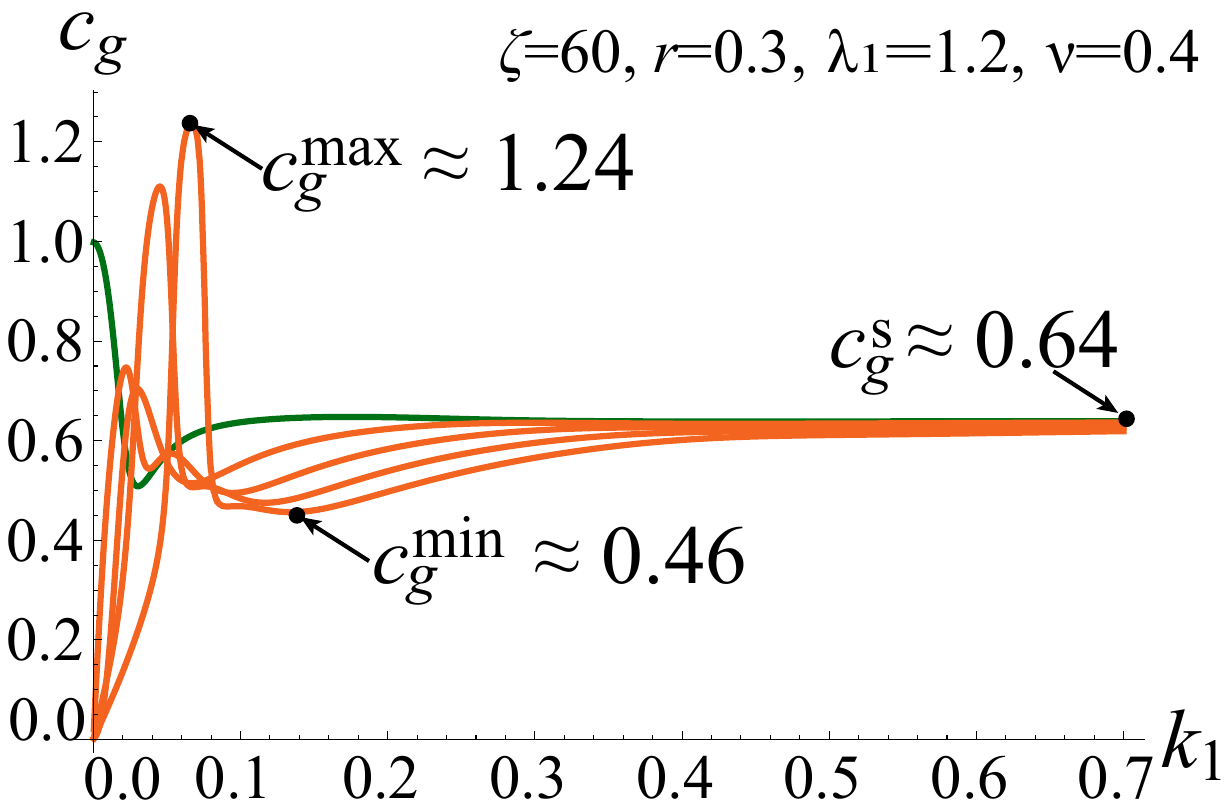}}\label{fig-cg-k1-model-group3-3}\caption{Group velocity curves of the first five modes for the three skin models, with the main parameters and characteristic group velocities annotated. In each panel, the group velocities $c_g^{\mathrm{s}}$, $c_g^{\mathrm{max}}$, and $c_g^{\mathrm{min}}$ correspond to the short-wavelength limit, maximum group velocity, and local minimum group velocity, respectively.}\label{fig:cg-k1-model3}
\end{figure}

\section{Convergence check}\label{Convergence check}

\renewcommand{\thefigure}{D\arabic{figure}}
\setcounter{figure}{0}
\renewcommand{\thetable}{D\arabic{table}}
\setcounter{table}{0}

\begin{figure}[htbp]
     \centering
      {\subfigure[]{\includegraphics[width=0.45\textwidth]{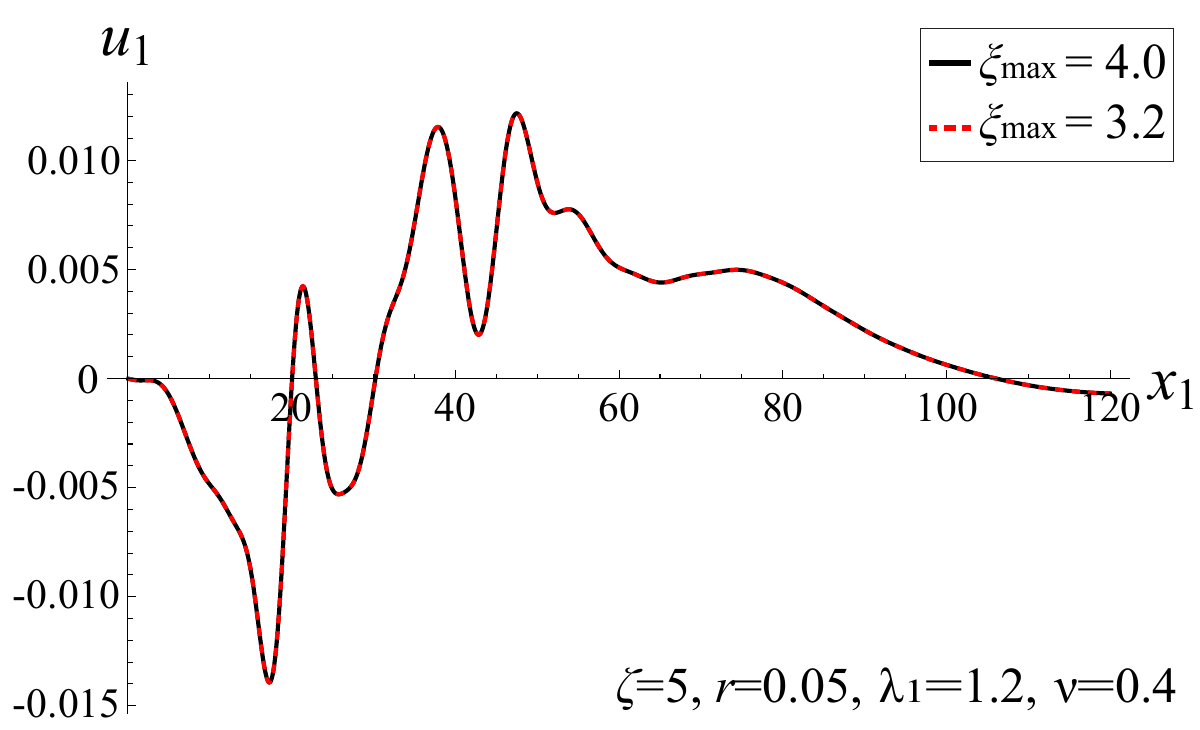}\label{figure-u1-upper-limit-group2-1}}\subfigure[]{\includegraphics[width=0.45\textwidth]{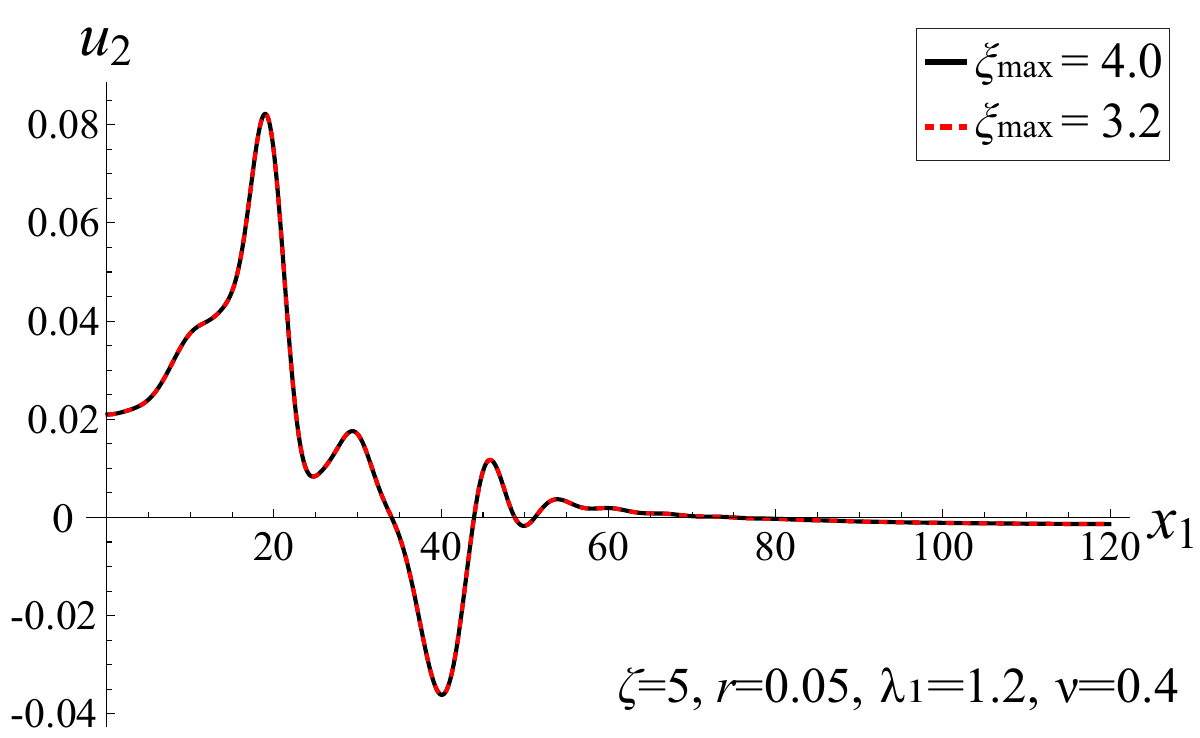}\label{figure-u2-upper-limit-group2-1}}\\\subfigure[]{\includegraphics[width=0.45\textwidth]{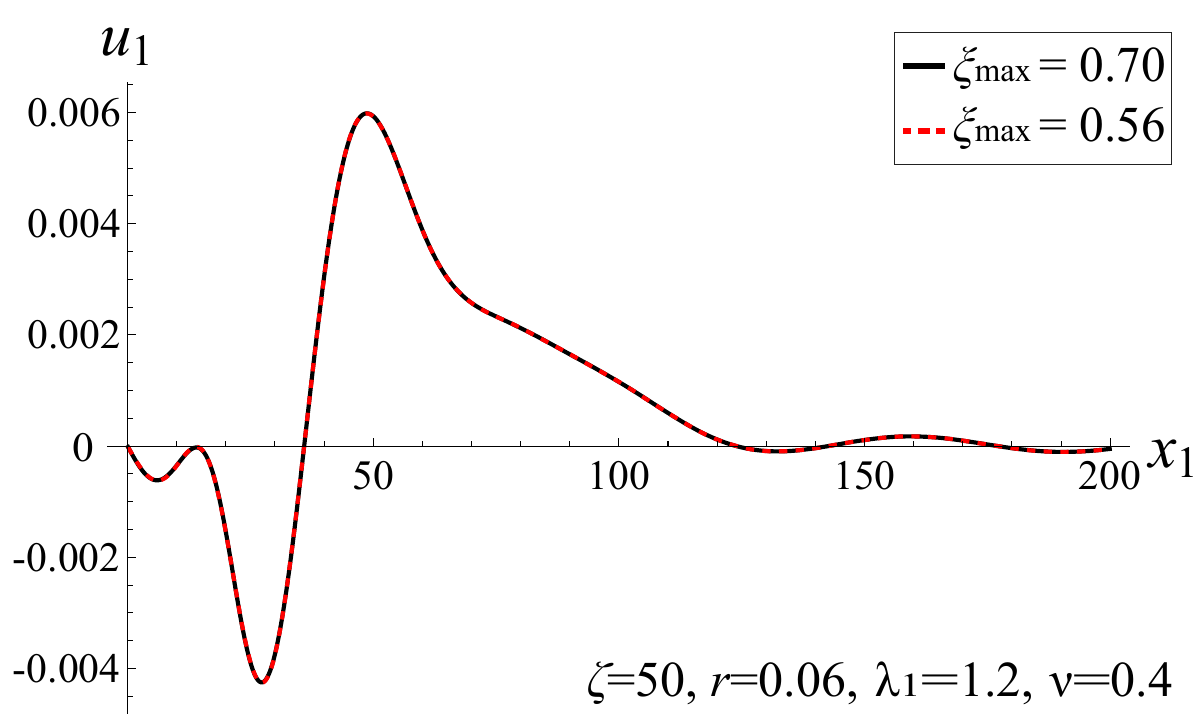}\label{figure-u1-upper-limit-group3-1}}\subfigure[]{\includegraphics[width=0.45\textwidth]{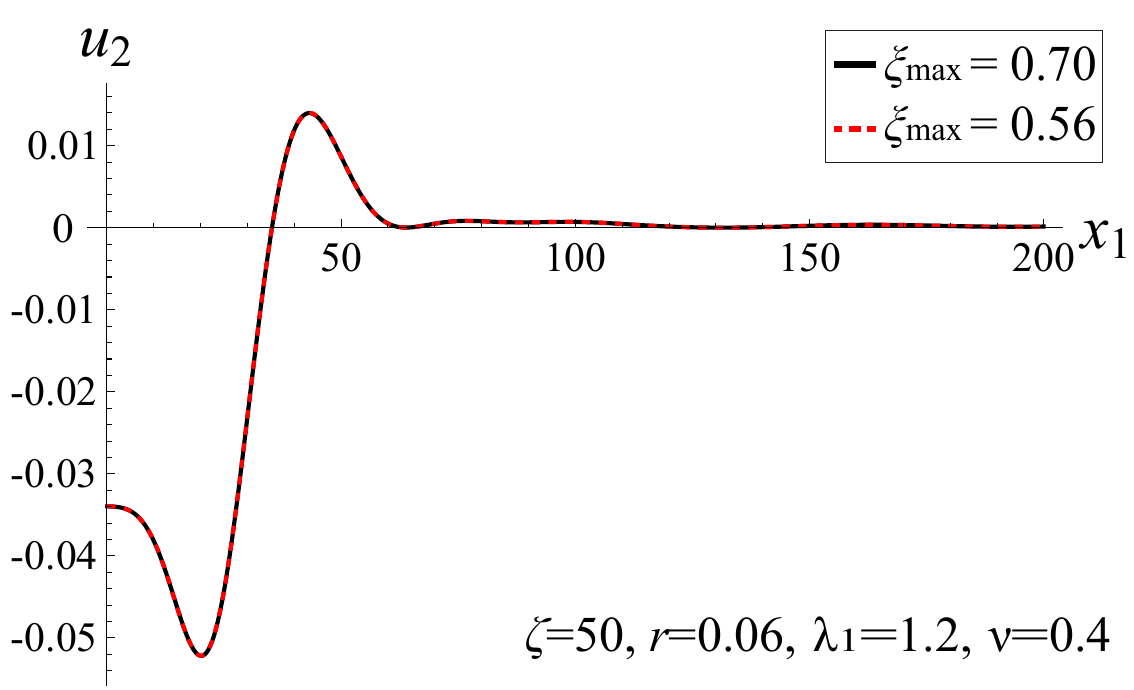}\label{figure-u2-upper-limit-group3-1}}}\caption{Comparison of the displacement distributions $u_1$ and $u_2$ on the upper surface at $t_1=80$ for the gingiva and skin models. The parameter values are specified in each panel.}\label{figure-upper-limit-error-group}
\end{figure}

According to equation \eqref{eq:NRMSE-formulation}, the NRMSE between the solutions obtained using four and five modes is calculated over the interval $(0,120)$ with $K=1500$ discretization points, and the results are reported in Table \ref{tab:NRMSE-of-mode-number}. We further illustrate the convergence with respect to the upper integration limits for the gingiva and skin models in Figure \ref{figure-upper-limit-error-group}.

\begin{table}[h]
\centering
\caption{NRMSEs of the displacement fields obtained using four and five modes.}
\vspace{3mm}
\label{tab:NRMSE-of-mode-number}
\begin{tabular}{lcc}
\hline
Model & NRMSE of $u_1$ (\%) & NRMSE of $u_2$ (\%)\\
\hline
Gingiva model used in Figures \ref{fig:figure-u1-x1-model-group2-1} and \ref{fig:figure-u2-x1-model-group2-1} & 0.88 & 0.67\\
Skin model used in Figures \ref{fig:figure-u1-x1-model-group3-1} and \ref{fig:figure-u2-x1-model-group3-1} & 0.58 & 0.63\\
\hline
\end{tabular}
\end{table}

\bibliographystyle{elsarticle-harv}
\bibliography{references}

\end{document}